\def\year{2023}\relax
\documentclass[letterpaper]{article} % DO NOT CHANGE THIS

\usepackage{aaai22}  % DO NOT CHANGE THIS
\usepackage{times}  % DO NOT CHANGE THIS
\usepackage{helvet}  % DO NOT CHANGE THIS
\usepackage{courier}  % DO NOT CHANGE THIS
\usepackage[hyphens]{url}  % DO NOT CHANGE THIS
\usepackage{graphicx} % DO NOT CHANGE THIS
\usepackage{natbib}  % DO NOT CHANGE THIS AND DO NOT ADD ANY OPTIONS TO IT
\usepackage{caption} % DO NOT CHANGE THIS AND DO NOT ADD ANY OPTIONS TO IT
\DeclareCaptionStyle{ruled}{labelfont=normalfont,labelsep=colon,strut=off} % DO NOT CHANGE THIS
\usepackage{algorithm}
\usepackage{algorithmic}
\usepackage{newfloat}
\usepackage{listings}
\floatstyle{ruled}
\newfloat{listing}{tb}{lst}{}
\floatname{listing}{Listing}
\usepackage[dvipsnames]{xcolor}
\usepackage{multirow}
\usepackage{booktabs}
\usepackage{amsfonts}
\usepackage{amsmath}
\usepackage[capitalize]{cleveref}

\newcommand{\header}[1]{{\noindent{\textbf{#1}}}}
\definecolor{ruby}{rgb}{0.6, 0.1, 0.3}
\definecolor{navy}{rgb}{0.1, 0.1, 0.8}
\definecolor{olive}{rgb}{0.1, 0.6, 0.2}

\usepackage[colorinlistoftodos,prependcaption,textsize=tiny]{todonotes}
\usepackage{xargs}
\newcommandx{\unsure}[2][1=]{\todo[linecolor=red,backgroundcolor=red!25,bordercolor=red,#1]{#2}}
\newcommandx{\change}[2][1=]{\todo[linecolor=blue,backgroundcolor=blue!25,bordercolor=blue,#1]{#2}}
\newcommandx{\infoneeded}[2][1=]{\todo[linecolor=OliveGreen,backgroundcolor=OliveGreen!25,bordercolor=OliveGreen,#1]{#2}}
\newcommandx{\lxnote}[2][1=]{\todo[linecolor=ruby,backgroundcolor=ruby!25,bordercolor=ruby,#1]{#2}}
\newcommandx{\atnote}[2][1=]{\todo[linecolor=Blue,backgroundcolor=Blue!25,bordercolor=Blue,#1]{#2}}
\newcommandx{\cvknote}[2][1=]{\todo[linecolor=Green,backgroundcolor=Green!25,bordercolor=Green,#1]{#2}}
\newcommandx{\joshnote}[2][1=]{\todo[linecolor=Yellow,backgroundcolor=Green!25,bordercolor=Green,#1]{#2}}
\newcommandx{\thiswillnotshow}[2][1=]{\todo[disable,#1]{#2}}

\newcommand{\foundation}[1]{\textit{#1}}

\usepackage{xspace}
\newcommand{\yta}{\texttt{YTA}\xspace}
\newcommand{\nta}{\texttt{NTA}\xspace}
\newcommand{\esh}{\texttt{ESH}\xspace}
\newcommand{\nah}{\texttt{NAH}\xspace}
\newcommand{\info}{\texttt{INFO}\xspace}
\newcommand{\NA}{\texttt{NA}\xspace}
\newcommand{\YA}{\texttt{YA}\xspace}

\usepackage{multibib}
\usepackage[autostyle]{csquotes}  

\newcommand{\topic}[1]{\textit{#1}}
\newcommand{\example}[1]{\textrm{#1}}

\usepackage{ctable}
\usepackage{booktabs} % For formal tables
\definecolor{twitterbg}{rgb}{0.39215686274509803, 0.5843137254901961, 0.9294117647058824}
\definecolor{newsbg}{rgb}{0.803921568627451, 0.3607843137254902, 0.3607843137254902}
\definecolor{redditbg}{rgb}{1.0, 0.5490196078431373, 0.0}

\newcolumntype{M}[1]{>{\arraybackslash}m{#1}}
\newcolumntype{P}[1]{>{\centering\arraybackslash}p{#1}}
\usepackage{color,colortbl}
\usepackage{tabularray}

\newcommand{\ourmodel}{{Mformer}\xspace}

\usepackage{subcaption}
\usepackage{lscape}

\newcommand{\answerYes}[1]{\textcolor{blue}{#1}} 
\newcommand{\answerNo}[1]{\textcolor{teal}{#1}} 
\newcommand{\answerNA}[1]{\textcolor{gray}{#1}}

\newcites{AP}{Appendix References}

\title{Measuring Moral Dimensions in Social Media with \ourmodel}
\author{
    Tuan Dung Nguyen, 
    Ziyu Chen,
    Nicholas George Carroll, 
    Alasdair Tran,
    Colin Klein, 
    Lexing Xie \\
}

\affiliations{
Australian National University\\
\{josh.nguyen, ziyu.chen, nicholas.carroll, alasdair.tran, colin.klein, lexing.xie\}@anu.edu.au
}

\begin{document}

\maketitle

\begin{abstract}
    The ever-growing textual records of contemporary social issues, often discussed online with moral rhetoric, present both an opportunity and a challenge for studying how moral concerns are debated in real life.
    Moral foundations theory is a taxonomy of intuitions widely used in data-driven analyses of online content, but current computational tools to detect moral foundations suffer from the incompleteness and fragility of their lexicons and from poor generalization across data domains.
    In this paper, we fine-tune a large language model to measure moral foundations in text based on datasets covering news media and long- and short-form online discussions.
    The resulting model, called \ourmodel, outperforms existing approaches on the same domains by 4--12\% in AUC and further generalizes well to four commonly used moral text datasets, improving by up to 17\% in AUC.
    We present case studies using \ourmodel to analyze everyday moral dilemmas on Reddit and controversies on Twitter, showing that moral foundations can meaningfully describe people's stance on social issues and such variations are topic-dependent.
    Pre-trained model and datasets are released publicly.
    We posit that \ourmodel will help the research community quantify moral dimensions for a range of tasks and data domains, and eventually contribute to the understanding of moral situations faced by humans and machines.
\end{abstract}

\maketitle

\begin{figure*}[t]
	\centering
	\includegraphics[width=1\linewidth]{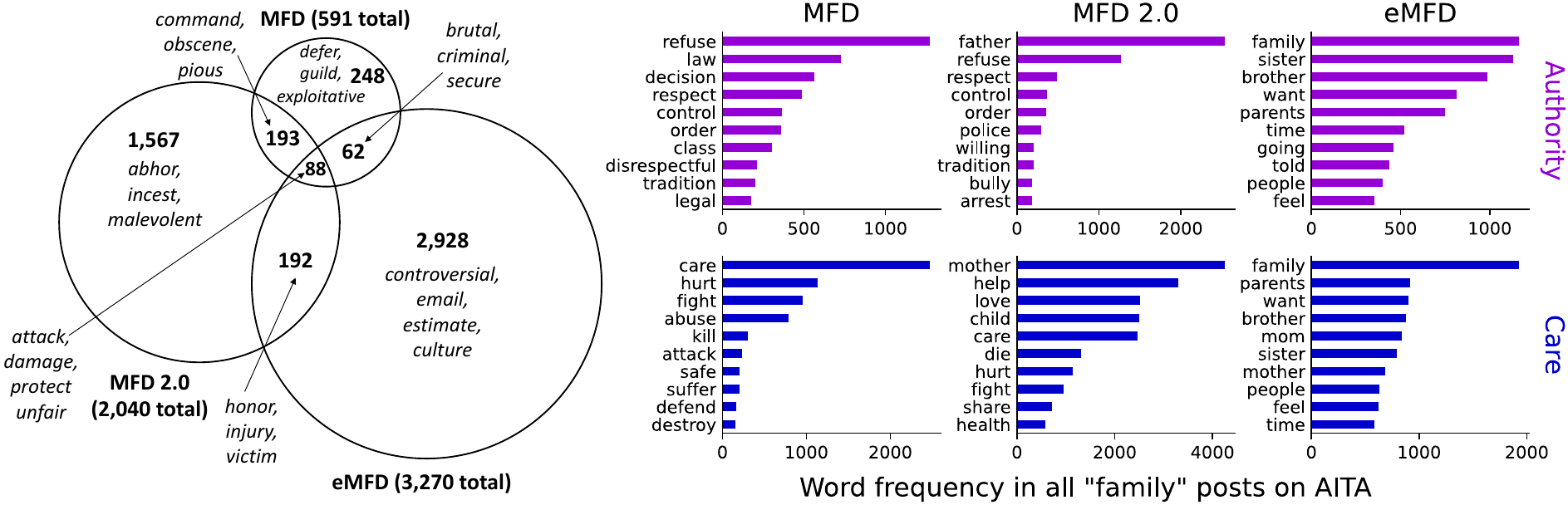}
	\caption{Three existing lexicons---MFD, MFD 2.0, and eMFD---used for word count in detecting moral foundations. \emph{Left}: Venn diagram depicting the sizes of these lexicons with example words. \emph{Right}: 10 most popular words for two moral foundations (\foundation{authority} and \foundation{care}) in each lexicon that are found in 6,800 \texttt{r/AmItheAsshole} posts of the topic \topic{family}. See \Cref{sec:wordcount_limits}.}
	\label{fig:mfds_aita_wordcount}
\end{figure*}

\section{Introduction}
\label{sec:intro}

Recent years have witnessed a growing interest in the study of moral content on social media. Many online discussions have a tendency to reflect aspects of morality, and researchers thus far have aimed to study how and to what extent moral dimensions vary throughout this vast domain. 
One particularly influential framework in the analysis of such content is moral foundations theory (MFT), which maps morality to five fundamental psychological dimensions called ``moral foundations'': \foundation{authority}, \foundation{care}, \foundation{fairness}, \foundation{loyalty} and \foundation{sanctity} \citep{haidtIntuitiveEthicsHow2004a,haidtRighteousMindWhy2012}. Prior research suggests that the variation in moral sentiment within and across cultures can be attributed to differences in the way these cultures realize and value each moral foundation. 
Notable works, including those on vaccine hesitancy \citep{weinzierlHesitancyFramingsVaccine2022}, social norms \citep{forbesSocialChemistry1012020} and news story framing \citep{mokhberianMoralFramingIdeological2020b}, have used this dimension-mapping approach to uncover large-scale patterns of moral belief and judgment. 

As human labeling does not scale to the size of modern corpora, MFT-based studies of online moral content must rely on tools to automatically detect moral foundations in text. However, existing methods, especially word count programs based on human-crafted lexicons, are surprisingly lacking in their consistency and ability to generalize to different domains (see \Cref{fig:mfds_aita_wordcount} for an  illustration). Variations across these methods can lead to significant changes in downstream findings based on such  measurements. In this work, we propose \ourmodel, a \underline{M}oral foundations classifier based on trans\underline{former}s fine-tuned on datasets from diverse domains, which is released publicly.\footnote{\url{https://github.com/joshnguyen99/moral_axes}}
Compared to a set of current approaches, {we find that simply using diverse datasets to fine tune works surprisingly well}---we observe that \ourmodel consistently achieves better accuracy on several datasets, with a relative AUC improvement of 4--17\%. %show its superior predictive performance compared to current, widely used methods across several benchmarks. 
Through two case studies involving moral stories on Reddit and controversies on Twitter, we demonstrate the effectiveness of \ourmodel in explaining non-trivial variations in people's moral stances and judgments across many social issues. The main contributions of this work are as follows:

% In this work, we examine the task of detecting moral foundations in text and provide a systematic evaluation of current, widely used methods. We advocate for the adoption of machine learning classifiers---which give much better predictive performance even in novel domains---as opposed to popular but oversimplistic word count programs. In two case studies on social media, we demonstrate the utility of MFT in capturing moral beliefs and judgments across many topics. The main contributions of this work are as follows.

\begin{itemize}
	\item We introduce \ourmodel, a moral foundations classifier based on a fine-tuned large language model {on text data from diverse domains} (\Cref{sec:mf_classifiers}).
	\item Through an in-depth analysis of word count programs, we show why and how they tend to fall short in labeling moral foundations in text (\Cref{sec:wordcount_limits}). On the other hand, \ourmodel consistently performs the best across several in- and out-of-domain datasets (\Cref{sec:mf_classifiers,sec:external_eval}).
	\item We demonstrate the utility of moral foundations in text analysis through two case studies involving (i) moral stories and judgments and (ii) stance toward several controversial topics (\Cref{sec:measurements}). We highlight the difference between downstream conclusions resulting from word count and those from \ourmodel. This suggests that many prior findings {relying on MFT measurements may warrant further scrutiny}.
\end{itemize}

\section{Related Work}
\label{sec:related_work}

Our work builds upon the literature on defining key moral dimensions, detecting them in text, and measuring a variety of patterns pertaining to morality within large corpora.

\subsection{Moral Foundations Theory}
\label{sec:related_work:MFT}

Moral foundations theory, developed in psychology, proposes a taxonomy over what are called ``moral intuitions'' \citep{haidtIntuitiveEthicsHow2004a,haidtRighteousMindWhy2012}. In an attempt to explain the similarities and differences in moral judgment across cultures, Haidt and colleagues postulated key categories of intuitions behind people's moral judgments. These so-called ``moral foundations'' fall into five spheres: \foundation{authority}, relating to traits such as deference to higher authorities to maintain group cohesion; \foundation{care}, upholding the virtues of nurturing and protection; \foundation{fairness}, involving equal treatment and reward; \foundation{loyalty}, relating to prioritizing one's group and alliances; and \foundation{sanctity}, including intuitions about maintaining the sacredness of the body and avoiding moral contamination. Anthropological evidence suggests these foundations are universal, although which traits constitute a virtue or a vice vary across cultures \citep{grahamMoralFoundationsTheory2013a}.

Most studies using MFT have focused on characterizing political ideology, especially in the U.S. context. For example, pertaining to the foundations \foundation{care} and \foundation{fairness}, liberals often support the commitment to justice and actions that uphold equality and minimize suffering. Conservatives, on the other hand, tend to value all five foundations somewhat equally, additionally upholding virtues such as loyalty to one's country and respecting authority and order \citep{grahamLiberalsConservativesRely2009a,mcadamsFamilyMetaphorsMoral2008,haidtWhenMoralityOpposes2007,vanleeuwenPerceptionsSocialDangers2009}. The theory has also been used to characterize the differences in moral sentiment surrounding socially significant topics such as stem cell research \citep{cliffordHowWordsWork2013}, vaccine hesitancy \citep{aminAssociationMoralValues2017}, euthanasia, abortion, animal cloning, and same-sex marriage \citep{kolevaTracingThreadsHow2012}.

\subsection{Automatically Detecting Moral Foundations}
\label{sec:related_work:MF_detection}

A requirement for large-scale studies of MFT is the automatic detection of moral foundations. While the gold standard for moral foundation labeling is human annotation, this requires extensive training and evaluation, which does not feasibly scale to large text collections. We categorize the automatic scoring of moral foundations into word count, embedding similarity, and supervised classification methods.

\emph{Word count methods} rely on a human-crafted lexicon, which is a mapping from words to foundations. When scoring a document, each token contained in the lexicon counts as one occurrence of the foundation that it maps to. Three lexicons, often called moral foundations dictionaries (MFDs), have been extensively used in the literature: MFD \citep{grahamMoralFoundationsDictionary2012}, MFD 2.0 \citep{frimerMoralFoundationsDictionary2019c}, and extended MFD \citep{hoppExtendedMoralFoundations2021}; we further review these in \Cref{sec:wordcount_limits}. Word count methods resemble LIWC, a popular program that scores psychological and emotional features in text \citep{tausczikPsychologicalMeaningWords2010}.

\emph{Embedding similarity methods} define the relevance between a document and a moral foundation as the cosine similarity between their word embeddings such as word2vec \cite{mikolovDistributedRepresentationsWords2013}. In particular, each foundation is represented by the average embedding of its ``seed words,'' and similarly each document is the average of its tokens.  \citet{mokhberianMoralFramingIdeological2020b}, for example, used this method to score moral foundations as features of news articles. Similar to Rocchio classification \citep{manningIntroductionInformationRetrieval2008}, embedding similarity essentially relies on the centroid of a set of seed words and is well-known to have limited expressivity. Furthermore, choosing a good set of seed words remains a major practical challenge.

\emph{Supervised classification methods} treat the detection of moral foundations as a text classification task. For example, \citet{hooverMoralFoundationsTwitter2020a} annotated a dataset of tweets and used it to train support vector machines for classification. Similarly,  \citet{tragerMoralFoundationsReddit2022} fine-tuned BERT \cite{devlinBERTPretrainingDeep2019}, a large language model (LLM), using annotated social media data for the same task and reported state-of-the-art performance. However, these models may not generalize well to new datasets \citep{liscioCrossDomainClassificationMoral2022}. {Recently, \citet{guoDataFusionFramework2023} trained moral foundation classifiers with a new loss function to incorporate domain variations. In comparison, our approach uses a standard LLM architecture trained on multiple data domains. This approach is arguably more straightforward as it requires no additional hyperparameters, does not demand explicit one-hot encoding of the domains, and thus offers improved adaptability when a new domain enters into the training dataset.}

Finally, some prior work made a distinction in the polarity of moral foundations, i.e., whether an example portrays a virtue or a vice of a foundation. As we explain later in \Cref{sec:mf_classifiers:dataset}, there are several conceptual and practical concerns for this approach. Thus, we decide to score only moral foundations, irrespective of polarity, in this work.

\subsection{Analyses of Text}
\label{sec:related_work:MFT_studies}

A growing body of work has adopted MFT to the analysis of moral rhetoric from online media.
For example, \citet{mokhberianMoralFramingIdeological2020b} studied the relationship between story framing and political leaning of several news sources. 
% After using embedding similarity to score moral foundations, a statistical analysis 
The study found that on topics such as immigration and elections, conservative-leaning sources tend to focus on the virtues of \foundation{sanctity} such as austerity and sacredness, while liberal-leaning sources emphasize the condemnation of the vices of \foundation{sanctity} like dirtiness and unholiness. 
% In another work, \citet{rezapourIncorporatingMeasurementMoral2021} studied the relationship between moral foundations and stance (in favor, against or neutral) on several political topics discussed on Twitter. The authors used the MFD to detect foundations and subsequently found a significant relationship between, for example, the topic abortion and moral values surrounding foundations such as \foundation{care}, \foundation{loyalty} and \foundation{sanctity}.
\citet{hoppGraphLearningApproachDetecting2020} analyzed movie scripts and identified relevant foundations in movie scenes exhibiting moral conflicts. Using network-theoretic methods alongside moral foundations, the authors were able to construct communities of characters with specific shared moral characteristics. In argument mining, \citet{kobbeExploringMoralityArgumentation2020} proposed to use moral foundations in the automatic assessment of arguments. The study found a significant correlation between moral sentiment (the presence of moral foundations) and audience approval of an argument. In another line of work, \citet{forbesSocialChemistry1012020} and \citet{ziemsMoralIntegrityCorpus2022} used MFT to categorize collections of moral ``rules of thumb'' collected from large crowds and assessed LLMs' moral sentiment predictions.

Empirical results based on moral foundations scoring currently face two important limitations. First, there is no unified and highly accurate method on which researchers rely to label their text corpora. Second, and as a result, conclusions drawn from prior studies likely suffer from low-quality scoring and may vary based on what method is chosen. A solution sufficiently addressing these two challenges will allow MFT-based analyses to scale and facilitate their reproducibility, comparison, and generalization.

%\section{Moral Foundations Via Word Count}
\section{Limitations of Moral Foundations Dictionaries}
\label{sec:wordcount_limits}

Word count methods based on lexicons remain the most popular for automatically detecting moral foundations, often serving as the default choice among researchers. Here we detail how they work and several of their limitations when applied to text corpora.

\header{Scoring via word count} Each lexicon, called a moral foundations dictionary (MFD), contains a list of words and the foundations they represent. For example, the word ``deceive'' can be mapped to the foundation \foundation{loyalty}. A document to be scored is first tokenized; then, for each token that is also in the MFD, the count for the foundation that it is mapped to is incremented. For instance, if the token is ``deceive,'' the count for \foundation{loyalty} is increased. 
These methods are easy to implement, involve no model training, and are interpretable since they directly show what words signify a foundation. 

\header{Available lexicons} Three versions of the MFD have been used extensively in the literature. The first MFD, released as a dictionary to be used in LIWC-like programs, contains nearly 600 words \citep{grahamMoralFoundationsDictionary2012}. Later, a new version with over 2K words called MFD 2.0 \citep{frimerMoralFoundationsDictionary2019c} was released in which the creators extended their expert-crafted word lists by querying a word embedding \citep{mikolovDistributedRepresentationsWords2013} for similar terms. Another recent variant, called the extended MFD (eMFD, 3.2K words) \citep{hoppExtendedMoralFoundations2021}, contains one-to-many mappings associated with moral foundation weights, between 0 and 1, for each word. We list some example words of these MFDs in \Cref{tab:mfd_example_words} in the appendix and show some overlap between them in \Cref{fig:mfds_aita_wordcount}.\footnote{The appendix is found at \url{https://arxiv.org/abs/2311.10219}.}

\header{Limitations}
%Word count methods are easy to implement, involve no model training and are interpretable through showing what words directly signify a foundation. 
%We find that, however, such scoring methods have several important limitations that can 
We want to understand the different MFD versions and what they each capture, but have since identified other limitations that will
affect downstream measurements and interpretations.

% 1. Lexicon sizes and their words:
% - Supposedly all words are morally relevant regarding the foundations they map to
% - Their overlap is very small
First, these lexicons have a fixed and limited vocabulary. See \Cref{fig:mfds_aita_wordcount} (left) above for some example words and a visualization. Most of the words in these lexicons are supposedly morally relevant, but their overlap is 
surprisingly small. For instance, the 281 words common to both MFD and MFD 2.0 only account for 47.5\% of MFD and 13.8\% of MFD 2.0. 
In eMFD, 89.5\%  of the words do not appear in either MFD or MFD 2.0 at all. Such lack of consensus in expert- and machine-created lexicons is concerning, and one can rightfully question whether the resulting scores are reliable. 

% 2. Unclear how to handle word variants and derivations
% - The set of all inflections of each word is incomplete. This means that one needs to trace every word to its lexeme and the process is unclear.
% - How to handle morphological derivations, especially with respect to disambiguating parts of speech. -> Refer to Rezapour et al. for their version of the MFD including POS 
Second, when using the MFDs, it is unclear how word variations should be handled. For example, the MFD 2.0 explicitly contains the verb ``desecrate'' and its inflected forms ``desecrates,'' ``desecrated'' and ``desecrating.'' However, this is not the case for the verb ``deify,'' which is in the lexicon while ``deified'' is not. A direct but possibly non-exhaustive way to address this is to lemmatize all tokens in a document before performing a lexicon lookup. Another related issue is how to disambiguate the parts of speech of some words in these dictionaries. For example, given the word ``bully'' in the MFD 2.0, should it be counted when it is a verb, noun, or even an adjective (which, in this case, means ``excellent'')? Researchers have attempted at disambiguation \citep{rezapourEnhancingMeasurementSocial2019}, 
but accounting for word senses alone does not mitigate the drawbacks of lacking inflections or the incompleteness of the vocabulary.
%most notably the ``extended MFD'' built on top of the MFD by .

% 3. Favor longer texts
{Third, longer documents tend to have a higher chance of having a dictionary match. In Appendix \Cref{fig:mfd_word_count_vs_length_correlation} (top panel), we show that the number of words found in each dictionary is highly correlated with how many words each text input has (at $r$=0.59, 0.72, and 0.98 respectively for MFD, MFD 2.0, and eMFD). When foundation scores are normalized by the input length, such strong positive relationships disappear (\Cref{fig:mfd_word_count_vs_length_correlation}, bottom panel), suggesting that length-normalized scoring might be preferable if one has to use dictionary-based methods. On the other hand, there are examples for which the detected moral foundation is due to one matching word in a long post---\Cref{appn:mfd_limits:input_length} shows an example that triggers the foundation \foundation{authority} because the word ``refuse'' appears once in a 140-word long post.}

% 4. Encode the bias of those who created the dictionaries
Finally, the hand coding of words to moral foundations by experts is a source of personal subjectivity and social bias. In the same analysis using Reddit posts, we discover some problematic associations. \Cref{fig:mfds_aita_wordcount} (right) presents the most frequently matched words that are related to two foundations: \foundation{authority} and \foundation{care}. Using  MFD 2.0, some associations such as ``father'' with \foundation{authority} and ``mother'' with \foundation{care} appear to reflect the bias of the lexicon creators when assigning moral intuitions to very general familial roles. {\Cref{appn:mfd_limits:family_bias} contains a systematic comparison of foundation scores for posts containing ``father'' and those with ``mother,'' showing that posts with ``mother'' scores higher for \foundation{care} (MFD and MFD2.0) and \foundation{loyalty} (MFD), and that posts with ``father'' score higher for \foundation{authority} (MFD 2.0) and lower on \foundation{care} (MFD 2.0). The same figure also shows that \ourmodel does not suffer from the same bias.}

% 5. Relate these limitations to sentiment analysis
{
These limitations are not restricted to moral foundations. Within sentiment analysis, where lexicons such as LIWC \citep{tausczikPsychologicalMeaningWords2010} are used, the same pitfalls are especially illuminating. First, word count explicitly ignores the context-dependent nature of language by relying on (normalized) frequency as a direct proxy to sentiment \citep{pangOpinionMiningSentiment2008,puschmannTurningWordsConsumer2018}, and by treating polysemous words equivalently \cite{schwartzDataDrivenContentAnalysis2015}. Second, since lexicons are static, the validity and accuracy of this method highly depend on its input's domain \citep{gonzalez-bailonSignalsPublicOpinion2015}. And third, word count can be shown to predict sentiment more poorly than simply word presence, suggesting that the length of an input may influence the overall score more substantially \citep{pangOpinionMiningSentiment2008}. Machine learning approaches, especially with the advent of LLMs, can overcome these challenges through their ability to learn contextualized patterns and superior cross-domain generalizability.
}

Overall, dictionary-based moral foundation scoring seems fragile, lacks consensus in what they capture, and has inherent biases in social stereotypes. {We caution the use of manually curated lexicons without further scrutiny, and recommend examining their coverage and accuracy before interpreting aggregate results.} Recognizing that developing a good lexicon takes significant effort and is difficult to get right, we adopt a data-driven approach for moral foundation scoring which avoids these limitations and can account for the nuances in language that exist beyond individual words.

%we believe that detecting moral foundations by counting words in some fixed lexicon is an oversimplistic approach that may lead to very coarse predictions. Although some workarounds exist (such as augmenting the MFDs with part-of-speech tags) they do not fully address the limitations of these methods. We hence advocate for more sophisticated, data-driven predictors that are grounded in machine learning. In the following sections, we describe our approach, called \ourmodel, based on fine-tuning a large language model using high-quality, public datasets. 
% In \Cref{sec:evaluation} we also present an extensive evaluation of MFormer's performance in comparison with existing methods, including word count.

% To illustrate these two points, we take 6,800 Reddit posts of the topic \topic{family} on \texttt{r/AmItheAsshole} \citep{nguyenMappingTopics1002022b}, containing over 2.9M tokens, and score them using the three lexicons. The result shows that many posts are considered to contain a foundation simply because they have only one word in the dictionary maps to that foundation. See \Cref{appn:mfd_limits} for some examples. As we show in \Cref{fig:mfds_aita_wordcount}, right, some associations such as ``father'' $\rightarrow$ \foundation{authority} and ``mother'' $\rightarrow$ \foundation{care} appear to reflect the bias of the creators of the MFD 2.0 when assigning moral intuitions to general familial roles.

% Alasdair's: maybe discuss the limitations of embedding similarity + logistic regression

%\section{Classifying Moral Foundations}
\section{Constructing and Evaluating \ourmodel}
\label{sec:mf_classifiers}

In this section, we describe \ourmodel, a language model fine-tuned from a wide range of data to score moral foundations in text. We introduce the datasets \ourmodel is trained on in \Cref{sec:mf_classifiers:dataset}. Then we describe the training procedure along with some baselines (\Cref{sec:mf_classifiers:classifiers}). Finally, we present evaluation details that highlight \ourmodel's efficacy (\cref{sec:mf_classifiers:evaluation}).

% we describe a new method for detecting moral foundations in text using machine learning, which we call \ourmodel. We first introduce the datasets on which \ourmodel is trained. Then we describe the method in more detail, along with some baselines. Finally, we present evaluation details that highlight \ourmodel's efficacy.

\subsection{Datasets}
\label{sec:mf_classifiers:dataset}

We first describe the dataset used to train and evaluate moral foundation classifiers. We combine three publicly released, high-quality data sources labeled with moral foundations.

\header{Twitter} \citep{hooverMoralFoundationsTwitter2020a}
\label{sec:mf_classifiers:dataset:twitter}
This dataset contains 34,987 tweets encompassing seven ``socially relevant discourse topics'': All Lives Matter, Black Lives Matter, 2016 U.S. Presidential election, hate speech, Hurricane Sandy, and \#MeToo. Annotators were trained to label the tweets with moral foundations and their sentiments (virtue and vice), with at least three annotations per tweet. We keep all tweets in this dataset for our use and determine that a tweet contains a foundation $f$ if at least one annotator labeled it with $f$. Further, for each foundation, we merge the labels for its virtue and vice into one: e.g., the raw labels ``purity'' and ``degradation'' are mapped into the same foundation \foundation{sanctity}. 

\header{News} \citep{hoppExtendedMoralFoundations2021}  
\label{sec:mf_classifiers:dataset:news}
This dataset was used to construct the eMFD lexicon (cf. \Cref{sec:related_work:MF_detection}). The authors pulled 1,010 news articles, most of which on politics, from the GDELT dataset \citep{leetaruGDELTGlobalData2013a} and employed workers to label these articles with moral foundations. Specifically, each annotator was assigned a foundation-article pair and then asked to highlight all sections in the article that contain this foundation. We segment every article into sentences and assign a moral foundation $f$ to a sentence if any part of it is contained within a highlighted section labeled with $f$. This yields 32,262 instances in total.

\begin{table}[]
	\centering
	\normalsize
	\begin{tabular}{lcccc}
		\toprule
		Source       & Twitter & News     & Reddit  & Total  \\
		\midrule
		Data Period       & '10--'17       & '12--'17  & '20--'21 & --      \\
		% Topic        &    --     &  --        & --        & --      \\
		\# Examples     & 34,987  & 34,262   & 17,886  & 87,135 \\
		Avg. \# Tokens  & 19.3    & 28.0     & 41.7    & 27.3   \\
		\# Annotators   & 854     & 13       & 27      & --      \\
		% Processing   & --       & Sentence & --       & --      \\
		\midrule
		\% Authority & 33.4    & 24.9     & 19.2    & 27.1   \\
		\% Care      & 40.6    & 24.8     & 26.5    & 31.5   \\
		\% Fairness  & 35.9    & 24.2     & 29.5    & 30.0   \\
		\% Loyalty   & 31.1    & 24.4     & 11.1    & 24.4   \\
		\% Sanctity  & 22.3    & 19.9     & 9.8     & 18.8 \\
		\bottomrule
	\end{tabular}
	\caption{\normalsize Three moral foundations datasets used to develop \ourmodel.}
    \label{table:dataset:mf_dataset_stats}
\end{table}

\header{Reddit} \citep{tragerMoralFoundationsReddit2022}
\label{sec:mf_classifiers:dataset:reddit}
This dataset contains 17,886 comments on 12 different subreddits roughly organized into three topics: U.S. politics, French politics, and everyday moral life. In annotation, the authors separated the foundation \foundation{fairness} into two classes: \foundation{equality} (concerns about equal outcomes for all individuals and groups) and \foundation{proportionality} (concerns about getting rewarded in proportion to one's merit). Another label, \foundation{thin morality}, was defined for cases in which moral concern is involved but no clear moral foundation is in place. We merge both \foundation{equality} and \foundation{proportionality} into their common class \foundation{fairness} and consider \foundation{thin morality} as the binary class 0 for all foundations, which results in the same five moral foundation labels. Finally, for each comment, we assign a binary label 1 for foundation $f$ if at least one annotator labeled this comment with $f$.

\header{A profile of the datasets}
\label{sec:mf_classifiers:dataset:profile_all}
We combine the three sources---Twitter, News, and Reddit---into one dataset, yielding 87,135 instances with 2.4M tokens. \Cref{table:dataset:mf_dataset_stats} presents summary statistics. Each example has on average 27.3 tokens, with Reddit comments the longest (41.7 tokens) and tweets the shortest (19.3 tokens). The foundations \foundation{care} and \foundation{fairness} have the most positive instances in total, each with at least 30\% of the dataset. Among the three sources, tweets tended to contain more foundations than Reddit comments. For example, over 31\% of tweets contain \foundation{loyalty} while only 11.1\% of Reddit comments do.  Finally, for each foundation, we split this dataset into a training and test set with ratio 9:1, stratified by that foundation. {In \Cref{appn:dataset}, we describe the datasets in more detail, including their annotation scheme and agreement rate, and how label disagreement and train-test splitting are handled. Finally, there exist other datasets labeled with foundations, such as {\it covid} and {\it congress} used by \citet{guoDataFusionFramework2023}. We choose not to include them due to their significantly smaller size---in the 1--2,000 range rather than 15,000+.
% , which is the case for all three datasets above. We posit that including them may result in a small increase in scoring performance but leave this investigation as future work.
}

\header{Capturing moral foundation polarity} Some prior work has additionally considered \emph{polarity}, i.e., whether a text instance conveys a \emph{virtue} or a \emph{vice} of a moral foundation, resulting in ten classes (two for each foundation). In this work, we decide against this approach---instead only aiming to score the relevance of a foundation regardless of polarity---for three reasons. First and conceptually, virtues and vices are very loosely-defined terms whose perception is subject to cultural differences \citep[see \S 2.4.4 for an example of \foundation{authority}]{grahamMoralFoundationsTheory2013a}. Second, while some previous work has treated the virtue/vice distinction as a sentiment analysis task \cite{hoppExtendedMoralFoundations2021}, we believe this is somewhat na\"{i}ve since it lacks a theoretical justification. Third and operationally, the assignment of virtues or vices by human annotators is another source of noise on top of the noise in moral foundation labels. This is coupled with the fact that not all available datasets/lexicons capture this polar distinction. We do not argue that virtues and vices are irrelevant; rather, we believe they deserve a more thorough theoretical and practical treatment, which is beyond this work's scope.

\subsection{Moral Foundations Classifiers}
\label{sec:mf_classifiers:classifiers}

\header{\ourmodel}
\label{sec:mf_classifiers:classifiers:roberta}
LLMs have achieved state-of-the-art performance across a range of NLP benchmarks. Our work is not the first to use LLMs for this task; for example, \citet{tragerMoralFoundationsReddit2022} fine-tuned BERT \cite{devlinBERTPretrainingDeep2019} to create moral foundation classifiers. However, we note that prior work primarily focused on setting up a baseline for future work. As such, a careful treatment of the fine-tuning process and evaluation is necessary to substantiate the adoption of such methods.

We choose the RoBERTa-base architecture \cite{liuRoBERTaRobustlyOptimized2019b} with 12 self-attention layers for this task. Each document is tokenized and then two special tokens, 
\texttt{<s>} and \texttt{</s>}, are added to the beginning and end of the document, respectively. A classifier module follows the final attention layer, where the 768-dimensional embedding of the \texttt{<s>} token goes through a fully connected layer with 768 neurons followed by $\tanh$ activation. Finally, this is linearly mapped to a two-dimensional output vector and then converted to probabilities via a softmax layer. In fine-tuning RoBERTa, we find the optimal learning rate and the number of training epochs by performing a grid search. We end up with five binary classifiers, each of which outputs a score between 0 and 1 for every input text. We call the final fine-tuned models \ourmodel, for \underline{M}oral \underline{f}oundations using trans\underline{former}s. More training details are found in \Cref{appn:supervised_classifiers:roberta}.

\header{Baselines}
\label{sec:mf_classifiers:classifiers:existing}
For comparison, we consider as baselines all methods described in \Cref{sec:related_work:MF_detection}: \emph{word count}, \emph{embedding similarity}, and \emph{supervised classifiers}.

For word count, we score a document based on the description in \Cref{sec:wordcount_limits}. We experiment with three lexicons: MFD, MFD 2.0, and eMFD. For MFD and MFD 2.0, we increment the foundation count by one every time its example word is encountered and then divide the count by the total number of tokens. This represents the frequency with which the foundation is found among the tokens. For eMFD, since the lexicon contains soft counts between 0 and 1, every time a word in the dictionary is found we add all scores to their corresponding foundations. Then, the five-dimensional vector of foundation scores for the document is normalized by the number of tokens that match the eMFD's entries. For all three lexicons, the foundation scores are in $[0, 1]$. More detail is found in \Cref{appn:mfd}.

For embedding similarity, with each foundation $f$ and a document $d$, the score for $d$ is defined as the cosine similarity between the embedding vectors for $f$ and $d$. To encode $f$ and $d$, we use the GloVe embedding \citep{penningtonGloVeGlobalVectors2014}, specifically the ``Twitter'' 200-dimensional version. The vector representation for $f$ is defined as the average of the vectors for the ``seed words'' that represent $f$. Similarly, the vector for $d$ is the average of the vector representations of all of its tokens. The range for foundation scores is $[-1, 1]$. For more detail, including the seed words describing each foundation, see \Cref{appn:embedding_sim}.

Finally, for supervised classifiers, we train a logistic regression model on a range of sparse and dense embeddings. We find that, unsurprisingly, the embedding with the best performance is Sentence-RoBERTa, which is based on RoBERTa fine-tuned for sentence similarity \citep{reimersSentenceBERTSentenceEmbeddings2019a}. {In \Cref{appn:supervised_classifiers:logistic_regression}, we provide more details of logistic regression and compare it with support vector machine as used in previous work \citep{hooverMoralFoundationsTwitter2020a}.}

{
\header{Alternative to binary classifiers}
\ourmodel is a collection of five binary classifiers each for one moral foundation and the corresponding set of RoBERTa weights. We also consider a weight-shared variant in which only one model is used but the final classification layer contains five neurons, each followed by a sigmoid activation. In other words, this multi-label model outputs five binary probabilities simultaneously predicting each foundation. Compared to \ourmodel, multi-label RoBERTa requires less storage and training resources. However, we find that this model performs uniformly worse than \ourmodel, achieving 10.7--19.3\% lower test AUC than its binary classification counterparts (see \Cref{appn:supervised_classifiers:multilabel_roberta}). 
}

\begin{table*}
	\centering
	\fontsize{8pt}{0pt}\selectfont
	\begin{tabular}{M{1mm} @{\hskip 4mm} P{8mm} M{130mm} @{\hskip 2mm}M{23mm}}
		%%%%%%%%%%%%%%%%%%%%%%%%%%%%%%
		% & &  & \includegraphics[width=0.135\textwidth]{figs/roberta_highest_scoring_testsest/headers.pdf} \\
		\toprule
		% AUTHORITY
		\multirow{3}{*}[2mm]{\rotatebox[origin=c]{90}{\normalsize \textbf{Authority}}} & \rotatebox[origin=c]{0}{Twitter} &  \cellcolor{twitterbg!80} \example{I am a proponent of civil disobedience and logic driven protest only; not non/ irrational violence, pillage \& mayhem! \#AllLivesMatter} &  \includegraphics[width=0.135\textwidth]{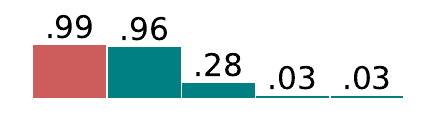} \\ 
		&  \rotatebox[origin=c]{0}{News} & \cellcolor{newsbg!80} \example{Earlier Monday evening, Pahlavi addressed a private audience and urged `civil disobedience by means of non-violence.'}  &   \includegraphics[width=0.135\textwidth]{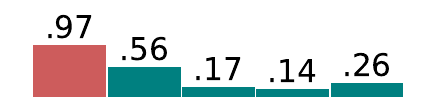}\\ 
		& \rotatebox[origin=c]{0}{Reddit} & \cellcolor{redditbg!80} \example{Our politicians are openly encouraging rioters to harm and even kill police now} &  \includegraphics[width=0.135\textwidth]{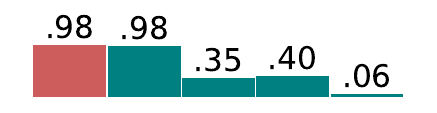} \\ 
		\midrule
		% CARE
		\multirow{3}{*}[-2mm]{\rotatebox[origin=c]{90}{\normalsize\textbf{Care}}} & \rotatebox[origin=c]{0}{Twitter} &  \cellcolor{twitterbg!80}  \example{\#BlackLivesMatter SHAME ON YOU! @SenSanders is the best hope for social justice and you hurt him, you hurt me, you hurt us all. SHAME!}
		&  \includegraphics[width=0.135\textwidth]{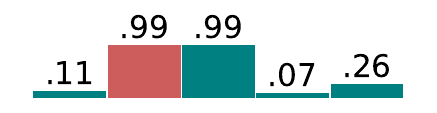} \\ 
		&  \rotatebox[origin=c]{0}{News} & \cellcolor{newsbg!80} \example{Just 10 days later, a gunman shot and killed three police officers in Baton Rouge, Louisiana, in what authorities called another `ambush-style' attack.}  &   \includegraphics[width=0.135\textwidth]{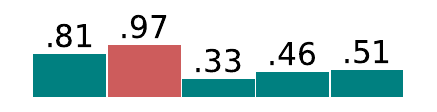}\\ 
		& \rotatebox[origin=c]{0}{Reddit} & \cellcolor{redditbg!80} \example{Wage slavery is indeed disgusting. Stay strong and safe comrade I wish you the best of luck. Educate agitate organize} &  \includegraphics[width=0.135\textwidth]{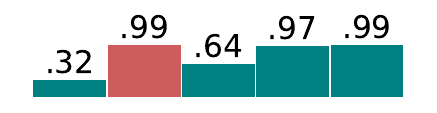} \\ 
		\midrule
		% FAIRNESS
		\multirow{3}{*}[0.5mm]{\rotatebox[origin=c]{90}{\normalsize \textbf{Fairness}}} & \rotatebox[origin=c]{0}{Twitter} &  \cellcolor{twitterbg!80} \example{It's about humanity \& equality \#ChapelHillShooting \#AllLivesMatter} &  \includegraphics[width=0.135\textwidth]{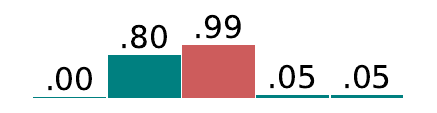} \\ 
		& \rotatebox[origin=c]{0}{News} & \cellcolor{newsbg!80}  \example{Victims of despotism are entitled to fairness and justice, and this is the message we are conveying to the whole world.}   &   \includegraphics[width=0.135\textwidth]{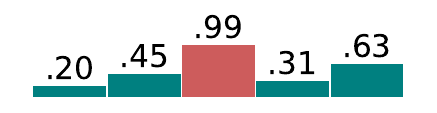}\\ 
		& \rotatebox[origin=c]{0}{Reddit} & \cellcolor{redditbg!80} \example{This money was taxed when it was earned, taxed when it was given and will be taxed when it is spent. Yikes} &  \includegraphics[width=0.135\textwidth]{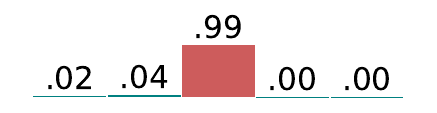} \\ 
		\midrule
		% LOYALTY
		\multirow{3}{*}[0.5mm]{\rotatebox[origin=c]{90}{\normalsize \textbf{Loyalty}}} & \rotatebox[origin=c]{0}{Twitter} &  \cellcolor{twitterbg!80} \example{Storify. Solidarity and support \#Ferguson \#Blacklivesmatter {[}URL{]}…}  &  \includegraphics[width=0.135\textwidth]{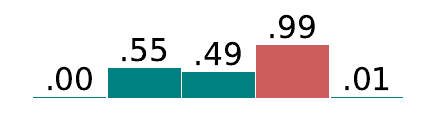} \\ 
		& \rotatebox[origin=c]{0}{News} & \cellcolor{newsbg!80}  \example{The small rally was aimed at offering unity and solidarity for all regardless of race, ethnicity, gender, religion, sexual orientation or identity.}   &   \includegraphics[width=0.135\textwidth]{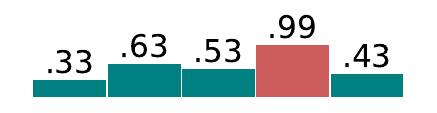}\\ 
		& \rotatebox[origin=c]{0}{Reddit} & \cellcolor{redditbg!80} \example{It's never unpatriotic to criticize the POTUS. They work \textit{for us}.} &  \includegraphics[width=0.135\textwidth]{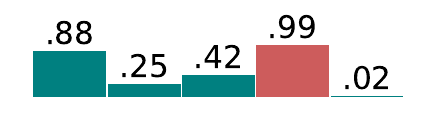} \\ 
		\midrule
		% SANCTITY
		\multirow{3}{*}[1mm]{\rotatebox[origin=c]{90}{\normalsize \textbf{Sanctity}}} & \rotatebox[origin=c]{0}{Twitter} &  \cellcolor{twitterbg!80} \example{we value the sacred human dignity of every single life! $\sim$@realDonaldTrump \#votervaluessummit \#ProLife \#MAGA}  &  \includegraphics[width=0.135\textwidth]{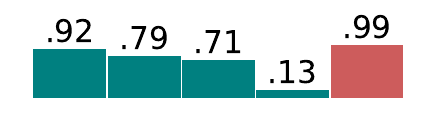} \\ 
		& \rotatebox[origin=c]{0}{News} & \cellcolor{newsbg!80}   \example{He said that sexuality was to adhere to `its God-given purpose.'}   &   \includegraphics[width=0.135\textwidth]{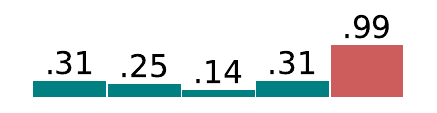}\\ 
		& \rotatebox[origin=c]{0}{Reddit} & \cellcolor{redditbg!80} \example{Incest is disgusting and he should seek help. What a gross f***. Sorry you're dealing with this but this is sick.} &  \includegraphics[width=0.135\textwidth]{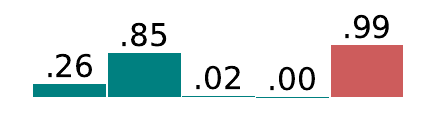} \\ 
		\bottomrule
		&&& \includegraphics[width=0.135\textwidth]{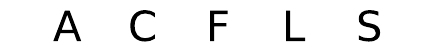} \\
	\end{tabular}
	\caption{Highest-scoring test examples for each foundation. The right-hand bar charts show \ourmodel's predicted scores.
		%Each bar chart on the right-hand column displays the RoBERTa scores for the five moral foundations: (from left to right) \foundation{authority}, \foundation{care}, \foundation{fairness}, \foundation{loyalty} and \foundation{sanctity}.
	}
	\label{table:roberta_highest_scoring_test}
\end{table*}

\subsection{Evaluation}
\label{sec:mf_classifiers:evaluation}

We evaluate classification methods presented in \Cref{sec:mf_classifiers:classifiers} using the hold-out test set in \Cref{sec:mf_classifiers:dataset}. The results show that \ourmodel outperforms all existing methods in scoring all foundations, often by a considerable margin.

\header{Evaluation metric}
\label{sec:mf_classifiers:evaluation:eval_metric} 
It is worth noting that this dataset is multi-labeled: each instance contains between zero and five foundations. Our goal is to build five classifiers each predicting the binary label of each foundation given an input. Two considerations are taken into account. First, all classifiers described in this section output a ``score'' representing the likelihood that a foundation exists in an input. Second, as shown in \Cref{table:dataset:mf_dataset_stats}, the dataset is unbalanced for all foundations with the percentage of positives being as low as 18.8\%. A suitable metric should be \emph{threshold-free} (it considers the scores and not just binary predictions), \emph{scale-invariant} (it considers prediction scores ranked on any scale), and take into account \emph{unbalanced class prior}. We therefore choose the area under the ROC curve (AUC) for evaluation. A useful statistical property of this metric is that the AUC of a classifier is equal to the probability that the classifier will rank a randomly chosen positive example higher than a randomly chosen negative example. While the range of the AUC is $[0, 1]$, a realistic lower bound is 0.5 which represents a classifier that randomly guesses positive half of the time \citep{fawcettIntroductionROCAnalysis2006}.

\begin{figure}[t]
	\centering
	\includegraphics[width=1\linewidth]{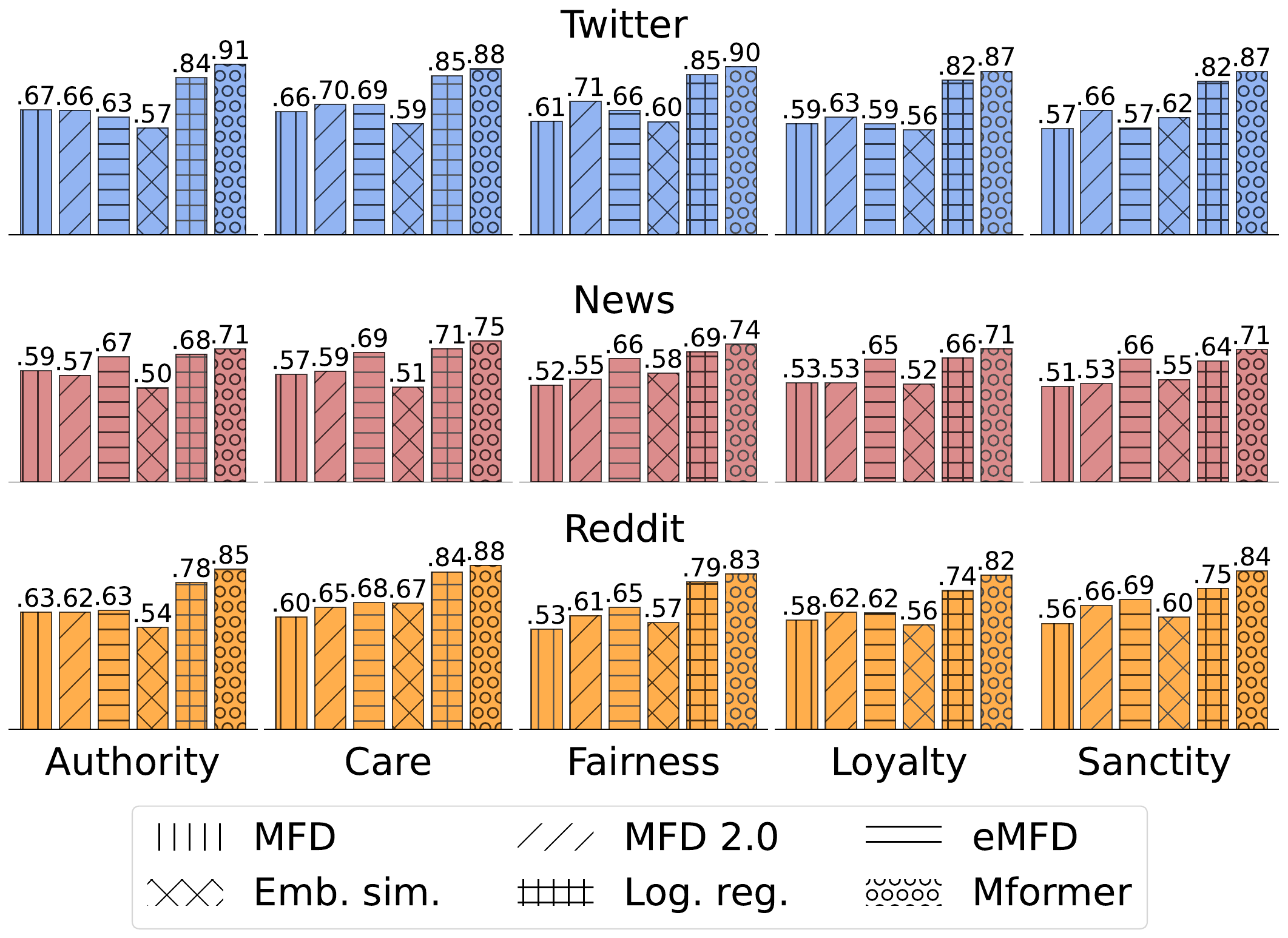}
	\caption{AUC on the Twitter (top row), news (middle row) Reddit (bottom row) portions of the test set for five moral foundation scoring methods: MFD, MFD 2.0, eMFD, embedding similarity, logistic regression, and \ourmodel.}
	\label{fig:roberta_mf:testset_auc}
\end{figure}

\header{In-domain evaluation}
\label{sec:mf_classifiers:evaluation:test_set}
We score all test examples using the methods described in \Cref{sec:mf_classifiers:classifiers} and report the test AUC in \Cref{table:evaluation:auroc_test_set} in the appendix. We find that embedding similarity shows the worst performance with AUC between 0.51 and 0.59 for all foundations, only slightly better than random guessing. Simple word count methods surprisingly perform better than embedding similarity, with each updated version of the MFD tending to improve from the previous. Logistic regression models further improve from word count, achieving the AUC between 0.76 and 0.81 for the Sentence-BERT embedding. The highest recorded test AUC for all foundations is by \ourmodel, where all foundations achieve an AUC between 0.83 and 0.85, a relative improvement of 4--12\% from logistic regression. When comparing the performance across foundations, we find that \foundation{care} and \foundation{fairness} are the easiest to score (both AUC = 0.85), while \foundation{loyalty} and \foundation{sanctity} are the more difficult but not sizably (both AUC = 0.83).

% \begin{figure*}
% 	\includegraphics[width=1\linewidth]{figs/Table2.png}
% 	\caption{Highest-scoring test examples for each foundation. The right-hand bar charts show \ourmodel's predicted scores.}
% 	\label{table:roberta_highest_scoring_test}
% \end{figure*}

Since the test sets are merged from three sources---Twitter, News, and Reddit---we examine the performance of these classifiers on each subset in \Cref{fig:roberta_mf:testset_auc}. The findings are similar: embedding similarity performs the worst, followed by word count (which gets better each newer version of the lexicon), then by logistic regression, and \ourmodel remains the best. Of all domains, tweets are the easiest to score: the AUC for \foundation{authority} and \foundation{fairness} goes up to 0.91 and 0.9, respectively. In contrast, sentences taken from news articles are the hardest: the AUC for all foundations only ranges from 0.71 to 0.75, with the lowest for \foundation{loyalty}. We suspect that the relatively low performance on news sentences is because of the way they were labeled: In \citep{hoppExtendedMoralFoundations2021}, annotators highlighted the \emph{sections} of an article containing a foundation, while we subsequently process these sections using \emph{sentence segmentation}. On the other hand, every tweet and Reddit comment was independently labeled for every foundation, which explains the quality of their labels.

Given its superior performance, we adopt \ourmodel as our final classifier. In \Cref{table:roberta_highest_scoring_test}, we show the highest-scoring examples for every foundation. The scores displayed in the right-hand column suggest that a lot of examples contain more than one moral foundation. For example, according to \ourmodel, the Reddit comment \example{``Wage slavery is indeed disgusting. Stay strong and safe comrade I wish you the best of luck. Educate agitate organize''} conveys \foundation{care}, \foundation{loyalty}, and \foundation{sanctity} with very high confidence. The coexistence and interplay of these foundations suggest complex moral constructs, which we will examine later in this paper.

\section{\ourmodel: Out-of-Domain Evaluation}
\label{sec:external_eval}

\begin{figure}[t]
	\centering
	\includegraphics[width=1\linewidth]{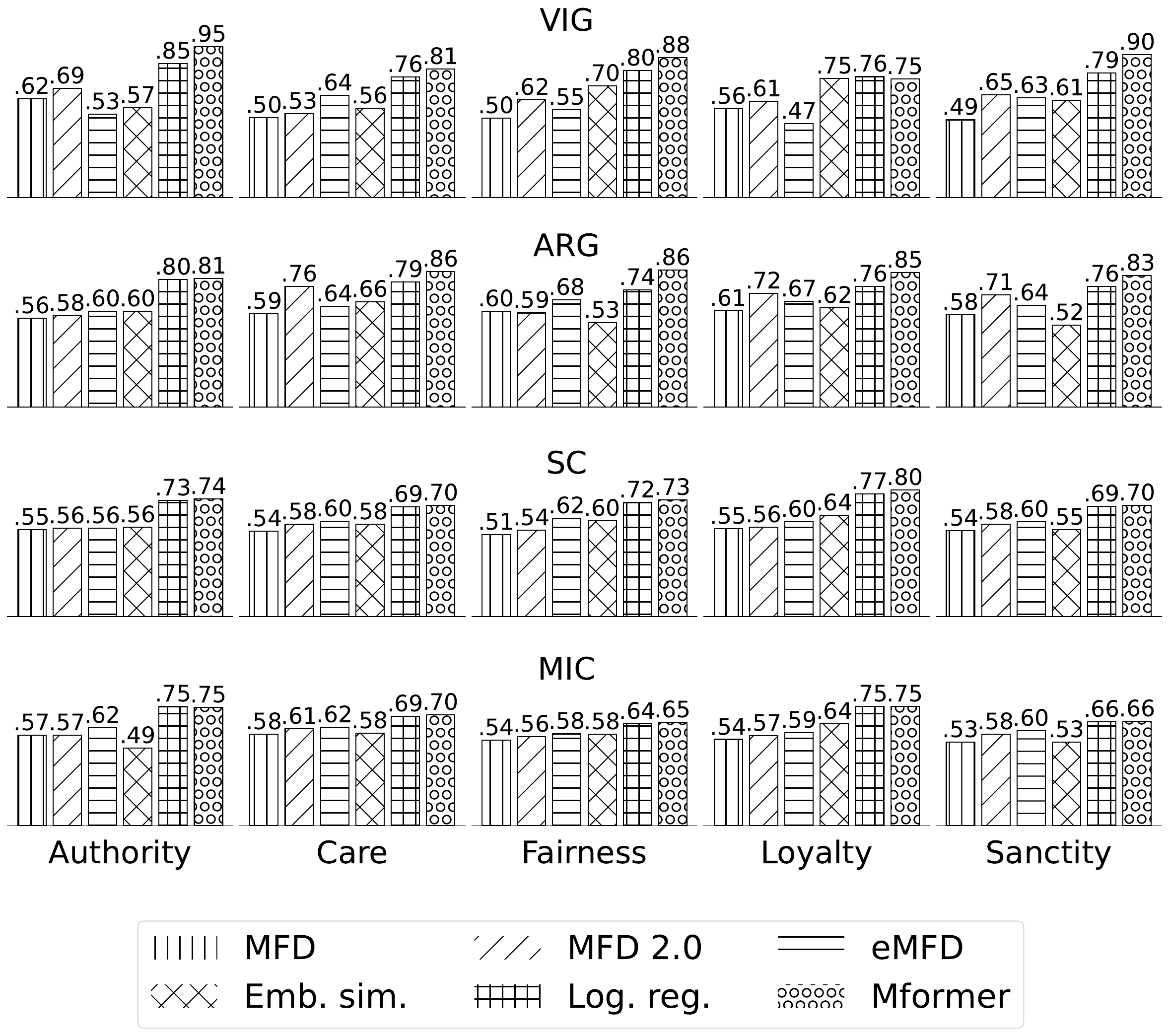}
	\caption{AUC on four external datasets for six moral foundations scoring methods: MFD, MFD 2.0, eMFD, embedding similarity, logistic regression, and \ourmodel.}
	\label{fig:external_aucs}
\end{figure}

As shown in \Cref{sec:mf_classifiers}, \ourmodel demonstrates superior predictive performance to existing methods on the test set. Here, we further highlight that \ourmodel also generalizes well to other data domains---one in the psychology literature and three in NLP---without any further fine-tuning. Specifically, in this section, we describe each dataset in detail and discuss \ourmodel's performance presented in \Cref{fig:external_aucs}. Cross-domain evaluation of moral foundations classifiers has been studied in \citet{liscioCrossDomainClassificationMoral2022}; however, the ``domains'' in their work are only restricted to the seven topics in the Twitter dataset (described in \Cref{sec:mf_classifiers:dataset}). Given the positive results recorded, we emphasize the potential of \ourmodel to be adopted to many analyses of moral rhetoric based on MFT without the costly training of a new model.

\header{Moral foundation vignettes (VIG)}
\label{sec:external_eval:VIG}
\citep{cliffordMoralFoundationsVignettes2015} This dataset contains 115 vignettes designed by the authors to assess humans' classification of moral foundations. Each vignette is a short description of a behavior that violates a foundation. An example for \foundation{fairness} is \example{``You see a politician using federal tax dollars to build an extension on his home.''} As presented in \Cref{fig:external_aucs}, \ourmodel performs very well on this dataset, achieving an AUC of 0.95 for \foundation{authority} and higher than the second-best method, logistic regression, by 7--15\%. The only surprising exception is \foundation{loyalty}, on which \ourmodel achieves an AUC of 0.75, slightly lower than logistic regression of 0.76 and equal to embedding similarity. Upon inspection, we find that some examples of \foundation{loyalty} tend to be misclassified as \foundation{authority} like the following vignette: \example{``You see a head cheerleader booing her high school's team during a homecoming game.''}

\header{Moral arguments (ARG)} 
\label{sec:external_eval:ARG}
This dataset contains 320 arguments taken from two online debate platforms \citep{wachsmuthComputationalArgumentationQuality2017}. \citet{kobbeExploringMoralityArgumentation2020} subsequently labeled each argument with moral foundations. On this dataset, we also observe very good results for \ourmodel with all AUC between 0.81 and 0.86, the highest among all methods and up to 17\% higher than the AUC for logistic regression. We find \foundation{authority} and \foundation{sanctity} relatively more difficult to classify. Some instances of \foundation{authority} tend to be confused with \foundation{care}; e.g., \example{``Some kids don't learn by spanking them.So why waste your time on that, when you can always take something valuable away from them.''} This is also observed for arguments containing \foundation{sanctity}---see \Cref{appn:external_datasets} for an example.

\header{Social chemistry (SC)}
\label{sec:external_eval:SC}
\citep{forbesSocialChemistry1012020}
This dataset contains 292K moral rules-of-thumbs (RoTs) labeled with moral foundations, social judgment and others. We use the test set and score all of its 29K instances. For \ourmodel, the AUC ranges between 0.70 and 0.80---highest among all methods---with specifically high AUC for \foundation{loyalty}. We also find that logistic regression comes close to \ourmodel, and is much better than word count methods which often perform marginally better than chance. As we explain in more detail in \Cref{appn:external_datasets}, the relatively low performance of \ourmodel, compared to that observed in VIG or ARG, may be attributed to this dataset's high level of label noise. As an example, the following RoT is predicted with a very high score for \foundation{care} but does not contain this ground-truth label: \example{``People should temper honesty with compassion, especially when it comes to family."}

\header{Moral Integrity Corpus (MIC)}
\label{sec:external_eval:MIC}
This dataset contains 99K annotated RoTs derived from 38K responses to questions on Reddit by chatbots to facilitate the study of their moral biases \citep{ziemsMoralIntegrityCorpus2022}. We use the test set with 11K examples for evaluation. Similar to SC, the AUC for \ourmodel on this dataset, ranging from 0.65 to 0.75, is lower than that on VIG or ARG, but remains the highest among all methods. We also suspect that this is largely due to label noise in the dataset as the RoTs were labeled similarly to those in SC. For instance, this RoT is predicted with a high score for \foundation{fairness} but does not contain this label: \example{``It's wrong to fight in an unjust war.''}

\section{Studying Moral Dilemmas and Controversies using \ourmodel}
\label{sec:measurements}

\begin{figure}[t]
	\centering
	\includegraphics[width=1\linewidth]{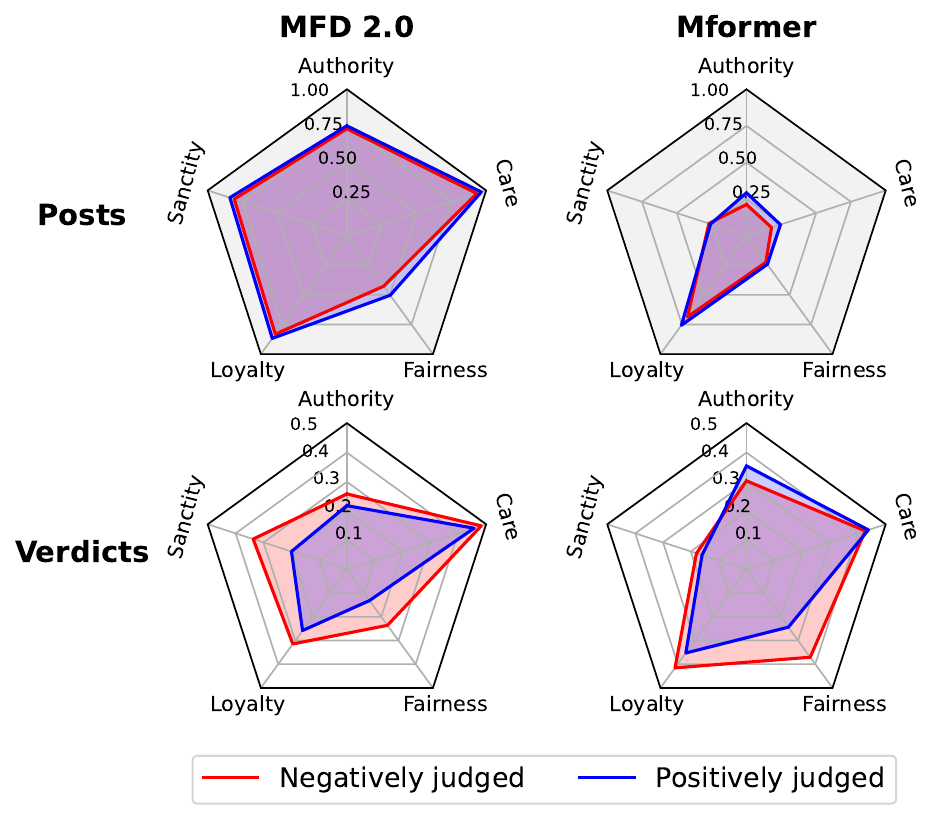}
	\caption{Posts (top) and verdicts (bottom) in the (\textit{family}, \textit{marriage}) topic pair on AITA. Each number in a radar plot indicates the proportion of posts (or verdicts) that contain the corresponding moral foundation. The moral foundations are detected by two methods: MFD 2.0 and \ourmodel. Red (resp. blue) indicates negative (resp. positive) verdicts.}
	\label{fig:aita:radar_topic_pairs}
\end{figure}

\begin{figure*}[t]
	\centering
	\includegraphics[width=1\linewidth]{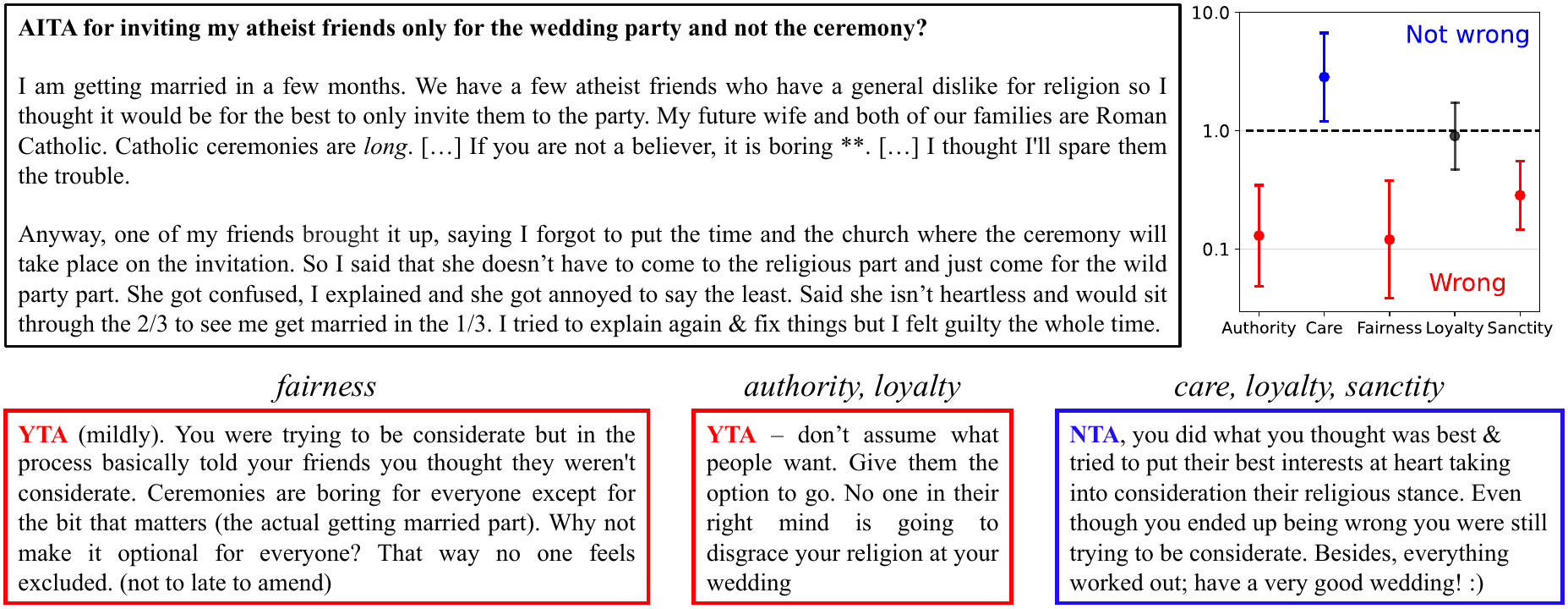}
	\caption{Example controversial thread on AITA. \emph{Top left}: the post, including its title in bold and body text. \emph{Top right}: odds ratios and 95\% CIs between the presence of a moral foundation in a judgment and the judgment's valence. Values above the dashed horizontal line indicate that a foundation is associated with positive valence (i.e., ``NTA'' or ``NAH''), while values below the dashed line indicate an association with a negative (i.e., ``YTA'' or ``ESH'') judgment. \emph{Bottom}: three judgments for this post. The foundations contained in each judgment are annotated at the top.}
	\label{fig:aita_filtered:odds_ratios}
\end{figure*}

So far, we have established that current methods used to score moral foundation may actually perform not much better than chance, and we have advocated for the adoption of \ourmodel based on its good performance across a number of domains without any further fine-tuning. In this section, we present some case studies, partly replicating some previous work, that highlight \ourmodel's efficacy in discovering patterns of morality in several socio-political domains.

% the efficacy of the MFT in discovering patterns of morality in several socio-political domains. We also highlight how sensitive downstream results can be when researchers replace word count with more accurate classifiers like \ourmodel to label large corpora of text.

%\subsection{Moral dimensions in moral dilemma topics}
\subsection{Moral Dimensions in Everyday Conflicts}
\label{sec:measurements:aita_radar}

Recently \citet{nguyenMappingTopics1002022b} analyzed over 100,000 moral discussions on \texttt{r/AmItheAsshole} (AITA), where users post an interpersonal conflict they have experienced and ask the community to judge if they are in the wrong. The authors used topic modeling to find the most salient topics of discussion in this community and found that topics and topic pairs are a robust thematic unit over AITA content. They then used the MFD 2.0 to label all posts and verdict comments to examine the patterns of framing and judgment pertaining to moral foundations across all topics and topic pairs.

Here we replicate the same study, this time using \ourmodel as a moral foundation labeling method instead of MFD 2.0. Our aim is to examine whether the findings in the previous study still hold when a better classifier is used. For more detail on the setting, see \Cref{appn:aita:radar}. Similar to \citet{nguyenMappingTopics1002022b}, we calculate the \emph{foundation prevalence}---defined as the proportion of posts/verdicts that contain each moral foundation---in each topic to examine the relative importance of these five moral foundations within every sphere of moral discussion.

% the result of which is presented in \Cref{fig:aita:all_posts_radar,fig:aita:all_verdicts_radar} in the appendix. 

%We find that the relative importance moral foundations in most topics changes when \ourmodel is used to label posts and verdicts instead of MFD 2.0. For example, posts in the topic \topic{family} were previously perceived to be primarily concerned with \foundation{care}, \foundation{loyalty} and \foundation{sanctity}; however, our results suggest that the dominant foundation is \foundation{loyalty}.

\Cref{fig:aita:radar_topic_pairs} presents the radar plots for foundation prevalence among all posts and verdicts in the (\topic{family}, \topic{marriage}) topic pair. Using MFD 2.0, we find that all foundations except \foundation{fairness} are salient among posts in this topic; however, the results by \ourmodel indicate that only \foundation{loyalty} is dominant. For verdicts, while the results by \ourmodel agree with the previous finding that the foundation \foundation{care} is significant, we also find that it is \foundation{loyalty}, not \foundation{sanctity}, that is of major concern among these judgments. Differences in prevalence also hold for most of the 47 topics found in that study, which we present in \Cref{fig:aita:all_posts_radar,fig:aita:all_verdicts_radar} in the appendix.

% When looking at topic pairs, we find that the difference in findings is even more significant. In \Cref{fig:aita:radar_topic_pairs}, we present the radar plots for foundation prevalence in all posts and verdicts in the (\topic{family}, \topic{marriage}) topic pair. Using MFD 2.0, we find that all foundations except \foundation{fairness} are salient among posts in this topic; however, the results by \ourmodel indicate that only \foundation{loyalty} is dominant. For verdicts, while the results by \ourmodel agree with the previous finding that the foundation \foundation{care} is significant, we also find that it is \foundation{loyalty}, not \foundation{sanctity}, that is of major concern among these judgments.

% These findings suggest that while word count may be a simple off-the-shelf method, the downstream results it produces may be misleading. We therefore advocate for the adoption of \ourmodel in labeling large samples of text.

These results highlight that important findings are subject to change when researchers use different scoring methods. We believe that \ourmodel is a good candidate for adoption as it performs better than all existing methods across numerous benchmarks (\Cref{sec:mf_classifiers,sec:external_eval}) and is robust to its internal binary thresholds (see \Cref{fig:aita:radar_topic_pairs_2levels} in the appendix).

%\subsection{Understanding conflicting moral judgments}
\subsection{Moral Dimensions in Opposing Judgments}
\label{sec:measurements:aita_filtered}

% We demonstrate the utility of moral foundations in characterizing conflicting judgments in controversial moral debates. To this end, we use the dataset by \citet{nguyenMappingTopics1002022b} containing over 100,000 moral discussions on \texttt{r/AmItheAsshole}. Each discussion contains a post---a description of a situation or action its author recently experienced---and judgments by other Reddit users in the form of comments. There are five types of judgments one can make: YTA (``you're the a*''), NTA (``not the a*''), ESH (``everyone sucks''), NAH (``no a*s here'') and INFO (``more information needded''). We consider a thread to be controversial if it received at least 50 top-level comments containing one of the four judgments: YTA, NTA, ESH and NAH. Further, the proportion of judgments with positive valence toward the author (YTA + ESH, dubbed YA) or those with negative valence (NTA + NAH, dubbed NA) must not exceed 70\% of all judgments. Only comments within 18 hours since a post was created are counted. This yields 2,135 posts and 466,485 judgments, with the median number of judgments per post being 110.

Here we present another study using AITA content. In contrast to \citet{nguyenMappingTopics1002022b}, who only looked at posts and their verdicts (i.e., highest-scoring judgments), we aim to analyze \emph{all judgments} within \emph{one post} to find any systematic differences in conflicting judgments---those that claim the author is in the wrong and those who think otherwise. To do so, we filter the dataset to contain only ``controversial'' posts with at least 50 judgments, which are split somewhat equally between the positive and negative valence. This yields 2,135 posts accompanied by 466,485 judgments. A detailed setting can be found in \Cref{appn:aita:aita_filtered}.

We use \ourmodel to score every post and judgment on five moral foundations. To convert the scores to binary labels, we set the highest-scoring 20\% of the posts to contain that foundation; the same applies to judgments. In other words, a post (or judgment) is said to contain foundation \foundation{loyalty} if it scores higher than at least 80\% of all posts (or all comments) on \foundation{loyalty}. This rather high threshold is motivated by our striving for high precision at the expense of recall. For each post and a foundation $f$, we calculate the odds that a positive (``not wrong'') judgment contains $f$, compared to the odds that such a positive judgment does not contain $f$. 
% If the odds ratio is above 1 (resp. below 1), the foundation $f$ is said to be associated with the positive (resp. negative) judgment. 

\Cref{fig:aita_filtered:odds_ratios} displays an example controversial thread on AITA. The post received 397 positive (the author is ``not wrong'') and 471 negative (``wrong'') judgments. The odds ratio plot at the top right suggests some distinct patterns: those that think the author is not wrong are 1.6 times more likely to focus on \foundation{care} (OR=1.57, 95\% CI=[1.08, 2.28]) by arguing that the author is simply looking out for their atheist friends and helping them avoid a religious event they might be uncomfortable with. On the other hand, those that judge the author to be in the wrong are 2.5 times more likely to emphasize the foundation \foundation{fairness} (OR=0.40, 95\% CI=[0.24, 0.66]) by stating that the author ought to be fair to all guests by inviting them to the ceremony as well. Negative judgments are also more likely to underlie \foundation{authority} (OR=0.41, 95\% CI=[0.27, 0.63]) and \foundation{sanctity} (OR=0.58, 95\% CI=[0.43, 0.77]); these judgments often argue that the author should not assume, on behalf of his friends, that they would not want to be at the ceremony just because it is religious and they are not. Finally, we find that the foundation \foundation{loyalty} is not significantly associated with positive or negative judgments (OR=0.96, 95\% CI=[0.72, 1.27]); it seems that within this situation, the author's loyalty to their friends is not being questioned as much as other moral values. 

Not all moral dilemmas give significant findings. Even after filtering less controversial posts, we only find, among the 2,135 controversial threads, 1,136 (53.2\%) of them to have significant results where at least one moral foundation is associated with a clear valence of wrong/not wrong. Nevertheless, the results suggest that conflicting judgments for moral dilemmas can be explained by their appeal to different moral foundations, which can be robustly detected by \ourmodel.

% \subsection{Characterizing stance on controversial topics}
\subsection{Moral Dimensions of Different Stances}
\label{sec:measurements:twitter_stance}

% In the final case study, we aim to analyze people's stance toward some controversial topics. 
Stance classification is concerned with determining whether a person is in favor of or against a proposition or a topic \cite{mohammadStanceSentimentTweets2017}. Here we explore the endorsement of moral foundations and its relationship with people's stance toward several controversial topics, as expressed through tweets. This study partially reproduces the analysis in \citet{rezapourIncorporatingMeasurementMoral2021}. We use a dataset of 4,870 tweets across six political topics: \topic{atheism}, \topic{climate change is a real concern}, \topic{Donald Trump}, \topic{feminist movement}, \topic{Hillary Clinton} and \topic{legalization of abortion} \citep{mohammadSemEval2016TaskDetecting2016}. Since we are only interested in polar stances (in favor or against), we remove all instances where the stance was labeled as ``none'' (i.e., either neutral or irrelevant to the topic). This results in 3,614 tweets in total, with the number of tweets per topic between 361 and 779. 
% We defer the detailed setting, some example tweets in each topic and a preliminary analysis to \Cref{appn:twitter_stance}.

% We refer to \Cref{appn:twitter_stance} for a detailed setup.

We score all tweets on every moral foundation using \ourmodel
%A preliminary analysis in \Cref{appn:twitter_stance:mwu} reveals some clear patterns between tweets in favor of and those against a topic. For example, tweets supporting \topic{atheism} score significantly higher than those against it on two foundations---\foundation{authority} and \foundation{fairness}---often focusing on criticizing the political authority by religious leaders or on condemning discrimination based on religion. 
and convert the raw predicted scores to binary labels by setting the highest-scoring 20\% of the tweets to contain each foundation. We then replicate the chi-square analysis by \citet[Section 5.3]{rezapourIncorporatingMeasurementMoral2021}, the results of which are presented in \Cref{tab:twitter_stance:chisquare} in the appendix. Apart from some similar findings, such as that no association is found between moral foundations and stance toward \topic{feminist movement}, we find many differences after using \ourmodel for detecting foundations instead of the MFD in the previous study. For instance, \citet{rezapourIncorporatingMeasurementMoral2021} found no significant correlation within the topic \topic{Donald Trump}, but our results show that this happens for three foundations: \foundation{authority} ($p < 0.001$), \foundation{care} ($p < 0.05$) and \foundation{loyalty} ($p < 0.001$). We suspect that these differences are largely due to the aforementioned shortcomings of word count methods: Most tweets are short, and so can easily fail to contain lexical entries, leading to zero counts for some foundations and hence false negatives.

\begin{figure}[t]
	\centering
	\includegraphics[width=1\linewidth]{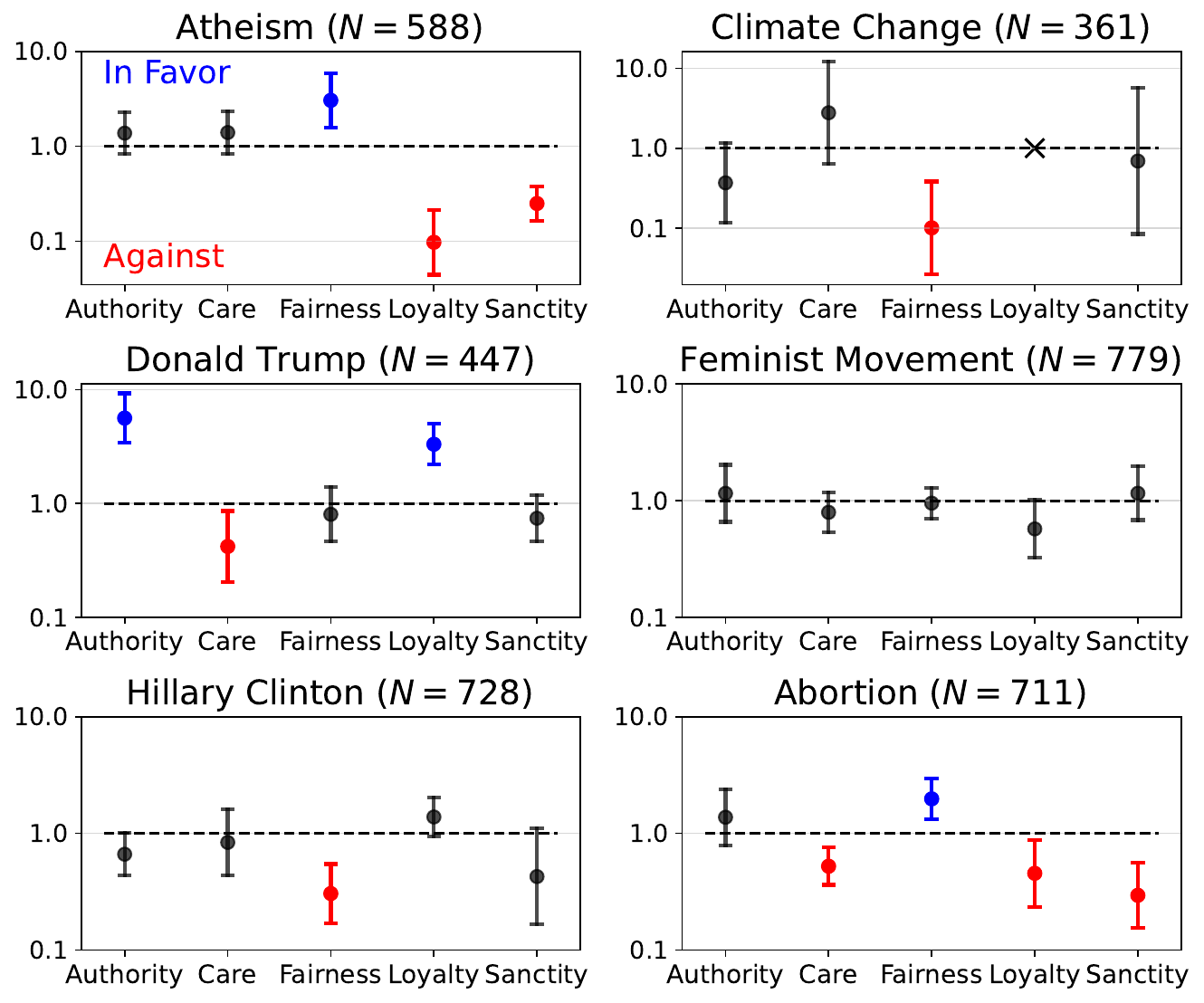}
	\caption{Odds ratios and 95\% CIs between the presence of a moral foundation in a tweet and the tweet's stance toward each topic. The ``X'' mark for \foundation{loyalty} in climate change is due to no tweets containing this foundation.}
	\label{fig:twitter_stance:odds_ratios}
\end{figure}

Finally, we estimate the effect size of these associations by calculating the odds ratio (OR) between two binary variables: whether a foundation is present in a tweet and whether the tweet's author is in favor of a topic. \Cref{fig:twitter_stance:odds_ratios} shows the ORs for each topic over all five foundations. Almost all topics show a clear distinction in moral foundations between tweets in favor of and those against it. For example, Twitter users supporting the \topic{legalization of abortion} are 2 times more likely to appeal to \foundation{fairness} (OR=1.98, 95\% CI=[1.32, 2.97]), such as in the tweet \example{``Good morning @JustinTrudeau. Do you plan to tell @WadePEILiberal that women on PEI deserve the right to choose? \#cdnpoli \#peipoli.''} On the other hand, those against this view are more likely to underlie the foundations \foundation{care} (OR=0.52, 95\% CI=[0.36, 0.75]), \foundation{loyalty} (OR=0.45, 95\% CI=[0.24, 0.88]), and, especially, \foundation{sanctity} (OR=0.29, 95\% CI=[0.15, 0.56]) as in the tweet \example{``U don't have to be a religious to be pro-life. All u have to believe is that every life is sacred.''} The strong association between the disapproval of abortion and the foundations \foundation{sanctity} and \foundation{care} is consistent with the finding in \citet{kolevaTracingThreadsHow2012}, although this prior work found mixed results when it comes to \foundation{fairness} and \foundation{loyalty}.

Another topic that allows us to compare with prior findings is \topic{climate change}. Previously, \citet{feinbergMoralRootsEnvironmental2013} showed that, on social media and news outlets, the moral rhetoric surrounding the environmental discourse primarily focuses on the \foundation{harm/care} foundation. While this may be true, we do not find significant evidence that an appeal to \foundation{care} signifies a positive or negative attitude toward climate change (OR=2.78, 95\% CI=[0.64, 12.07]). This could be due to the fact that \foundation{care} is salient in arguments on both sides of the discourse, but each side portrays different framing patterns around the foundation, which is an interesting topic of examination for future work. Nevertheless, our results show that \foundation{fairness} is highly associated with the negative stance toward this topic (OR=0.10, 95\% CI=[0.03, 0.38]). Tweets against climate change often focus on the unfair treatment of those who hold ``alternative'' viewpoints, such as in \example{``Climate deniers is a term used to silence those pointing out the hypocrisy in the fanatical zeal on \#climatetruth.''} 

% Another example is the topic \topic{Donald Trump} where proponents are much more likely to underlie the foundations \foundation{authority} (5.62 times) and \foundation{loyalty} (3.31 times), such as in the tweet \example{``@realDonaldTrump You have our backs, and we have yours. You have my vote. Boycotting anyone who opposes your patriotism. [...]''} Those who oppose the former U.S. president, however, dominate in their adherence to the foundation \foundation{care} (2.38 times), often criticizing his mistreatment of certain groups such as in the tweet: \example{``Witnessing the destruction of the \#middleclass If you are black, white Hispanic, or any other creed support \#Trump \#usa \#4thjuly.''}

% This analysis highlights the utility of the MFT in analyzing political stance on social media. Viewpoints in favor of a topic can typically differ from those against that topic based on the moral foundations they appeal to. We also show that labeling tweets with foundations using word count methods may lead to false negative results, such as in the topic \topic{Donald Trump} described above.

\section{Conclusion}
\label{sec:conclusion}

In this paper, we examine tools for characterizing moral foundations in social media content. We show that empirical findings based on MFT are specifically dependent on the scoring methods with which researchers label their data. Furthermore, we find that these methods, especially word count programs, often perform poorly and are biased in several ways. We instead propose \ourmodel, a language model fine-tuned on diverse datasets to recognize moral foundations, as an alternative classifier. We highlight the superior performance of \ourmodel compared to existing methods across several benchmarks spanning different domains. Using \ourmodel to analyze two datasets on Reddit and Twitter, we demonstrate its utility in detecting important patterns of moral rhetoric, such as conflicting judgments for the same moral dilemma that depend on the specific foundations upon which participants rely.

% in two datasets as that conflicting judgments for the same moral dilemma can be explained by their appeal to different moral foundations; or that people's stance on several controversial topics, like the legalization of abortion, as expressed through tweets, can also be due to the foundations they underscore.

% based on fine-tuning a language model using annotated data, as an alternative and more accurate moral foundation detector. Using \ourmodel, we discover several patterns of moral beliefs such as that conflicting judgments for the same moral dilemma can be explained by their appeal to different moral foundations; or that people's stance on several controversial topics, like the legalization of abortion, as expressed through tweets, can also be due to the foundations they underscore.

% Labeling content---e.g., posts on Reddit or tweets on Twitter---with moral foundations typically relies on automatic approaches, and we show that word count programs based on human-crafted lexicons can produce problematic predictions due to their heuristic nature. We instead introduce \ourmodel, based on fine-tuning a language model using annotated data, as an alternative and more accurate moral foundation detector. Using \ourmodel, we discover several patterns of moral beliefs such as that conflicting judgments for the same moral dilemma can be explained by their appeal to different moral foundations; or that people's stance on several controversial topics, like the legalization of abortion, as expressed through tweets, can also be due to the foundations they underscore.

\header{Limitations} Labeling moral foundations is an inherently subjective task.
%, as evidenced by the low inter-annotator agreement rates found in previous studies \citep{hooverMoralFoundationsTwitter2020a,tragerMoralFoundationsReddit2022} and by the label noise in some datasets presented in \Cref{sec:external_eval}. 
%This has a direct effect on models trained on these datasets, and we believe that an effort to correct label noise will be beneficial. With regard to the findings in \Cref{sec:measurements}, 
We acknowledge that the scope of morality goes beyond what is observable on social media, and internet samples may not be representative of moral life as a whole.  However, we think that social media datasets are large enough in size that one is able to draw important conclusions about how certain communities describe their moral concerns and judgments. {Potential misuse of the proposed model and method could include content targeting and spreading mis-information.}

\header{Ethical considerations} All datasets used in this paper are publicly available from prior work. In all analyses, we ensure that personal or potentially self-identifying information, such as usernames or URLs, is removed. The findings we present are descriptive insofar as morality is relevant to social issues debated online and we do not make normative claims within any of the examined domains.

\header{Broader perspectives} We primarily focus on the MFT because it is popular, which is in turn due to the availability of large annotated datasets. Even within this framework, the focus on how exactly each moral foundation is portrayed (e.g., as a vice or a virtue) is potentially important and can yield more fine-grained, novel results. In addition, alternative categorizations of moral beliefs based on competing theories exist and are worthy of examination. Among these, morality-as-cooperation is a rising candidate \citep{curryMoralityCooperationProblemCentred2016a,curryItGoodCooperate2019a}, providing another set of dimensions with a different theoretical foundation.

{Recognizing the prevalence of word count methods in detecting moral foundations, we believe a thorough evaluation is warranted. This work establishes that, similar to other contexts such as sentiment analysis, lexicon-based word count programs often ignore contextual information, do not generalize well due to domain dependency, and may reveal social biases due to the way words are hand-chosen. Through a careful treatment of Mformer from training to (cross-domain) evaluation, we show that machine learning-based methods have potential to overcome these challenges.}

Social media is a prolific resource for studying many aspects of morality, such as what moral dimensions are emphasized on both sides of a controversial issue. Findings from these studies can inform us about important social norms that guide debate on these issues, and can have practical implications for automated content moderation within online discussion forums as well as for the understanding of moral conflicts by machines. This work aims to caution researchers interested in this direction about the limitations of the available tools for measuring moral dimensions, and provide a more robust and reproducible alternative that has been evaluated across a range of benchmarks. %. A thorough understanding of their limitations ensures that results are robust, reproducible and have meaningful implications. 

\section*{Acknowledgements}

This work is supported by the CSIRO CRP Program and the Australian Research Council Project DP190101507. The authors would like to thank members of the Humanising Machine Intelligence Project at ANU for their feedback.

% \newpage
	
{\small \bibliography{main-bib.bib}}

\newpage

\section*{Paper Checklist}
\begin{enumerate}
	\item For most authors...
	\begin{enumerate}
		\item  Would answering this research question advance science without violating social contracts, such as violating privacy norms, perpetuating unfair profiling, exacerbating the socio-economic divide, or implying disrespect to societies or cultures?
		\answerYes{Yes. This work introduces a robust, highly accurate tool for measuring moral foundations in text, which can help understand moral sentiment across different social contexts. To the best of our knowledge, the findings are descriptive and do not violate any of the above.}
		\item Do your main claims in the abstract and introduction accurately reflect the paper's contributions and scope?
		\answerYes{Yes}
		\item Do you clarify how the proposed methodological approach is appropriate for the claims made? 
		\answerYes{Yes, see \Cref{sec:mf_classifiers} in the main text and \Cref{appn:dataset,appn:supervised_classifiers} for more detail on our methodology.}
		\item Do you clarify what are possible artifacts in the data used, given population-specific distributions?
		\answerYes{Yes}
		\item Did you describe the limitations of your work?
		\answerYes{Yes. We have discussed these in \Cref{sec:conclusion}.}
		\item Did you discuss any potential negative societal impacts of your work?
		\answerYes{Yes. \ourmodel and the findings in this paper are useful in characterizing moral sentiment on social media. We believe they do not pose any direct negative societal impacts.}
		\item Did you discuss any potential misuse of your work?
	    \answerYes{Yes. This was discussed in \Cref{sec:conclusion} under ``Limitations''.}	
        %\answerNo{No. \ourmodel is used to measure the relevance of each moral foundation to a document. We do not think there exists any significant misuse of this tool.}
		\item Did you describe steps taken to prevent or mitigate potential negative outcomes of the research, such as data and model documentation, data anonymization, responsible release, access control, and the reproducibility of findings?
		\answerYes{Yes. In \Cref{sec:conclusion}.}
		\item Have you read the ethics review guidelines and ensured that your paper conforms to them?
		\answerYes{Yes.}
	\end{enumerate}
	
	\item Additionally, if your study involves hypotheses testing...
	\begin{enumerate}
		\item Did you clearly state the assumptions underlying all theoretical results?
		\answerYes{Yes, the hypothesis tests used in \Cref{sec:measurements:aita_filtered,sec:measurements:twitter_stance} are described in more detail in \Cref{appn:aita:aita_filtered} and \Cref{appn:twitter_stance:ORs}, respectively.}
		\item Have you provided justifications for all theoretical results?
		\answerYes{Yes}
		\item Did you discuss competing hypotheses or theories that might challenge or complement your theoretical results?
		\answerYes{Yes}
		\item Have you considered alternative mechanisms or explanations that might account for the same outcomes observed in your study?
		\answerYes{Yes}
		\item Did you address potential biases or limitations in your theoretical framework?
		\answerYes{Yes}
		\item Have you related your theoretical results to the existing literature in social science?
		\answerYes{Yes, MFT is a prominent theory in moral psychology, as described in \Cref{sec:related_work:MFT}.}
		\item Did you discuss the implications of your theoretical results for policy, practice, or further research in the social science domain?
		\answerYes{Yes. The results based on \ourmodel's predictions may challenge the validity of the claims by prior work, as we discussed in \Cref{sec:measurements}.}
	\end{enumerate}
	
	\item Additionally, if you are including theoretical proofs...
	\begin{enumerate}
		\item Did you state the full set of assumptions of all theoretical results?
		\answerNA{NA}
		\item Did you include complete proofs of all theoretical results?
		\answerNA{NA}
	\end{enumerate}
	
	\item Additionally, if you ran machine learning experiments...
	\begin{enumerate}
		\item Did you include the code, data, and instructions needed to reproduce the main experimental results (either in the supplemental material or as a URL)?
		\answerYes{Yes. The main paper and supplemental material describe the dataset and training instructions in detail. Upon publication, these will be released publicly.}
		\item Did you specify all the training details (e.g., data splits, hyperparameters, how they were chosen)?
		\answerYes{Yes, see \Cref{sec:mf_classifiers:classifiers:roberta} in the main text and \Cref{appn:supervised_classifiers:roberta}.}
		\item Did you report error bars (e.g., with respect to the random seed after running experiments multiple times)?
		\answerNA{NA}
		\item Did you include the total amount of compute and the type of resources used (e.g., type of GPUs, internal cluster, or cloud provider)?
		\answerYes{Yes, see \Cref{appn:supervised_classifiers:roberta}.}
		\item Do you justify how the proposed evaluation is sufficient and appropriate to the claims made? 
		\answerYes{Yes, the justification for the AUC metric is found in \Cref{sec:mf_classifiers:evaluation:eval_metric}.}
		\item Do you discuss what is ``the cost'' of misclassification and fault (in)tolerance?
		\answerNA{NA}
		
	\end{enumerate}
	
	\item Additionally, if you are using existing assets (e.g., code, data, models) or curating/releasing new assets...
	\begin{enumerate}
		\item If your work uses existing assets, did you cite the creators?
		\answerYes{Yes. See \Cref{sec:mf_classifiers:dataset}.}
		\item Did you mention the license of the assets?
		\answerYes{Yes. All datasets are publicly available.}
		\item Did you include any new assets in the supplemental material or as a URL?
		\answerNo{No.}
		\item Did you discuss whether and how consent was obtained from people whose data you're using/curating?
		\answerNA{NA.}
		\item Did you discuss whether the data you are using/curating contains personally identifiable information or offensive content?
		\answerYes{Yes. For the Reddit dataset used in \Cref{sec:measurements:aita_radar,sec:measurements:aita_radar}, we follow the process of removing URLs and potential self-identifying information by its curators \citet{nguyenMappingTopics1002022b}.}
		\item If you are curating or releasing new datasets, did you discuss how you intend to make your datasets FAIR? % (see \citet{fair})?
		\answerNA{NA}
		\item If you are curating or releasing new datasets, did you create a Datasheet for the Dataset? %(see \citet{gebru2021datasheets})? 
		\answerNA{NA}
	\end{enumerate}
	
	\item Additionally, if you used crowdsourcing or conducted research with human subjects...
	\begin{enumerate}
		\item Did you include the full text of instructions given to participants and screenshots?
		\answerNA{NA}
		\item Did you describe any potential participant risks, with mentions of Institutional Review Board (IRB) approvals?
		\answerNA{NA}
		\item Did you include the estimated hourly wage paid to participants and the total amount spent on participant compensation?
		\answerNA{NA}
		\item Did you discuss how data is stored, shared, and deidentified?
		\answerNA{NA}
	\end{enumerate}
	
\end{enumerate}

\onecolumn

\appendix
\renewcommand\thefigure{\thesection.\arabic{figure}}
\counterwithin{table}{section}

%\begin{figure}[t]
%	\centering
%	\includegraphics[width=1\linewidth]{figs/attention_scores_roberta_one_example.pdf}
%	%		\caption{Attention scores extracted from five RoBERTa classifiers for moral foundations.
%		%		\label{fig:roberta_mf:attention_scores}
%\end{figure}

% \usepackage{multirow}
\section{Moral Foundations Dictionaries}
\label{appn:mfd}

\begin{table}[]
	\centering
	\caption{Some example words in three lexicons used for scoring moral foundations: MFD, MFD 2.0 and eMFD. For eMFD, the words are sorted by their corresponding foundation weights so that highest-scoring words appear first. For the intersection of these lexicons, see \Cref{fig:mfds_aita_wordcount}.}
	\small
	\begin{tabular}{ccp{0.75\linewidth}}
		\toprule
		Foundation                 & Lexicon & Example words                                                                                                                                                                   \\ 
		\midrule
		\multirow{3}{*}[-5mm]{Authority} & MFD     & abide, command, defiance, enforce, hierarchical, lawful, nonconformist, obey, obstruct, protest, rebel, subvert, tradition, urge, violate                                       \\
		& MFD 2.0 & allegiance, boss, chaotic, dictate, emperor, father, hierarchy, illegal, lionize, matriarch, overthrow, police, revere, subvert, uprising                                       \\
		& eMFD    & elect, throwing, ambitions, sponsored, rebellion, protested, voluntarily, denounced, immigrant, separatist, banning, counter, interfere, nationals, protesting                  \\ 
		\midrule
		\multirow{3}{*}[-5mm]{Care}      & MFD     & abuse, benefit, compassion, cruel, defend, fight, guard, harmonic, impair, killer, peace, protect, safeguard, violence, war                                                     \\
		& MFD 2.0 & afflict, brutalize, charitable, discomfort, empathized, genocide, harshness, inflict, mother, nurse, pity, ravage, torture, victim, warmhearted                                 \\
		& eMFD    & tortured, pocket, cruel, harsh, sexually, raping, hostility, income, persecution, stranded, knife, drivers, imprisonment, killed, punishments                                   \\ 
		\midrule
		\multirow{3}{*}[-5mm]{Fairness}  & MFD     & balance, bigot, constant, discriminate, egalitarian, equity, favoritism, honesty, impartial, justifiable, prejudge, reasonable, segregate, tolerant, unscrupulous               \\
		& MFD 2.0 & avenge, bamboozle, conniving, defraud, equity, freeload, hypocrisy, impartial, liar, mislead, proportional, reciprocal, swindle, trust, vengeance                               \\
		& eMFD    & rigged, undermining, disproportionately, compensation, discrimination, punished, steal, throwing, flawed, instances, excessive, restrictive, inability, discriminatory, tariffs \\ 
		\midrule
		\multirow{3}{*}[-5mm]{Loyalty}   & MFD     & ally, clique, comrade, deceive, enemy, familial, guild, immigrate, joint, membership, nationalism, patriotic, spying, terrorism, united                                         \\
		& MFD 2.0 & allegiance, backstab, coalition, fellowship, group, homeland, infidelity, kinship, nation, organization, pledge, sacrifice, tribalism, unpatriotic, wife                        \\
		& eMFD    & disrupt, wing, dictatorship, betrayal, loyalty, legislators, outsiders, renegotiate, fearful, credibility, pocket, couples, exploit, rage, retribution                          \\ 
		\midrule
		\multirow{3}{*}[-5mm]{Sanctity}  & MFD     & austerity, chaste, decency, exploitation, filthy, germ, holiness, immaculate, lewd, obscene, pious, repulsive, sickening, taint, virgin                                         \\
		& MFD 2.0 & abhor, befoul, catholic, exalt, fornicate, hedonism, impurity, leper, marry, nunnery, organic, pandemic, repulsive, trashy, waste                                               \\
		& eMFD    & raping, sexually, swamp, rigged, exploit, tissue, objects, rape, infections, sex, uphold, wing, victimized, smoking, sacred                                                     \\ 
		\bottomrule
	\end{tabular}
	\label{tab:mfd_example_words}
\end{table}

Here we provide more details on the three moral foundations dictionaries (MFDs) described in \Cref{sec:related_work:MF_detection}. Some example words in each dictionary are presented in \Cref{tab:mfd_example_words}. We also describe in detail how a document is scored on five moral foundation dimensions using these dictionaries.

\subsection{Moral Foundations Dictionary (MFD)}
\label{appn:mfd:mfd}

\citet{grahamMoralFoundationsDictionary2012} released the original version of the MFD to accompany their research on moral foundations \citep{haidtRighteousMindWhy2012,grahamMoralFoundationsTheory2013a}. The MFD contains author-compiled 325 entries mapping a word prefix to a foundation and a sentiment. An example mapping is \texttt{venerat* -> AuthorityVirtue}, indicating that any word beginning with ``venerat''---such as ``venerating,'' ``venerated'' or ``veneration''---will be mapped to the virtue of \foundation{authority}. The authors also included another category called ``morality general,'' which we exclude from our method. Finally, as \citet{mokhberianMoralFramingIdeological2020b} have previously converted all prefixes to eligible words, so we use their version of the MFD for word count.\footnote{The dictionary is available online at \url{https://github.com/negar-mokhberian/Moral_Foundation_FrameAxis}.} This contains 591 words in total. 

Since we do not model moral sentiments (virtue and vice), we merge the virtue and vice for each foundation together, giving us five original foundations. To score each document, we first perform tokenization using \texttt{spaCy} \cite[model \texttt{en\_core\_web\_md}, version 3.1.0]{honnibalSpaCyIndustrialstrengthNatural2020}. Then we go over each token to check whether it or its lemma is contained in the MFD. If there is a match, we increment the count of the corresponding foundation by one. For example, if the document contains the word ``venerated,'' the count \foundation{authority} is incremented by one. The result is five counts where each count represents the number of times we encounter a foundation in this document. Finally, these counts are divided by the total number of tokens to give frequencies, or \emph{scores}. For instance, a score of 0.05 for \foundation{loyalty} means that 5\% of the tokens in this document are mapped to \foundation{loyalty}.

Note that a prefix/word can be mapped to more than one foundation. For example, ``exploit'' is mapped to \foundation{care} and \foundation{sanctity}. In cases like this, we increment the count of all eligible foundations.

\subsection{Moral Foundations Dictionary 2.0 (MFD 2.0)}
\label{appn:mfd:mfd2}

In an attempt to extend the word lists in the original MFD, \citet{frimerMoralFoundationsDictionary2019c} started with a list of prototypical (author-compiled) words for each foundation. Then, each list was extended by adding words that are close to the prototypical words; here, ``closeness'' refers to the cosine distance between word embeddings in the word2vec space \citep{mikolovDistributedRepresentationsWords2013}. The newly found words then went under manual validation by the authors who decided whether to keep or remove each word for every foundation. The result is a collection of 2,104 words in the lexicon. We also merge the virtue and vice of each foundation into one, resulting in five unique labels. The scoring of documents is done identically to MFD. 

The scoring of a document is done identically to MFD. Specifically, we merge both sentiments for each foundation, resulting in five total categories. We tokenize the document and check for each token whether it or its lemma is contained in the lexicon. Each time a match occurs, the count for the foundation that the token maps to is incremented by one. Similarly, in cases where a token maps to more than one foundation, all foundations are incremented. Finally, the five counts are normalized by the number of tokens to give foundation frequencies.

\subsection{Extended Moral Foundations Dictionary (eMFD)}
\label{appn:mfd:emfd}

\citet{hoppExtendedMoralFoundations2021} aimed to construct a dictionary in a data-driven fashion. The authors pulled 1,010 news articles from the GDELT dataset \cite{leetaruGDELTGlobalData2013a} and employed over 854 online annotators to label these articles with moral foundations. In particular, each annotator was assigned an article and a specific foundation. The annotator was asked to highlight the parts of the article that signify that foundation. This yielded a total of 73,001 raw notations.

Instead of mapping a word to some specific foundation, the authors consider a mapping to \emph{all} foundations, each with a specific \emph{weight}. The higher the weight, the more ``relevant'' the word is to a foundation. Each weight is between 0 and 1 and represents the frequency with which a word is highlighted in the context of that foundation. For example, the score associated with \foundation{care} for the word ``suffer'' is 0.32, indicating that when annotators were assigned to label \foundation{care} (among all eligible articles) and saw the word ``suffer,'' they highlighted this word 32\% of the time. This is a more nuanced way of associating words with MFs, allowing for flexibility in terms of association strength. Overall, the lexicon contains 3,270 words, each of which contains five scores for the foundations.

To score a document, we follow the procedure detailed in \citep{hoppExtendedMoralFoundations2021}. We start with a 5-dimensional vector of zeros for the foundations. After tokenizing a document, for every token that is in the dictionary, we add the 5-dimensional score vector for that token to the document's scores. Finally, we divide the document vector by the number of tokens that matched the lexicon's entries. The result is five scores for the document, each one between 0 and 1 and equal to the average foundation score contributed by all matching tokens.

\section{Scoring moral foundations using embedding similarity}
\label{appn:embedding_sim}

\begin{table}[t]
	\centering
	\caption{Keywords used to create their concept vectors for each foundation using the embedding similarity method.}
	\small
	\begin{tabular}{cp{0.8\linewidth}}
		\toprule
		Foundation  & Keywords \\
		\midrule
		Authority & authority, obey, respect, tradition, subversion, disobey, disrespect, chaos \\
		Care & kindness, compassion, nurture, empathy, suffer, cruel, hurt, harm \\
		Fairness &  equality, egalitarian, justice, nondiscriminatory, prejudice, inequality, discrimination, biased, proportional, merit, deserving, reciprocal, disproportionate, cheating, favoritism, recognition \\
		Loyalty & loyal, solidarity, patriot, fidelity, betray, treason, disloyal, traitor \\
		Sanctity & purity, sanctity, sacred, wholesome, impurity, depravity, degradation, unnatural \\   
		\bottomrule
	\end{tabular}
	\label{table:classifiers:ddr_keywords}
\end{table}

\emph{Embedding similarity}, or distributed dictionary representations (DDR) \citep{gartenDictionariesDistributionsCombining2018}, is another method used in prior work to score moral foundations in text. Basically, moral foundations and documents are represented by vectors in high-dimensional Euclidean spaces (embeddings). The foundation score for each pair of document and foundation is defined as the cosine similarity between their vector representations.

We start with a \emph{word embedding}, which maps every word in the dictionary to a dense vector in $k$ dimensions. Examples of widely used embeddings are word2vec \citep{mikolovDistributedRepresentationsWords2013} and GloVe \citep{penningtonGloVeGlobalVectors2014}. For a foundation $f$ represented by a set of $n_f$ keywords with their word embeddings $\{w_i \in \mathbb{R}^k : i=1,\ldots, n_f\}$. The vector representation for $f$ is the average of all word vectors
\begin{equation*}
	\label{eq:ddr:foundation_embedding}
	c_f = \frac{1}{n_f} \sum_{i=1}^{n_f} w_i.
\end{equation*}
Documents can be encoded similarly. Given a document $d$ as a sequence of $n_d$ tokens $(w_i \in \mathbb{R}^k)_{i=1}^{n_d}$, the vector representation of the document is
\begin{equation*}
	\label{eq:ddr:document_embedding}
	c_d = \frac{1}{n_d} \sum_{i=1}^{n_d} w_i.
\end{equation*}
To score document $d$ with respect to foundation $f$, we take the cosine similarity between their vector representations:
\begin{equation*}
	\label{eq:ddr:cosine_similarity}
	s_{d, f} = \frac{\sum_{j=1}^{k} [c_d]_j [c_f]_j}{\sqrt{\sum_{j=1}^{k} [c_d]_j^2} \sqrt{\sum_{j=1}^{k} [c_f]_j^2}},
\end{equation*}
where $[\cdot]_j$ denotes the $j$th component of a vector. The score $s_{d, f}$ is in the range $[-1, 1]$, and the higher the score, the more ``similar'' $c_d$ and $c_f$ are.

In \Cref{sec:mf_classifiers:classifiers:existing}, we use the GloVe word embedding \citep{penningtonGloVeGlobalVectors2014}, specifically the ``Twitter'' version which was trained using a corpus of 27 billion tokens and contains 1.2 million unique vectors in $k = 200$ dimensions.\footnote{The embedding can be found at \url{https://nlp.stanford.edu/projects/glove}.} Documents are tokenized similarly to that in the word count methods described above and all tokens are lowercased. For every token that does not exist in the embedding, we use the zero vector to represent it. For the sets of words describing a foundation, we use the same prototypical words in \citep[Appendix B, Table 20]{tragerMoralFoundationsReddit2022}, which is also presented in \Cref{table:classifiers:ddr_keywords}. The word lists describing the sub-categories \foundation{equality} and \foundation{proportionality} are merged into one that describes \foundation{fairness}.

\section{Limitations of Word Count Methods for Scoring Moral Foundations}
\label{appn:mfd_limits}

In this section, we give some examples in the \texttt{r/AmItheAsshole} dataset where the prediction of moral foundation using word counts is problematic. Note that the scoring is only done on the body text, not including the title.

\subsection{Effect of Input Length}
\label{appn:mfd_limits:input_length}

%\begin{figure}[t]
%	\centering
%	\includegraphics[width=1\linewidth]{figs/mfd_mfd2_emfd_post_length_vs_matched_words.pdf}
%	\caption{Correlation between a post's length and the number of words in that post that are found in an MFD lexicon. We use the MFD, MFD 2.0 and eMFD to score 6,800 posts of the topic \topic{family} on \texttt{r/AmItheAsshole} \citep{nguyenMappingTopics1002022b}. Two-sided Pearson correlation coefficients ($r$, reported on top of each plot) are all statistically significant with $p < 10^{-10}$. Error bars represent 95\% CIs on the predictions of a linear regression model.}
%	\label{fig:mfd_word_count_vs_length_correlation}
%\end{figure}

\begin{figure}
	\centering
	\begin{subfigure}[b]{1.0\textwidth}
		\centering
		\includegraphics[width=0.9\linewidth]{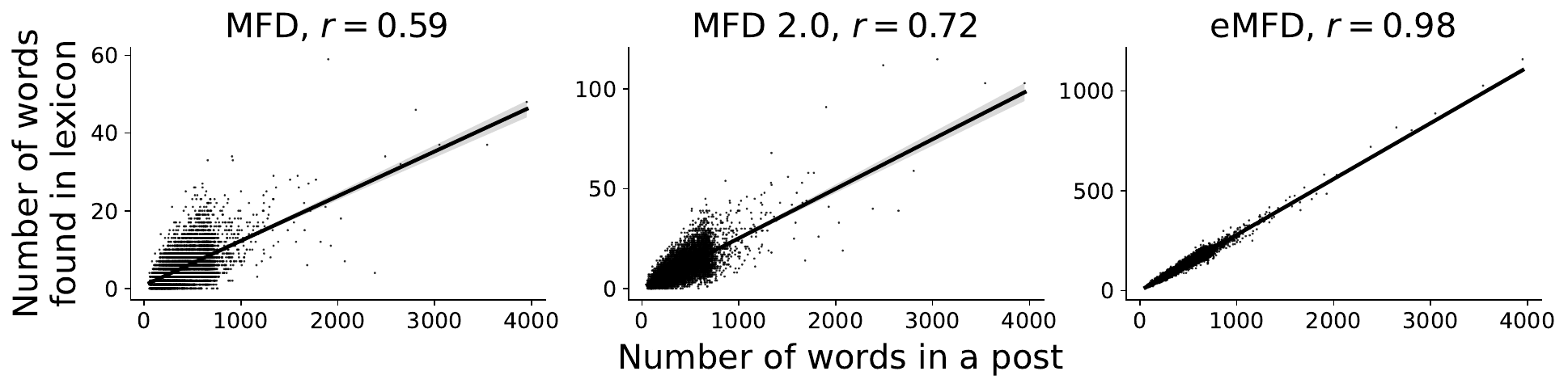}
	\end{subfigure}
	\hfill
	\begin{subfigure}[b]{1.0\textwidth}
		\centering
		\includegraphics[width=0.9\linewidth]{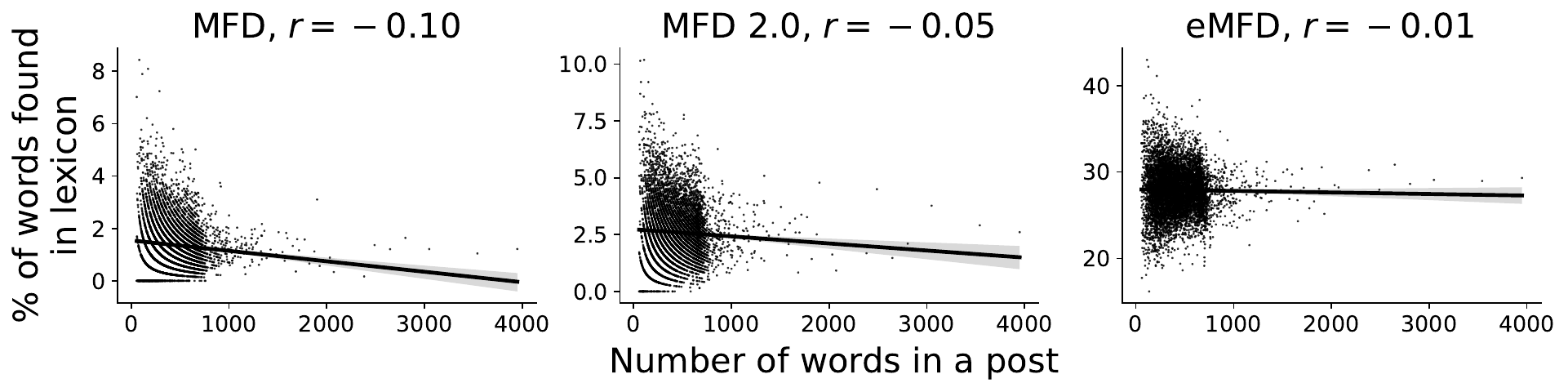}
	\end{subfigure}
	\caption{Correlation between a post's length and the number of words in that post that are found in an MFD lexicon. We use the MFD, MFD 2.0 and eMFD to score 6,800 posts of the topic \topic{family} on \texttt{r/AmItheAsshole} \citep{nguyenMappingTopics1002022b}. The y-axis in the top panel represents the number of words within each post that are contained in each lexicon, whereas the y-axis in the bottom panel is the same count normalized by the number of words in that post. Two-sided Pearson correlation coefficients ($r$, reported on top of each plot) are all statistically significant with $p < 10^{-10}$. Error bars represent 95\% CIs on the predictions of a linear regression model.}
	\label{fig:mfd_word_count_vs_length_correlation}
\end{figure}

\begin{figure}
	\centering
	\includegraphics[width=1\linewidth]{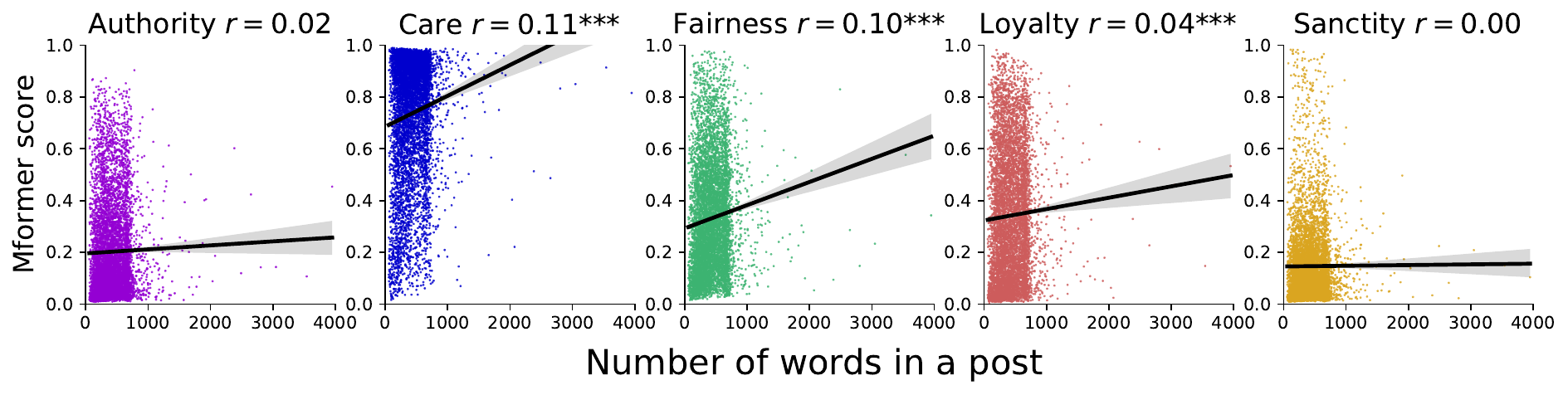}
	\caption{{Correlation between a post's length and its moral foundation scores predicted by \ourmodel. Two-sided Pearson correlation coefficients are reported on top of each plot, where ***$p < 0.001$. Error bars represent 95\% CIs on the predictions of a linear regression model.}}
	\label{fig:mformer_word_count_vs_length_correlation}
\end{figure}

In the following example, we score the post using MFD 2.0. The lexicon detects one word (in bold) that signifies the foundation \foundation{authority}, which leads to the prediction that this post contains the mentioning of \foundation{authority}.

\begin{figure}[t]
	\centering
	\includegraphics[width=1\linewidth]{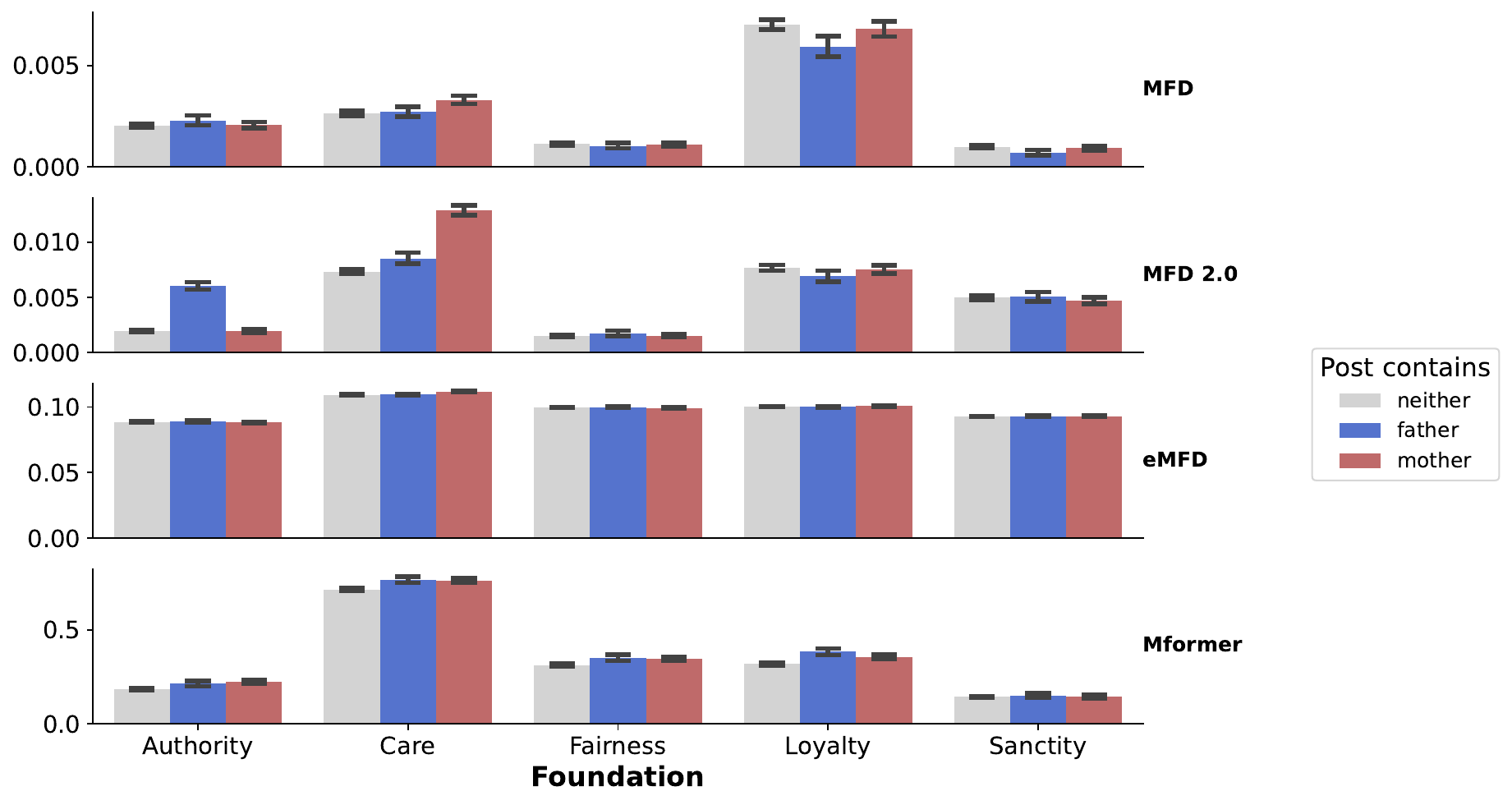}
	\caption{{Mean foundation scores (and 95\% CI) for 6,800 posts of the topic \topic{family} on \texttt{r/AmItheAsshole} \citep{nguyenMappingTopics1002022b}. The bar colors represent posts that contain the word ``father'' (blue, $N = 1,472$), those that contain the word ``mother'' (red, $N = 2,180$) and those that contain neither of these words (grey, $N = 3,960$). Each post is scored by word count (including with MFD, MFD 2.0 and eMFD) and \ourmodel.}}
	\label{fig:mfd_word_count_family_bias}
\end{figure}

\begin{displayquote}
	\example{\underline{Title}: AITA for giving up on my addict sister?}
	
	\example{\underline{Body text}: My sister is 2 years older than me and we used to be close but she moved out a few years ago to live with her abusive boyfriend and she got addicted to meth soon after. She kicked out the boyfriend a few months ago and went to rehab and then came to live with us (me and my parents) and about 3 days after she was released from rehab, she went back to her (ex)boyfriend and soon got addicted again. I’m at the point where if she calls me for help or money, I don’t answer and I tell her it’s her own problem. I know she’s going to let it kill her because she \textbf{refuses} to let anyone help her despite my parents’ multiple attempts. I honestly don’t care if she lets it kill her at this point.}
\end{displayquote}

\subsection{Bias on Familial Roles by Word Count Methods}
\label{appn:mfd_limits:family_bias}

{
Here we provide more evidence of social bias which can be revealed from the scoring results by word count methods. \Cref{fig:mfds_aita_wordcount} and \Cref{sec:wordcount_limits} in the main text show that when MFD 2.0 is used to score content in the topic \topic{family}, there is a clear association of familial roles to moral foundations, such as ``father'' to \foundation{authority} and ``mother'' to \foundation{care}.

We find that this distinction directly results in a difference in the distributions of foundation scores for these posts. In \Cref{fig:mfd_word_count_family_bias}, we score the same 6,800 posts using MFD, MFD 2.0, eMFD and \ourmodel and compare the mean score for three types of posts: those containing the word ``father,'' those containing the word ``mother'' and those containing neither. As seen in bar plots on the second row for MFD 2.0, posts containing ``father'' on average score dominantly higher for \foundation{authority} than those containing ``mother.'' This pattern is the opposite for \foundation{care}, where posts containing ``mother'' score much higher than the those containing ``father.''

We do not find any significant difference in the mean foundation scores between these groups when using \ourmodel, an evidence that our model is unlikely to suffer from the same problem. This does not conclusively show that \ourmodel successfully avoids all biases, but at least in this context it does not seem to associate particular words or roles with moral foundations directly.
}

\section{The Moral Foundations Dataset}
\label{appn:dataset}

In this section, we describe in more detail the collection and processing of the moral foundations dataset in \Cref{sec:mf_classifiers:dataset}. Since this dataset is used to train moral foundation classifiers, it must contain examples as text and five binary labels, one for each foundation. The goal is to build binary classifiers to predict whether each foundation is present in an input.

\subsection{Moral Foundations Twiter Corpus}
\label{appn:dataset:twitter}

This dataset was introduced by \citet{hooverMoralFoundationsTwitter2020a} and released publicly.\footnote{\url{https://osf.io/k5n7y}} It contains a total of 34,987 tweets\footnote{\citet{hooverMoralFoundationsTwitter2020a} reported 35,108 tweets in their dataset, but the released corpus contains only 34,987 unique tweets.} 
encompassing seven ``socially relevant discourse topics'': All Lives Matter, Black Lives Matter, 2016 U.S. Presidential election, hate speech, Hurricane Sandy, and \#MeToo \citet{hooverMoralFoundationsTwitter2020a}. Thirteen annotators were carefully trained to label the tweets with moral foundations and their sentiments (virtue and vice), with at least three annotations per tweet. Annotators were allowed to choose more than one foundation that match a tweet, or no foundation at all (in this case, the authors defined the label as ``non-moral''). 

The format of this dataset is a tweet plus all labels given by each annotator. For example, the tweet 
\begin{quote}
	``There is race war being engineered and financed right in front your eyes in America. Wake up! \#alllivesmatter''
\end{quote}
was labeled by three annotators, who gave it the following sets of foundations: \{\foundation{harm}\}, \{\foundation{care}, \foundation{fairness}\} and \{\foundation{harm}\}. We first merge the moral sentiments (virtue and vice) for all labels, so that \foundation{care} and \foundation{harm} both become \foundation{care}. Then for each tweet, we determine that it contains a foundation $f$ if at least one annotator labeled it with $f$. So, the labels for the above examples are \foundation{care} and \foundation{fairness} as at least one annotator gave it one of these foundation labels.

We keep all tweets in this dataset for our use. Finally, we perform train-test splitting for each foundation. Specifically, for foundation $f$ we randomly choose 10\% of the dataset as the test set, ensuring that the training and test sets are stratified.

\subsection{Moral Foundations News Corpus}
\label{appn:dataset:news}

This dataset was collected and used to create the eMFD lexicon \citep{hoppExtendedMoralFoundations2021}.\footnote{\url{https://osf.io/vw85e}} As described in \Cref{appn:mfd:mfd2} above, the dataset contains 1,010 unique news articles and 73,001 raw highlights produced by 854 annotators. Each highlight represents the annotation that the corresponding part of the article contains a specific foundation.

We follow the preprocessing by \citet{hoppExtendedMoralFoundations2021} for this raw data. We only keep annotators who spent at least 45 minutes on their task and exclude one annotator who spent too long. Only 63,958 annotations from 557 annotators remain. Then, only documents that had been labeled by at least two annotators who differed in their assigned foundation are kept, yielding 47,650 highlights for 992 articles.

From the remaining articles and highlights, we create a moral foundation-labeled dataset by segmenting the articles into sentences. Specifically, we use \texttt{spaCy} to extract all sentences within each article. Then we label a sentence $s$ with a foundation $f$ if $s$ was highlighted (in part or in whole) with $f$ by at least one annotator. For example, if a highlight associated with foundation \foundation{sanctity} contains a sentence---either entirely or partially---then that sentence will have the binary label 1 for \foundation{sanctity}. This yields 32,262 sentences in total from 992 articles.

Similarly, we perform stratified train-test splitting for each moral foundation. Due to the annotation setting, not every article was labeled with all foundations. Therefore, sentences for which no annotator was tasked with labeling using a particular foundation are excluded from the training and test sets for that foundation.

\subsection{Moral Foundations Reddit Corpus}
\label{appn:dataset:reddit}

Introduced by \citet{tragerMoralFoundationsReddit2022}, this dataset contains a total of 17,886 comments\footnote{The number of comments reported in \citep{tragerMoralFoundationsReddit2022} is 16,123.}  extracted from 12 different subreddits roughly organized into three topics: U.S. politics, French politics, and everyday moral life \citep{tragerMoralFoundationsReddit2022}.\footnote{\url{https://huggingface.co/datasets/USC-MOLA-Lab/MFRC}} A total of 27 trained annotators were tasked with labeling these comments with moral foundations. The authors followed a recent work by \citet{atariMoralityWEIRDHow2022} which proposes to separate the foundation \foundation{fairness} into two classes: \foundation{equality} (concerns about equal outcome for all individuals and groups) and \foundation{proportionality} (concerns about getting rewarded in proportion to one's merit). In addition, another label, \foundation{thin morality}, was defined for cases in which moral concern is involved but no clear moral foundation is in place. Therefore, the total number of binary labels is seven. 

For the labels, to be consistent with the other two sources, we merge both \foundation{equality} and \foundation{proportionality} into their common class \foundation{fairness} and consider \foundation{thin morality} as the binary class 0 for all foundations. This results in the same five moral foundation labels. Then, a comment receives a binary label 1 for a foundation if at least one annotator labeled this comment with this foundation. 

We similarly perform train-test splitting for each moral foundation.

\subsection{Further Discussions on Moral Foundations Datasets}
\label{appn:dataset:labeling}

\begin{table}[]
	\caption{{Summary of the definitions of moral foundations used to train annotators of three datasets, Twitter \citep{hooverMoralFoundationsTwitter2020a}, News \citep{hoppExtendedMoralFoundations2021} and Reddit \citep{tragerMoralFoundationsReddit2022}. These datasets are described in more detail in \Cref{appn:dataset:twitter,appn:dataset:news,appn:dataset:reddit}. For News, the full definitions and examples can be found in \citep[Supplemental Materials]{hoppExtendedMoralFoundations2021}. For Reddit, the foundation \foundation{fairness} was split into two classes, \foundation{equality} and \foundation{proportionality}; we report the definitions for both here.}}
	\begin{tabular}{c | p{0.26\linewidth} | p{0.26\linewidth} | p{0.26\linewidth}}
		\toprule
		Foundation & Twitter \citep[see ``Annotation'']{hooverMoralFoundationsTwitter2020a}                                                                                                                                                                            & News  \citep[see Section 4 of ``Supplemental Materials'']{hoppExtendedMoralFoundations2021}                                                                                                                                                                                                                                                                                                                                                                                                                                                                                  & Reddit  \citep[see Section 3.1]{tragerMoralFoundationsReddit2022}                                                                                                                                                                                                                                                                                                                                                                                                                                                                                                        \\
		\midrule
		Authority  & Prescriptive concerns related to submitting to authority and tradition   and prohibitive concerns related to not subverting authority or tradition.                                  & (...) Authority recognizes that leaders and followers represent a   mutual relationship that fosters group success. Subversion is a consequence   of leaders or followers disrupting the healthy relationship of social   hierarchies. (...)                                                                                                                                                                                                                                   & Intuitions about deference toward legitimate authorities and   high-status individuals. It underlies virtues of leadership and respect for   tradition, and vices of disorderliness and resenting hierarchy.                                                                                                                                                                                                                                                                                               \\
		\midrule
		Care       & Prescriptive concerns related to caring for others and prohibitive   concerns related to not harming others.                                                                         & Care is often exemplified through nurturing, assisting, and protecting   others. (...) Harm can be considered any time of distress or pain inflicted   on another. (...)                                                                                                                                                                                                                                                                                                       & Intuitions about avoiding emotional and physical damage or harm to   another individual. It underlies virtues of kindness, and nurturing, and   vices of meanness, violence and prejudice.                                                                                                                                                                                                                                                                                                                 \\
		\midrule
		Fairness   & Prescriptive concerns related to fairness and equality and prohibitive   concerns related to not cheating or exploiting others.                                                      & (...) Fairness is when the give and take between two parties is   equivalent. Reciprocity, unity, collaboration, cooperation, solidarity, and   proportionality are key concepts of fairness. (...) Cheating is when one party   violates the norms of justice by reaping the benefits without contribution or   consideration of the other party. Selfishness, freeloader, cheater, slacker,   liar, and manipulator describe the actions of one party against another. (...) & (Proportionality) Intuitions about individuals getting rewarded in proportion   to their merit (i.e., effort, talent, or input). It underlies virtues of   meritocracy, productiveness, and deservingness, and vices of corruption and   nepotism. (Equality) Intuitions about egalitarian treatment and equal outcome   for all individuals and groups. It underlies virtues of social justice and equality,   and vices of discrimination and prejudice. \\
		\midrule
		Loyalty    & Prescriptive concerns related to prioritizing one’s ingroup and   prohibitive concerns related to not betraying or abandoning one’s ingroup.                                         & (...) Loyalty includes showing group pride, patriotism, and a   willingness to sacrifice for the group. (...) Betrayal is demonstrated when   group members put themselves prior to the group or damage the group image and   identity. (...)                                                                                                                                                                                                                                  & Intuitions about cooperating with in-groups and competing with   out-groups. It underlies virtues of patriotism and self-sacrifice for the   group, and vices of abandonment, cheating, and treason.                                                                                                                                                                                                                                                                                                       \\
		\midrule
		Sanctity   & Prescriptive concerns related to maintaining the purity of sacred   entities, such as the body or a relic, and prohibitive concerns focused on the contamination of such entities. & (...) Sanctity relates to things that are noble, pure and elevated, such as the human body which should not be harmed. (...) Degradation deals   with ideas of physical disgust, particularly when it has to do with the body.   (...)                                                                                                                                                                                                                                       & Intuitions about avoiding bodily and spiritual contamination and degradation. It underlies virtues of sanctity, nobility, and cleanliness and vices of grossness, impurity, and sinfulness. \\
		\bottomrule           
	\end{tabular}
	\label{tab:dataset:mf_definitions}
\end{table}

{
The moral foundations dataset described in \Cref{sec:mf_classifiers:dataset} and this appendix section is aggregated from three different sources. Here, we provide more details comparing these datasets in terms of the definition of each moral foundation and agreement rate.

\Cref{tab:dataset:mf_definitions} presents the definition of each moral foundation used in the annotation process for each dataset, Twitter, News and Reddit. We find that the way each foundation is described is consistent among the three sources. In all settings annotators were informed about moral foundations both in actions that uphold them (virtues) and those that violate them (vices). We also find consistency in their content. For example, all definitions of the foundation \foundation{loyalty} are related to one's priority towards one's ingroup where patriotism is a common example. The major difference among these datasets is with regard to how detailed the examples for each foundation are. News, for instance, includes several examples of each foundation, such as ``seeking societal harmony while risking personal well-being'' or ``verbally, physically, or symbolically attacking outgroup members because they are outgroup members'' for \foundation{loyalty} \citep[see Section 4, Subsection 7 in the Supplemental Materials]{hoppExtendedMoralFoundations2021}. In addition, as described in \Cref{appn:dataset:reddit} above, in the Reddit dataset the foundation \foundation{fairness} is split into two concepts, \foundation{equality} and \foundation{proportionality}, both of which are presented in \Cref{tab:dataset:mf_definitions}. We find that these definitions only represent more specific cases of \foundation{fairness} and do not contradict this concept in any particular way. Overall, based on the consistency with which a moral foundation is portrayed throughout the three domains, we think that aggregating them does not pose any challenge from a definitional perspective.

With respect to inter-annotator agreement rate, we note that \citet{hooverMoralFoundationsTwitter2020a} and \citet{tragerMoralFoundationsReddit2022} report these metrics for this in the Twitter and Reddit datasets, respectively. (We do not have this rate for the News dataset because example sentences are extracted from original news articles.) In particular, both works report Fleiss's Kappa \citep{fleissMeasuringNominalScale1971} and its prevalence- and bias-adjusted variant PABAK \citep{simKappaStatisticReliability2005} for each foundation. Both works also report relatively medium PABAK values, suggesting the subjective nature of this annotation task. In using these datasets for training supervised classifiers, to account for label noise several methods have been proposed, such as by taking the majority vote. For these datasets, we decide to assign a foundation $f$ to an example if at least one annotator gave it this label. The reason for this choice is because of the inherent subjectivity of this task and the reliance on ``moral intuitions rather than [...] deliberations'' \citep{hoppExtendedMoralFoundations2021}, which we believe make it more likely for annotators to ``miss'' a true label than mislabeling a negative example (higher false negative rate than false positive rate). Further, taking the majority vote will decrease the size of the training set significantly; a favor for data quantity at the expense of higher label noise can be justified 
}

\section{Training Moral Foundation Classifiers}
\label{appn:supervised_classifiers}

In this section, we provide more detail on training/fine-tuning moral foundation classifiers described in \Cref{sec:mf_classifiers:classifiers:existing}. Specifically, we use two models, logistic regression and RoBERTa for sequence classification \cite{liuRoBERTaRobustlyOptimized2019b}, and describe the hyperparameter tuning setting for each model.

\subsection{Logistic Regression}
\label{appn:supervised_classifiers:logistic_regression}

For each foundation $f$, we train an $\ell_2$-regularized logistic regression model to estimate $p(f \mid d)$, the probability that a document $d$ is labeled with $f$. We choose to tune $C$, the hyperparameter equal to the inverse of the $\ell_2$ regularization strength, in $\{ 10^{-7}, 10^{-6}, \ldots, 10^{6}, 10^{7} \}$. The best value of $C$ is one that achieves the highest average validation AUC over 10 folds in the training set. We use four document embeddings for this model:
\begin{itemize}
	\item Tf-idf \citep{manningIntroductionInformationRetrieval2008}: We tokenize every training example using \texttt{spaCy}. Then we lemmatize and lowercase every token. We filter out all tokens that are in the set of standard English stop words, then remove all tokens with non-alphabetic characters and those with fewer than 3 characters. The vocabulary discovered from the training set contains 12,586 unique tokens, corresponding to the dimensionality of the embedding.
	\item SpaCy \citep{honnibalSpaCyIndustrialstrengthNatural2020}: We also use the built-in static word embedings by \texttt{spaCy}. Specifically, after tokenizing a document, each token within the document is associated with a word vector with $200$ dimensions. The vector representation for the document is the average of these word embeddings.
	\item GloVe \citep{penningtonGloVeGlobalVectors2014}: Similar to the \texttt{spaCy} embedding, the GloVe embedding of a document is the average of all of its token embeddings. We follow the same setting described in \Cref{appn:embedding_sim} above, where we use the ``Twitter'' 200-dimensional version of GloVe and use the zero vector to represent all out-of-vocabulary tokens.
	\item Sentence-RoBERTa \citep{reimersSentenceBERTSentenceEmbeddings2019a}: This is RoBERTa \cite{liuRoBERTaRobustlyOptimized2019b} fine-tuned for a sentence similarity task. To encode a document, we perform a forward pass through this network. The output provided by the final layer (before fitting it through a classification head) contains embeddings for all of the document's tokens. The vector representation for the document is taken as the average of all token vectors. We specifically choose the RoBERTa-large architecture, which outputs a 1,024-dimensional embedding.
\end{itemize}

%\begin{figure}[t]
%	\centering
%	\includegraphics[width=0.6\linewidth]{figs/svm_vs_logreg.pdf}
%	\caption{{Performance comparison between linear SVM (trained using the tf-idf embedding) and logistic regression (trained using the Sentence-RoBERTa embedding).}}
%	\label{fig:appn:supervised_classifiers:svm_vs_logreg}
%\end{figure}

\begin{figure}[t]
	\begin{minipage}[t]{0.48\textwidth}
		\centering
		\includegraphics[width=1\linewidth]{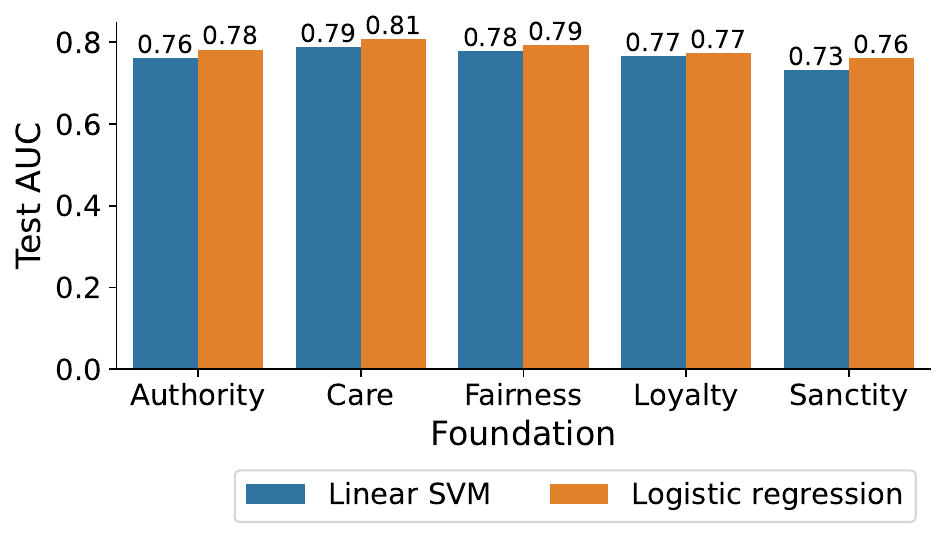}
		\caption{{Performance comparison between linear SVM (trained using the tf-idf embedding) and logistic regression (trained using the Sentence-RoBERTa embedding).}}
		\label{fig:appn:supervised_classifiers:svm_vs_logreg}
	\end{minipage}
	\hfill
	\begin{minipage}[t]{0.48\textwidth}
		\centering
		\includegraphics[width=1\linewidth]{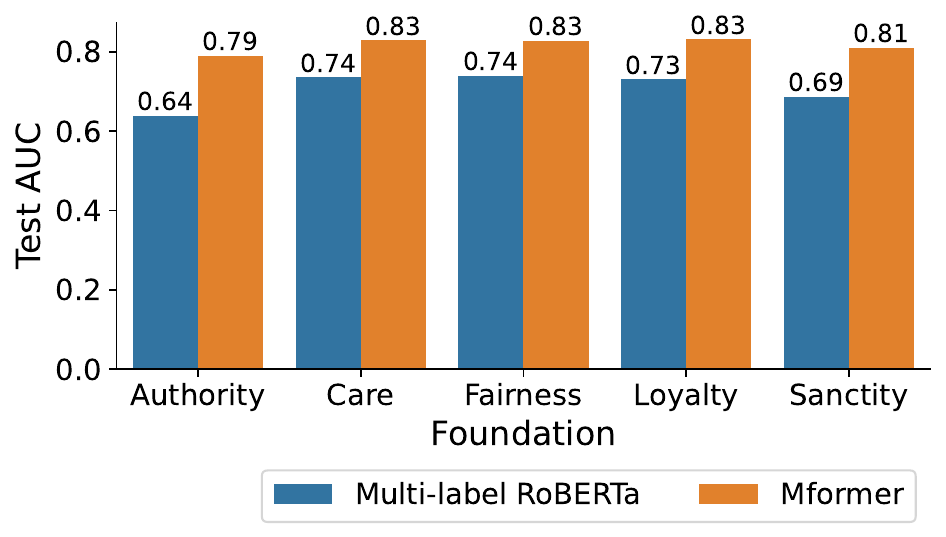}
		\caption{{Performance comparison between two RoBERTa variants: multi-label and Mformer. For the multi-label variant, RoBERTa's final classification layer contains 5 neurons, each followed by a \emph{sigmoid} activation to represent the binary probability for each class. For Mformer, each foundation is associated with one version of RoBERTa binary classifier.}}
		\label{fig:roberta_eval:roberta_multilabel_vs_binary}
	\end{minipage}
\end{figure}

{
\header{Alternatives to logistic regression} 
In addition to logistic regression, we also consider support vector machines (SVM), another type of linear classifier that can achieve robustness by finding a ``maximum-margin'' decision function. This method was used by \citet{hooverMoralFoundationsTwitter2020a} in setting up a prediction baseline for the Twitter dataset.

In particular, we embed each training example using tf-idf similar to above. Different from \citet[see ``Methodology'']{hooverMoralFoundationsTwitter2020a}, we do not consider only words belonging to MFD or MFD 2.0 but consider the entire vocabulary (12,586-dimensional), which already contains words in these lexicons, for our embedding, leading to a more versatile representation.  Then, for each moral foundation, we train a linear SVM with $\ell_2$ regularization using the binary presence of that foundation as the supervised signal. Following \citet{hooverMoralFoundationsTwitter2020a}, the regularization hyperparameter is set to $C = 1.0$.

Using the same test set, we compare linear SVM and the highest-performing logistic regression models (using the Sentence-RoBERTa embedding) in \Cref{fig:appn:supervised_classifiers:svm_vs_logreg}. The results show that logistic regression achieves a higher test AUC than SVM for all foundations, although the difference can be marginal, especially for \foundation{loyalty}. We therefore only report the results by logistic regression in the main text for further evaluation (cf. \Cref{fig:roberta_mf:testset_auc,fig:external_aucs}).
}

\subsection{RoBERTa}
\label{appn:supervised_classifiers:roberta}

The moral foundation classifiers described and used in this paper are RoBERTa \citep{liuRoBERTaRobustlyOptimized2019b}. This language model is a direct extension of BERT \citep{devlinBERTPretrainingDeep2019} with a more advanced and robust training procedure. RoBERTa has two versions: RoBERTa-base, which has a hidden size of 768, and RoBERTa-large with a hidden size of 1,024. We choose the former architecture for this task.

For each moral foundation, we fine-tune RoBERTa on a binary text classification task: whether the foundation is present in the input or not. We first tokenize a document using HuggingFace's built-in tokenizer. Then, the sequence of tokens is truncated to a maximum of 510 tokens if it is longer than that. Two special tokens, \texttt{<s>} and \texttt{</s>} are added to the beginning of the end of the sequence, respectively, ensuring the maximum sequence length is 512. These tokens are often called the CLS and SEP tokens.

In fine-tuning RoBERTa, we start with the weight checkpoints released on HuggingFace.\footnote{\url{https://huggingface.co/roberta-base}} After the final self-attention layer, the 768-dimensional embedding of the token \texttt{<s>} is fed through a linear layer with 768 neurons, followed by the $\tanh$ activation. Finally, this goes through another linear layer with two outputs, followed by the softmax transformation to estimate the probability of the two classes. We use the typical categorical cross-entropy loss to compare the probabilities with the ground-truth binary label of the input. All RoBERTa weights are optimized using the AdamW optimizer \citep{loshchilovDecoupledWeightDecay2018}, where we set its parameters to $\epsilon = 10^{-8}$, $\beta_1 = 0.9$, $\beta_2 = 0.99$. We also add an $\ell_2$ regularization term for the weights with the regularization strength $\lambda = 0.01$. We use a batch size of 16 during optimization. {Finally, in fine-tuning we use one NVIDIA TITAN V GPU with 12GB of VRAM.}

%\begin{figure}[t]
%	\centering
%	\includegraphics[width=0.6\linewidth]{figs/roberta_multilabel_vs_binary.pdf}
%	\caption{{Performance comparison between two RoBERTa variants: multi-label and Mformer. For the multi-label variant, RoBERTa's final classification layer contains 5 neurons, each followed by a \emph{sigmoid} activation to represent the binary probability for each class. For Mformer, each foundation is associated with one version of RoBERTa binary classifier.}}
%	\label{fig:roberta_eval:roberta_multilabel_vs_binary}
%\end{figure}

With respect to hyperparameter tuning, we control two quantities: the learning rate $\alpha \in \left\{ 10^{-6}, 10^{-5}, 3 \times 10^{-5} \right\}$ and the number of training epochs in $\{2, 3, 4, 5, 6\}$. To do so, we randomly split the training set into a training and a validation portion of relative size 9:1. The splitting is done in a stratified manner, ensuring the same proportion of the positive class in the two subsets. We then perform a grid search over all combinations of $\alpha$ and the number of epochs and choose the combination with the highest validation AUC as the final hyperparameters. Finally, we combine the training and validation subsets into the original training set and fine-tune RoBERTa using the best hyperparameters to give the final models.

We call the final five fine-tuned classifiers \ourmodel.

\subsection{Multi-Label RoBERTa}
\label{appn:supervised_classifiers:multilabel_roberta}

{We also consider a version of RoBERTa which performs multi-label classification. In particular, different from \ourmodel which has five distinct models, this version only has one RoBERTa model but the final classification layer contains 5 neurons instead of 2. Each of the five neurons is followed by a sigmoid activation, which outputs a probability that the input contains each moral foundation. Note that the five outputs by multi-label RoBERTa do not necessarily add up to 1, because each output represents the probability for each foundation. Compared to \ourmodel, multi-label RoBERTa only requires one copy of RoBERTa, thereby reducing the space requirement by 5 times.
	
Note that our training dataset is multi-label, and that there exist examples for which some labels for moral foundations are missing. For these reasons, train-test splitting must be done differently. We employ the iterative stratification method by \citet{sechidisStratificationMultilabelData2011,szymanskiNetworkPerspectiveStratification2017}, setting 10\% of the dataset for testing and splitting the training set into a training and validation portion of ratio 9:1. In fine-tuning multi-label RoBERTa, we explore the same hyperparameters as above and choose the model with the highest validation AUC.

We find that the multi-label variant to \ourmodel leads to suboptimal performance. As shown in \Cref{fig:roberta_eval:roberta_multilabel_vs_binary}, which compares these two alternatives using the AUC on the same test set, \ourmodel performs significantly better than multi-label RoBERTa, obtaining an increase of 12.0--24.0\% in AUC. We therefore decide to use \ourmodel instead of multi-label RoBERTa in later analyses.
}

\begin{table*}[t]
	\centering
	\small
	\caption{AUC for moral foundation classifiers (\Cref{sec:mf_classifiers:classifiers}) evaluated on the hold-out test sets (\Cref{sec:mf_classifiers:dataset}) .}
	\begin{tabular}{c|ccc|c|cccc|c}
		\toprule
		\multicolumn{1}{c|}{\multirow{2}{*}{Foundation}} & \multicolumn{3}{c|}{Word count} & \multirow{2}{*}{\begin{tabular}[c]{@{}c@{}}Embedding \\ similarity \end{tabular}} & \multicolumn{4}{c|}{Logistic regression} & \multirow{2}{*}{\ourmodel} \\
		\multicolumn{1}{c|}{}                            & MFD      & MFD 2.0    & eMFD    &                                                                        & tf-idf   & spaCy   & GloVe  & S-RoBERTa  &                          \\ 
		\midrule
		Authority                                        & 0.64     & 0.63       & 0.64    & 0.52                                                                   & 0.75     & 0.72    & 0.72   & 0.78       & 0.85                     \\
		Care                                             & 0.62     & 0.66       & 0.69    & 0.55                                                                   & 0.78     & 0.77    & 0.77   & 0.81       & 0.85                     \\
		Fairness                                         & 0.56     & 0.64       & 0.66    & 0.58                                                                   & 0.77     & 0.76    & 0.76   & 0.79       & 0.84                     \\
		Loyalty                                          & 0.57     & 0.59       & 0.60    & 0.51                                                                   & 0.76     & 0.74    & 0.74   & 0.77       & 0.83                     \\
		Sanctity                                         & 0.54     & 0.60       & 0.59    & 0.59                                                                   & 0.71     & 0.73    & 0.71   & 0.76       & 0.83                     \\ 
		\bottomrule
	\end{tabular}
	\label{table:evaluation:auroc_test_set}
\end{table*}

\section{Evaluation of the Fine-Tuned RoBERTa Classifiers}
\label{appn:roberta_eval}

\begin{figure}[t]
	\centering
	\includegraphics[width=0.6\linewidth]{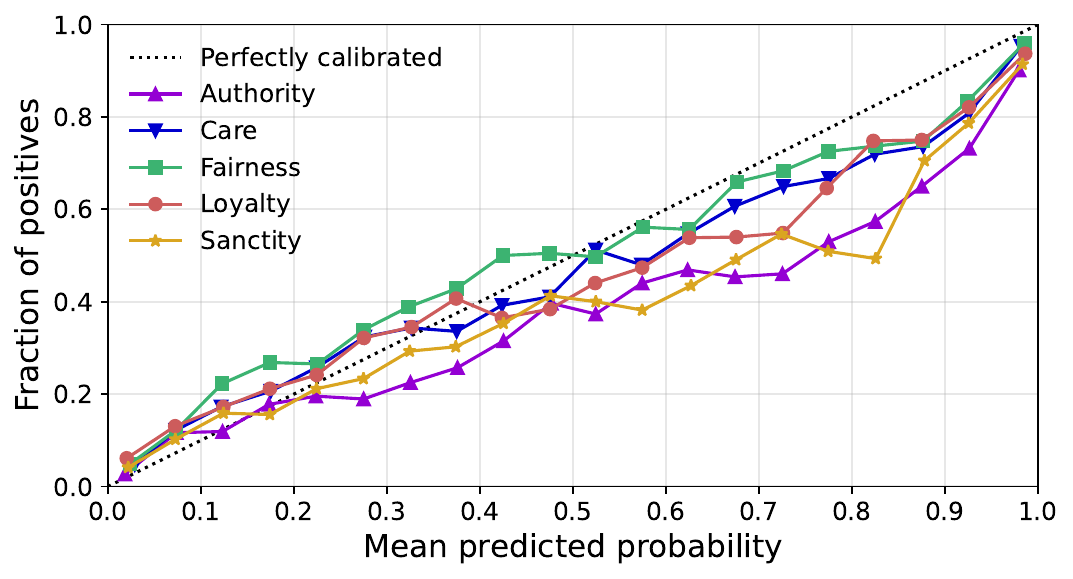}
	\caption{Calibration curves for the five \ourmodel classifiers.}
	\label{fig:roberta_eval:calibration}
\end{figure}

In this section we provide more detail on evaluation of \ourmodel models in \cref{sec:mf_classifiers:evaluation}.

\begin{figure}[t]
	\centering
	\includegraphics[width=1\linewidth]{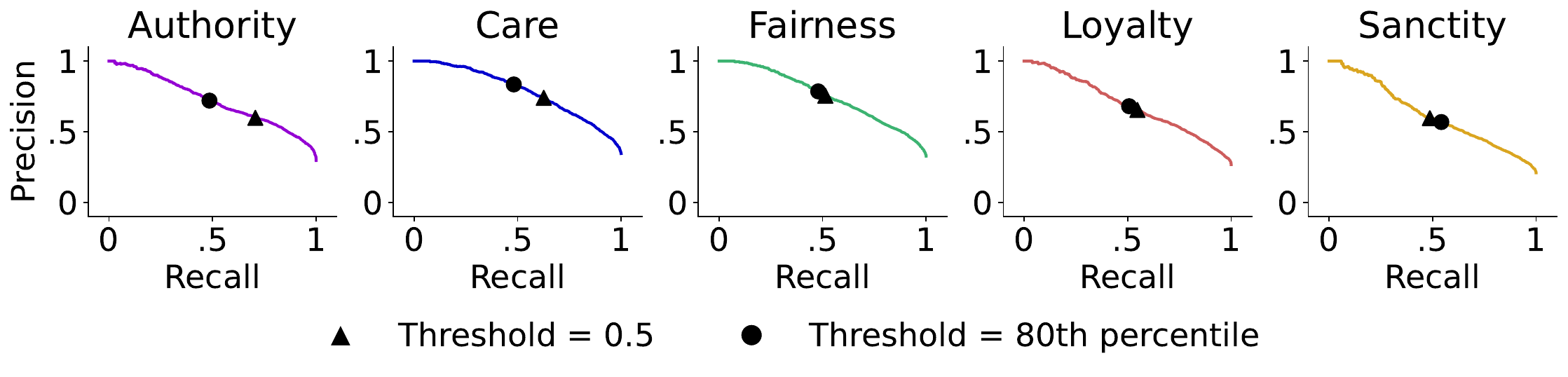}
	\caption{Precision-recall curves for the five \ourmodel moral foundation classifiers. Two thresholding values for the prediction scores are displayed: 0.5 (black triangles) and the 80th percentile of all predicted scores on the test set (black circles).}
	\label{fig:roberta_eval:pr_curves}
\end{figure}

\subsection{Calibration}
\label{appn:roberta_eval:calibration}

To assess whether the scores predicted by \ourmodel closely approximate actual probabilities, we present a calibration plot in \Cref{fig:roberta_eval:calibration}. For each moral foundation, we collect \ourmodel scores for all test examples. Then, the scores are discretized into 20 equal-width bins. The x-axis represents the average of the scores within each bin and the y-axis represents the fraction of examples in each bin that are in fact positive (i.e., contain the foundation). A perfectly calibrated classifier should have a diagonal calibration curve, as, for instance, a score of 0.7 implies that 70\% of examples predicted at that level are positive.

It is known that classifiers with outputs produced by sigmoid or softmax functions are well calibrated, which is the case in \Cref{fig:roberta_eval:calibration} since most curves are close to the diagonal line. We find that the calibration curves for \foundation{care} and \foundation{fairness} are the best, while those for \foundation{authority} and \foundation{sanctity} tend to deviate from the perfectly calibrated curve as predicted scores get higher.

\begin{table}[!htb]
	\small
	\centering
	\begin{minipage}{.48\linewidth}
		\caption{Precision at different thresholding levels.}
		\centering
		\begin{tabular}{ccccccc}
			\toprule
			Foundation & 95th & 90th & 80th & 70th & 60th & 50th \\ 
			\midrule
			Authority  & 0.94 & 0.86 & 0.72 & 0.64 & 0.57 & 0.51 \\
			Care       & 0.98 & 0.95 & 0.84 & 0.74 & 0.65 & 0.58 \\
			Fairness   & 0.98 & 0.92 & 0.79 & 0.69 & 0.61 & 0.54 \\
			Loyalty    & 0.92 & 0.84 & 0.68 & 0.59 & 0.52 & 0.45 \\
			Sanctity   & 0.88 & 0.72 & 0.57 & 0.48 & 0.41 & 0.36 \\
			\bottomrule
		\end{tabular}
		\label{table:roberta_eval:thresh_precision}
	\end{minipage}%
	\begin{minipage}{.48\linewidth}
		\centering
		\caption{Recall at different thresholding levels.}
		\begin{tabular}{ccccccc}
			\toprule
			Foundation & 95th & 90th & 80th & 70th & 60th & 50th \\ 
			\midrule
			Authority  & 0.16 & 0.29 & 0.48 & 0.64 & 0.77 & 0.85 \\
			Care       & 0.14 & 0.27 & 0.48 & 0.64 & 0.74 & 0.83 \\
			Fairness   & 0.15 & 0.28 & 0.48 & 0.63 & 0.74 & 0.82 \\
			Loyalty    & 0.17 & 0.31 & 0.51 & 0.66 & 0.77 & 0.85 \\
			Sanctity   & 0.21 & 0.34 & 0.54 & 0.68 & 0.79 & 0.86          \\
			\bottomrule
		\end{tabular}
		\label{table:roberta_eval:thresh_recall}
	\end{minipage} 
	\vspace{0.3cm}
	
	\begin{minipage}{.48\linewidth}
		\centering
		\caption{F-1 at different thresholding levels.}
		\begin{tabular}{ccccccc}
			\toprule
			Foundation & 95th & 90th & 80th & 70th & 60th & 50th \\ 
			\midrule
			Authority  & 0.27 & 0.43 & 0.58 & 0.64 & 0.66 & 0.64 \\
			Care       & 0.25 & 0.42 & 0.61 & 0.68 & 0.69 & 0.68 \\
			Fairness   & 0.26 & 0.43 & 0.60 & 0.66 & 0.67 & 0.65 \\
			Loyalty    & 0.29 & 0.46 & 0.58 & 0.62 & 0.62 & 0.59 \\
			Sanctity   & 0.34 & 0.47 & 0.56 & 0.56 & 0.54 & 0.51 \\
			\bottomrule
		\end{tabular}
		\label{table:roberta_eval:thresh_f1}
	\end{minipage} 
	\begin{minipage}{.48\linewidth}
		\centering
		\caption{Accuracy at different thresholding levels.}
		\begin{tabular}{ccccccc}
			\toprule
			Foundation & 95th & 90th & 80th & 70th & 60th & 50th \\ 
			\midrule
			Authority  & 0.75 & 0.77 & 0.79 & 0.78 & 0.76 & 0.71 \\
			Care       & 0.70 & 0.74 & 0.79 & 0.79 & 0.77 & 0.73 \\
			Fairness   & 0.72 & 0.76 & 0.79 & 0.78 & 0.76 & 0.71 \\
			Loyalty    & 0.77 & 0.80 & 0.80 & 0.78 & 0.74 & 0.69 \\
			Sanctity   & 0.83 & 0.83 & 0.82 & 0.78 & 0.72 & 0.65 \\
			\bottomrule
		\end{tabular}
		\label{table:roberta_eval:accuracy}
\end{minipage} 
\end{table}

\subsection{Choosing Classification Thresholds}
\label{appn:roberta_eval:thresholding}

In this paper far we have shown that \ourmodel (fine-tuned RoBERTa) outperforms existing methods---namely, word count, embedding similarity, and logistic regression---on predicting moral foundations in text based on the AUC metric. In most downstream analyses, binary labels are required from the raw prediction scores. Determining a threshold above which a prediction becomes positive reflects the tradeoff between precision and recall, which we show in \Cref{fig:roberta_eval:pr_curves} for each moral foundation.

The good calibration curves (\Cref{fig:roberta_eval:calibration}) described above suggest that a threshold of 0.5 is reasonable.  In \Cref{fig:roberta_eval:pr_curves}, the precision and recall for each foundation is represented as black triangles. Another widely-chosen method, especially when predicting in a novel domain, is to set the top $x$th percentile of all scores as the classification threshold, where $x \in [0, 1]$. A higher value of $x$ is typically justified by the researchers' preference for higher precision at the cost of low recall. To illustrate the precision-recall tradeoff, we present four metrics---precision, recall, F-1 and accuracy---for different levels of $x$ in \Cref{table:roberta_eval:thresh_precision,table:roberta_eval:thresh_recall,table:roberta_eval:thresh_f1,table:roberta_eval:accuracy}. We find that when $x = 80$ or $x=70$, both F-1 and accuracy remain relative high. Therefore, we choose either of these values when binary labels are needed in \Cref{sec:measurements}.

\section{External Moral Foundations Datasets and Evaluation}
\label{appn:external_datasets}

This section provides more details on the four external datasets used to examine the generalizability of our model, \ourmodel, described in \Cref{sec:external_eval}.

\subsection{Datasets}
\label{appn:external_datasets:datasets}

\subsubsection{Moral foundations vignettes dataset (VIG)}

This dataset contains 132 vignettes describing behaviors that violate a moral foundation \citep{cliffordMoralFoundationsVignettes2015}. Each vignette is short (14--17 words) and in the \example{``You see [behavior]''} format. An example that violates the foundation \foundation{authority} is \example{``You see a woman refusing to stand when the judge walks into the courtroom.''} In addition to the five moral foundations, two classes are considered: \foundation{liberty}, relating to freedom of choice, and \foundation{social norms}, describing actions that are unusual but are not considered morally wrong. In total, each class contains 16 to 17 examples except for the foundation \foundation{care}, which contains 32 examples where half are about actions causing physical harm to humans and the other half to nonhuman animals.

We consider this dataset to contain ``gold labels'' for three reasons. First, the vignettes were carefully generated by the authors to clearly violate only one foundation. Second, each vignette was independently rated by a large pool of annotators (approximately 30 respondents per example) and in every case, at least 60\% of annotators agreed on the intended foundation. Third, all vignettes which were believed by 20\% or more of annotators to violate an unintended foundation were excluded. In other words, the dataset only contains examples with high-confidence moral foundation labels.

We preprocess this dataset as follows. First, we merge the two sub-cases for \foundation{care}, namely physical harm to humans and to nonhuman animals, into simply \foundation{care}. Then, we remove the 17 instances violating \foundation{liberty}. Finally, for the 16 instances of \foundation{social norms}, we consider them to contain no foundation at all, i.e., the five binary labels for moral foundations are all 0. We decide against considering \foundation{liberty} examples to contain no foundation, as it could be shown to correlate with \foundation{fairness} and \foundation{authority} \citep{haidtRighteousMindWhy2012} which might unnecessarily complicate our evaluation later on. On the other hand, \foundation{social norms} cases were clearly designed not to violate any explicit moral foundation at all, hence they are good candidates for negative instances. The final dataset contains 115 vignettes with 16--17 examples for each foundation except for \foundation{care}, which has 32 examples.

\subsubsection{Moral arguments dataset (ARG)}

\citet{kobbeExploringMoralityArgumentation2020} aimed to study the effect of moral sentiment in the analysis of arguments, especially in the context of argument mining. To do so, the authors pulled 320 arguments taken from the online debate platforms \texttt{createdebate.com} and \texttt{convinceme.net}; the dataset is called Dagstuhl ArgQuality Corpus \citep{wachsmuthComputationalArgumentationQuality2017,habernalWhatMakesConvincing2016}. Different from the vignettes above, the arguments are much more diverse in length (min = 11, max = 148, median = 64 tokens per example) and in topic (e.g., evolution vs. creation, plastic water bottle ban, etc.). Then the two authors manually labeled these arguments with every moral foundation.

Since this dataset already contains the necessary binary labels for moral foundations, we thus perform no preprocessing and use all 320 instances. Of all arguments, nearly a third (96/320) were labeled with no foundation, over half (179/320) contain one foundation, and the maximum number of foundations per example is 3 (with 3/320 examples). An example argument containing \foundation{fairness} is \example{``Religion in the past has caused many wars. It encourages racism, sexism, and homophobia. It is something that gives us prejudice. It makes us hate one another. [T]he time has come to put a stop to it.''}

\subsubsection{Social chemistry 101 dataset (SC)}

This dataset contains 292K rules-of-thumb (RoTs), defined as ``descriptive cultural
norms structured as the judgment of an action'' \citep{forbesSocialChemistry1012020}. Each RoT was created by an online annotation worker based on a given \emph{situation}. For example, the situation \example{``Not wanting to be around my girlfriend when she’s sick''} prompted an annotator to give the RoT \example{``It's kind to sacrifice your well-being to take care of a sick person.''} Then, the RoTs were labeled with categorical attributes such as social judgments, cultural pressure, legality, and moral foundations. Similar to the previous two datasets, in this case, every RoT received five binary labels for the foundations.

We choose the test set, which contains 29,239 RoTs (10\% of the entire dataset) to evaluate the moral foundation classifiers. In this set, the RoTs are relatively short (median = 11 tokens per example). Over 60\% of the RoTs contain only one foundation, while 21\% contain no foundation at all. An example RoT labeled with \foundation{authority} is \example{``It is controlling to tell a partner what they can and cannot do with their body.''}

\subsubsection{Moral integrity corpus (MIC)}

Similar to SC, this dataset features 99K RoTs, which were created based on \emph{prompts} such as \example{``Do you ever smoke marijuana illegally?''} \citep{ziemsMoralIntegrityCorpus2022}. However, unlike SC, here the RoTs were generated by humans based on \emph{prompt-reply} pairs partially created by chatbot systems. In particular, with the prompt-reply pair ``Do you ever smoke marijuana illegally?''--``I smoke it to relax. It helps my anxiety,'' an annotator generated the RoT ``It is bad to use harmful substances.'' Subsequently, each RoT was labeled on several dimensions including the severity of violating that RoT, the degree to which other people agree with the RoT, and the moral foundations involved in the RoT.

We also choose the test set, containing 11,375 examples. The examples are also short (median = 9 tokens per instance) and 67\% contain one foundation. An example containing \foundation{loyalty} is \example{``It is wrong for a person to betray their friends.''}

\subsection{Scoring Moral Foundations}

In all four datasets described above (VIG, ARG, SC, and MIC), we score every instance using five methods described in \Cref{sec:related_work:MF_detection}. The summary of these methods is as follows:
\begin{itemize}
	\item Word count: We use the three dictionaries---MFD, MFD 2.0, and eMFD---for scoring every example. Exact detail about how examples are scored can be found in \Cref{appn:mfd}. Even though MFD 2.0 tends to perform better than MFD in most cases, we also want to examine MFD, especially regarding its limited vocabulary.
	\item Embedding similarity: \Cref{appn:embedding_sim} describes how scoring is done. We use the same word embedding and seed words.
	\item Supervised classifiers: We use the five \ourmodel models described in \Cref{appn:supervised_classifiers}. No further fine-tuning is done on any external datasets. We decide not to use any logistic regression models because \ourmodel is shown to outperform them in all previous benchmarks on the moral foundations test set.
\end{itemize}

% \begin{figure}[t]
% 	\centering
% 	\includegraphics[width=1\linewidth]{figs/external_auc_5_methods.pdf}
% 	\caption{AUC for scoring each moral foundation on four datasets: VIG, ARG, SC and MIC. Five scoring methods are used: MFD, MFD 2.0, eMFD, embedding similarity and \ourmodel.}
% 	\label{fig:external:auc_all}
% \end{figure}

\subsection{Evaluation}

Once we obtain the moral foundation scores, we also use the AUC to compare them against the ground-truth binary labels. The result is presented in \Cref{fig:external_aucs} in the main text. Some analysis of the AUC for \ourmodel has been discussed in \Cref{sec:external_eval} in the main text.

We find that \ourmodel classifiers give the best performance in all cases, often by a large margin compared to the second-highest scorer (e.g., 0.95 vs. 0.69 obtained by MFD 2.0). The only tie in the highest AUC is between \ourmodel and embedding similarity on predicting the foundation \foundation{loyalty} on the VIG dataset (both AUC = 0.75). 

We also find that, despite being based on dense word embeddings and not restricted to a small lexicon, embedding similarity is not clearly better than simple word count methods based on the MFDs. Sometimes, its performance is even worse: for example, when predicting \foundation{fairness} on the ARG dataset, the AUC for embedding similarity is 0.53---somewhat equal to random guessing---compared to 0.60 by MFD.

When comparing the three lexicons for word count, we find that MFD 2.0 often performs better than MFD. This is expected since MFD 2.0 is an extension of MFD with more than three times as many words. The eMFD also tends to improve from MFD 2.0, with some exceptions such as when predicting \foundation{loyalty} in ARG. However, this improvement is much smaller than what \ourmodel makes.

In the following, we provide further evaluation of \ourmodel models for each dataset.

\subsubsection{VIG}

We find that \ourmodel performs very well on this dataset. Given that the vignettes were carefully designed to elicit one particular moral foundation, and the fact the majority of a large number of annotators agreed with the ground-truth labels, the results are in strong favor of our models. We find that \ourmodel performs well when predicting the foundations \foundation{authority}, \foundation{fairness}, and \foundation{sanctity} in particular, with the recorded AUC between 0.88 and 0.95. However, the performance on \foundation{care} is not as impressive (AUC = 0.81). Upon inspection, we find that some examples for \foundation{care} tend to be misclassified as \foundation{fairness}; an example is \example{``You see a man quickly canceling a blind date as soon as he sees the woman.''} (As \citet{cliffordMoralFoundationsVignettes2015} reported, for this this example 16\% of respondents also thought it is about \foundation{fairness}.) The foundation hardest to score is \foundation{loyalty}, with an AUC of 0.75. Similarly, we find that some examples tend to be misclassified as \foundation{authority} as in this vignette: \example{``You see a head cheerleader booing her high school's team during a homecoming game.''}

We consider this good evidence in support of adopting \ourmodel as a new classifier of moral foundations. However, limitations exist, including the fact that the dataset is small; the vignettes are relatively short and simple; the scenarios are only about actions that exhibit a \emph{violation} moral foundation and not those that constitute a \emph{virtue}.

\subsubsection{ARG}

Similar to VIG, we think that this dataset contains high-quality labels since \citet{kobbeExploringMoralityArgumentation2020} directly studied moral foundations in arguments and the annotation process was executed carefully, although with fewer annotators. The results for \ourmodel on this dataset is also very good with all AUC between 0.81 and 0.86. The foundations \foundation{care} (AUC = 0.86), \foundation{fairness} (AUC = 0.86) and \foundation{loyalty} (AUC = 0.86) are the easiest to classify. For \foundation{sanctity} (AUC = 0.83), we find that low-scoring examples often score high on \foundation{care}. The following example was only labeled with \foundation{sanctity} but is scored low on this foundation while very high on \foundation{care}:
\begin{displayquote}
	\example{Americans spend billions on bottled water every year. Banning their sale would greatly hurt an already struggling economy. In addition to the actual sale of water bottles, the plastics that they are made out of, and the advertising on both the bottles and packaging are also big business. In addition to this, compostable waters bottle are also coming onto the market, these can be used instead of plastics to eliminate that detriment. Moreover, bottled water not only has a cleaner safety record than municipal water, but it easier to trace when a potential health risk does occur. [URL]}
\end{displayquote}
Similarly, we also find some examples for \foundation{authority} score low on this foundation but high on \foundation{care}, as in the argument: \example{``Some kids don't learn by spanking them..So why waste your time on that, when you can always take something valuable away from them.''}

\subsubsection{SC}

This is the largest dataset for evaluation with 29K examples (RoTs). While \ourmodel still performs better than all other methods, we find the AUC on SC, between 0.70 and 0.80, to be relatively low compared to that on VIG and ARG. Of all five foundations, \foundation{loyalty} (AUC = 0.80) is the easiest to score, followed by \foundation{authority} (AUC = 0.74) and \foundation{fairness} (AUC = 0.73).

We suspect that the relatively lower performance of \ourmodel is due to the fact that there exists some considerable label noise in this dataset. First, we note that moral foundations are one of many categorical attributes studied by \citet{forbesSocialChemistry1012020}, beside cultural pressure and social judgment for example. Since annotators were not exclusively tasked with identifying foundations in an RoT, it is possible that the foundation labels were noisily chosen. Second, while the examples in VIG and ARG were annotated by more than one annotator and subsequently went through manual inspection, this is not the case for SC's RoTs. Each RoT was only labeled by one worker in SC, and since this annotation task is known to be subjective and has low inter-annotator agreement \citep{hooverMoralFoundationsTwitter2020a,tragerMoralFoundationsReddit2022,kobbeExploringMoralityArgumentation2020}, we believe the noise level is significant and deserves a formal inspection. Third, the annotators for VIG and ARG went through careful training while the construction of SC tended to simplify this process: annotators were only asked ``Which moral foundations best fit this RoT?'' and given short descriptions and examples of each foundation. This could have had a significant effect on the labels.

Here we give some examples of label noise in SC. An RoT which was only labeled with the foundation \foundation{care} but scores very high on \foundation{fairness} reads:	\example{``You shouldn't call people homophobic slurs.''} While this RoT also scores high on \foundation{care}, it lacks a true label for \foundation{fairness}, and the false positive prediction for \foundation{fairness} should be true positive if the label was correct. An example of false negative prediction in the presence of an incorrect label is this RoT: \example{``It's fun to go swimming on a hot day.''} This example was labeled with foundations \foundation{authority} and \foundation{care} but it scores very low on all foundations according to \ourmodel.

\subsubsection{MIC}

The results for MIC are similar to SC. In particular, we find a relatively low AUC for \ourmodel, even though it performs better than all other methods. The AUC range is 0.65--0.75, with the highest performance achieved when scoring the foundations \foundation{authority} and \foundation{loyalty}. We also believe that the MIC dataset contains significant label noise due to the same reasons as SC.  A false positive example under label inaccuracy is \example{``It's interesting to learn about the shows you watch,''} which was labeled by a worker to contain the foundation \foundation{sanctity} but scores very low using \ourmodel. A false negative example is \example{``People should unite and fight against terrorism,''} which scores high on \foundation{loyalty} and should contain \foundation{loyalty} but does not according to the annotators.

\section{Analyzing Moral Discussions on Reddit using the Moral Foundations Theory}
\label{appn:aita}

In this section, we describe in more detail our study of moral stories and judgments on \texttt{r/AmItheAsshole} (AITA), a popular subreddit. Users on AITA post about a non-violent moral conflict they recently experience and invite the community to judge if they were in the right. This online community has garnered the attention of social NLP and computational social science researchers who study moral story framing and judgment from different perspectives. For example, \citet{zhouAssessingCognitiveLinguistic2021b} analyzed the relationship between linguistic features pertaining to moral role (e.g., victim or aggressor) and moral valence. \citet{botzerAnalysisMoralJudgment2022} studied language models' ability to classify moral judgments based on a described story. Recently, \citet{nguyenMappingTopics1002022b} mapped the AITA moral domain, consisting of over 100K stories, into a set of 47 topics encompassing several aspects of daily life such as \topic{marriage}, \topic{work}, \topic{appearance} and \topic{religion}. The authors found that topics are an important covariate in studying everyday moral dilemmas, and demonstrated that the framing (how a story is told) and judgment (how people perceive a story's author) vary across topics and topic pairs in non-trivial ways.

In this section, we perform two analyses. First, we replicate a study of AITA content using moral foundations by \citet{nguyenMappingTopics1002022b}. We show that the detection of foundations using the MFD 2.0, which was used in that study, significantly differs from our method, \ourmodel. This leads to non-trivial differences in the downstream results and, hence, interpretation. Second, we demonstrate the utility of the MFT in studying conflicting judgments within a thread, in which different moral valence appeal to moral foundations in distinct ways. These findings echo the prior work's argument that MFT is a useful framework to study moral judgments; however, researchers aiming to adopt the theory for studying moral content should be aware of word count's limitations.

\subsection{Moral Foundation Prevalence in Topics and Topic Pairs}
\label{appn:aita:radar}

%\begin{itemize}
%	\item Describe the dataset, especially about removing INFO posts
%	\item Summarize prior findings using MFD 2.0
%	\item Describe new scoring method
%	\item Compare the results
%\end{itemize}

In the first analysis, we replicate of the study performed by \citet{nguyenMappingTopics1002022b}. Specifically, we aim to see the difference in their downstream results of moral foundation prevalence when the posts and judgments on AITA are scored by our more performant \ourmodel. In the below, we describe this study in more detail.

\subsubsection{The AITA subreddit and dataset}

AITA is organized into discussion \emph{threads}. Each thread contains a \emph{post} made by an \emph{original poster} (OP) and \emph{comments} made by Reddit users. A post, which describes a moral conflict its OP experiences, contains a short \emph{title} of the form ``AITA (am I the a**h****)\ldots'' or ``WIBTA (would I be the a**h****)\ldots'', followed by a \emph{body text} which gives more detail to the story. Other users make comments below a post detailing their judgment and reasoning.

Five moral judgments are used on AITA: \yta (``you're the a**h***''), \nta (``not the a**h***''), \esh (``everyone sucks'') and \nah (``no a**h****s here'') and \info (``more information needed''). Each time a user makes a judgment, they must give one of these acronyms. Reddit users can upvote or downvote the post and comments, and the \emph{score} of a post/comment is the difference between its upvotes and downvotes. After 18 hours since a post was made, a Reddit bot assigns the judgment in the highest-scoring comment as the post's \emph{verdict}. 
% For an example thread on AITA, see \citep[Figure A1]{nguyenMappingTopics1002022b}.

We use the dataset released by \citet{nguyenMappingTopics1002022b},\footnote{\url{https://doi.org/10.5281/zenodo.6791835}} which contains 102,998 threads, including about 13M comments, from 2014 to 2019. For each post, we also use its highest-scoring comment as the final verdict. We remove all threads for which the final verdict is \info, since we are only interested in those with a concrete verdict. Finally, we place the rest four judgments into two groups: \YA (containing \yta and \esh, representing judgments with \emph{negative valence} against the OP) and \NA (containing \nta and \nah, representing judgments with \emph{positive valence} toward the OP). The final dataset contains 94,970 post-verdict pairs.

\subsubsection{Labeling posts and verdicts with moral foundations}

To predict the presence of moral foundations within a post or verdict, we use two methods:

\begin{itemize}
	\item Word count with MFD 2.0: this was used by \citet{nguyenMappingTopics1002022b}. Specifically, if a post/verdict contains a word that maps to foundation $f$ according to MFD 2.0, then the post/verdict is considered to contain $f$.
	\item \ourmodel: we score every post and verdict on the five foundations using \ourmodel in \Cref{sec:mf_classifiers:classifiers:roberta}. Then, for each foundation $f$, we assign a positive label to the instances in the top 20\% of the scores. That is, to contain foundation $f$, a post/verdict must score higher than at least 80\% of the dataset. 
\end{itemize}

\subsubsection{Measuring moral foundation prevalence in topics  and topic pairs}

\begin{figure}[t]
	\centering
	\includegraphics[width=0.7\linewidth]{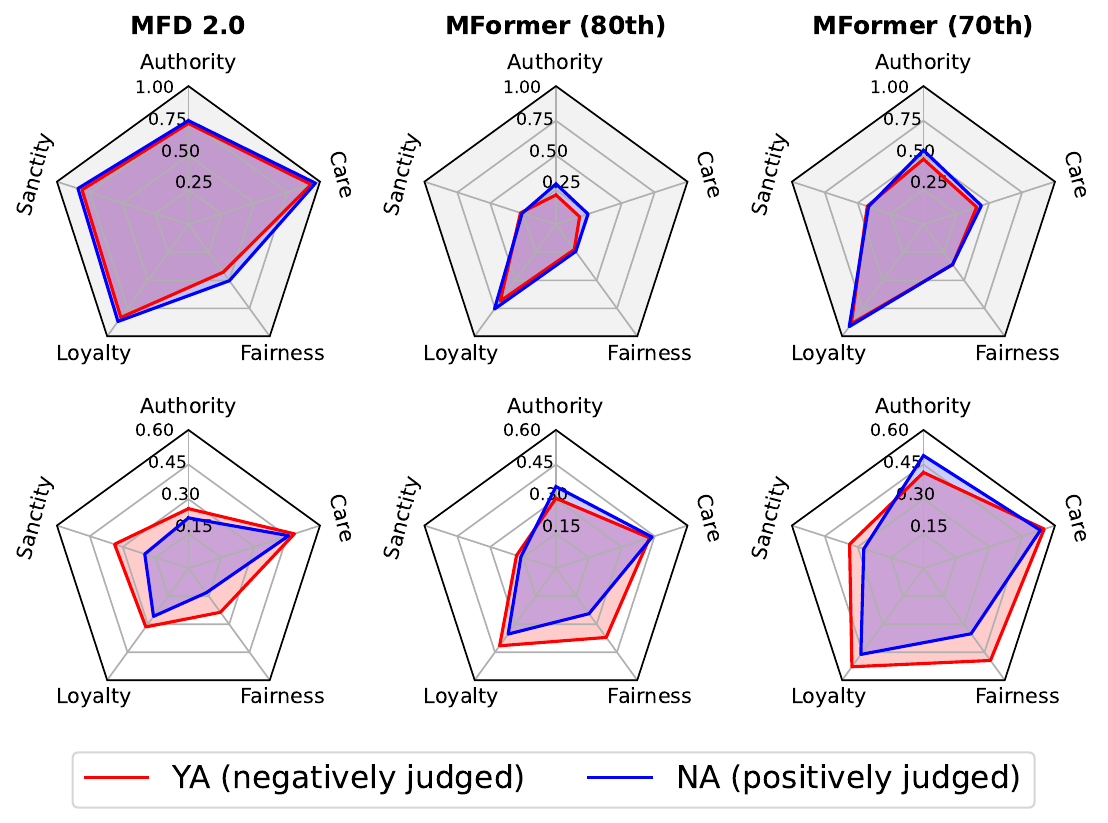}
	\caption{Posts (top) and verdicts (bottom) in the (\textit{family}, \textit{marriage}) topic pair on AITA. Each number in a radar plot indicates the proportion of posts (or verdicts) that contain each moral foundation. The moral foundations are detected by two methods: MFD 2.0 and \ourmodel. For \ourmodel, two thresholding values are displayed (80th and 70th percentiles). Red (resp. blue) indicates YA (resp. NA) valence.}
	\label{fig:aita:radar_topic_pairs_2levels}
\end{figure}

In \citet{nguyenMappingTopics1002022b}, the authors used topic modeling to map the domain AITA to 47 fine-grained topics such as \topic{family}, \topic{religion} and \topic{appearance}. They demonstrated three important points. First, a post is typically best described by two nominal topics such as \topic{family} and \topic{money}, hence the focus on \emph{topic pairs} as the thematic unit of the data. Second, topic pairs are an important covariate in AITA analysis, in which the authors found that every topic pair appeals to moral foundations in different ways. For example, when talking about \topic{family} and \topic{marriage} (\Cref{fig:aita:radar_topic_pairs}, first column), an OP tends to focus on the foundations \foundation{authority}, \foundation{care}, \foundation{loyalty} and \foundation{sanctity} but relatively downplays the aspect of \foundation{fairness}. On the other hand, verdicts in this topic pair only primarily focus on \foundation{care}. And third, moral foundations can help characterize the difference between positive (\NA) and negative (\YA) verdicts within each topic pair. Specifically, the study found that \YA judgments typically appeal to all moral foundations more often than do \NA judgments.

\begin{figure}[t]
	\centering
	\includegraphics[width=1\linewidth]{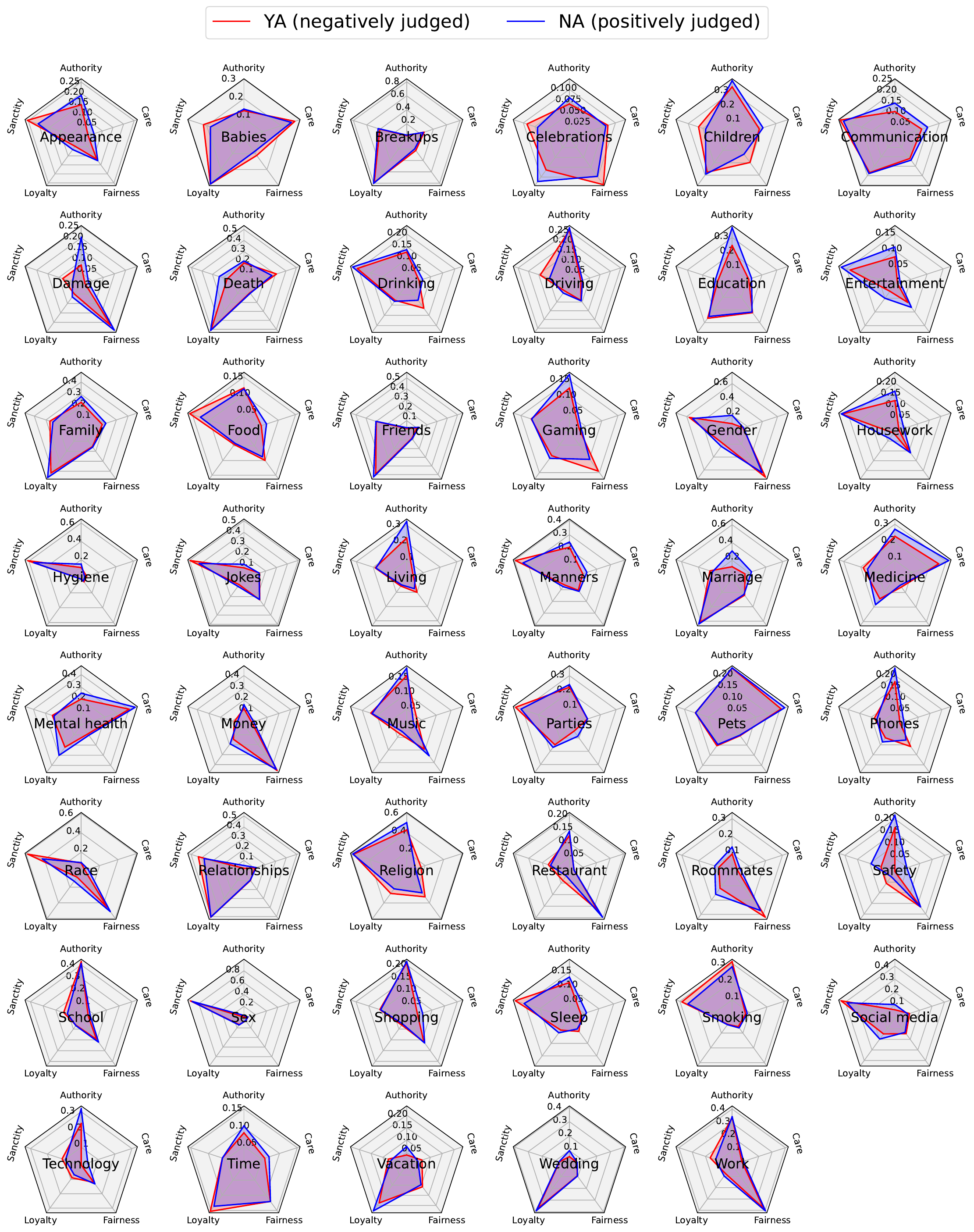}
	\caption{Prevalence of each moral foundation in each of the 47 topics on AITA. In each radar plot for a topic, each vertex represents the proportion of posts in that topic that contain the corresponding moral foundation. The moral foundations are predicted using our \ourmodel model. Blue (resp. red) pentagons correspond to NA-judged (resp. YA-judged) posts. These radar plots are reproduced from \citep[Fig. G2]{nguyenMappingTopics1002022b}, which was made using MFD 2.0.}
	\label{fig:aita:all_posts_radar}
\end{figure}

\begin{figure}[t]
	\centering
	\includegraphics[width=1\linewidth]{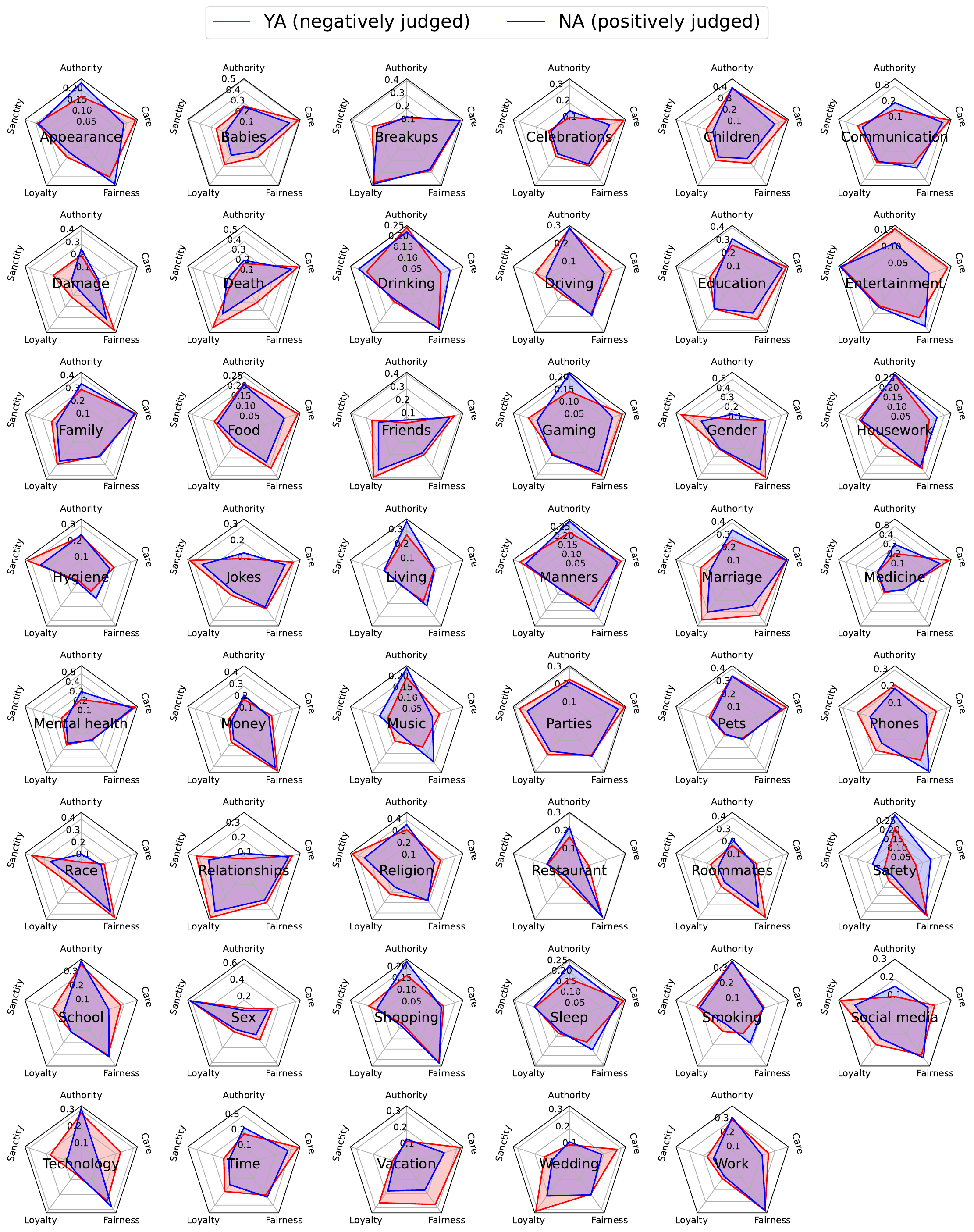}
	\caption{Prevalence of each moral foundation in each of the 47 topics on AITA. In each radar plot for a topic, each vertex represents the proportion of verdicts in that topic that contain the corresponding moral foundation. The moral foundations are predicted using our \ourmodel model. Blue (resp. red) pentagons correspond to NA (resp. YA) judgments. These radar plots are reproduced from \citep[Fig. G1]{nguyenMappingTopics1002022b}, which was made using MFD 2.0.}
	\label{fig:aita:all_verdicts_radar}
\end{figure}

We perform the same measurements, with the only difference in the moral foundation labels: instead of using MFD 2.0, we replace them with \ourmodel-predicted labels. Then, for each topic pair, we count the number of posts that contain a foundation $f$ and divide that by the total number of posts in that topic pair. This ratio gives the \emph{prevalence} for $f$, which is interpreted as the frequency with which a post in this topic pair is about $f$. We do the same for the verdicts. Finally, we also separated the posts and verdicts into \YA and \NA classes to investigate any difference between them. The prevalence of moral foundations within each topic is presented in \Cref{fig:aita:all_posts_radar} (for posts) and \Cref{fig:aita:all_verdicts_radar} (for verdicts). Additionally, \Cref{fig:aita:radar_topic_pairs} in the main text displays the foundation prevalence for posts and verdicts in the topic pair (\topic{family}, \topic{marriage}).

\subsubsection{Results}

We first compare the differences in foundation prevalence reported by \citet{nguyenMappingTopics1002022b}---who used the MFD 2.0 to infer the foundations within each post and verdict---and that produced using \ourmodel, which is in \Cref{fig:aita:all_posts_radar,fig:aita:all_verdicts_radar} in this paper. In most topics, the prevalence of moral foundations changes dramatically. For example, \citet{nguyenMappingTopics1002022b} showed that among posts in the topic \topic{family}, most of them are concerned with \foundation{care}, \foundation{loyalty}, and \foundation{sanctity}. However, in our result, we find that the only dominating foundation is \foundation{loyalty}, whereas the rest four foundations are somewhat equal in their prevalence. In another example, within verdicts in the topic \topic{roommates}, \citet{nguyenMappingTopics1002022b} found that foundation \foundation{care} is the most relevant; we, on the other hand, find that it is \foundation{fairness} that verdicts tend to highlight the most.

This difference is also observable, even to a larger extent, when topic pairs are considered. In addition to our discussion of \Cref{fig:aita:radar_topic_pairs} in the main text, here we also show that the relative importance of moral foundations produced by \ourmodel is not sensitive to the cutoff level of 80th percentile. In \Cref{fig:aita:radar_topic_pairs_2levels}, we present in the last column the same radar plots but at the 70th-percentile cutoff level, i.e., for each post/verdict to contain a foundation, it must score higher than at least 70\% of all posts/verdicts. The pentagons' shapes on the second and third columns of \Cref{fig:aita:radar_topic_pairs_2levels} do not change, indicating that they are not sensitive to the binary classification threshold of \ourmodel's predictions.

\subsection{Characterizing Conflicting Judgments in Highly Controversial Discussions}
\label{appn:aita:aita_filtered}

%\begin{itemize}
%	\item Describe AITA Filtered condition
%	\item Describe how scoring of an original post and its judgments is done
%	\item Present the result: an example thread with conflicting judgments + total number of threads that could be characterized by MFT
%\end{itemize}

In the second, and novel, analysis, we aim to investigate in more detail the \emph{positive} (\NA) and \emph{negative} (\YA) judgments on AITA from the perspective of the MFT. Specifically, we hypothesize that in a controversial, highly conflicting thread, there is a systematic difference in the way \NA judgments are made compared to \YA judgments when it comes to which moral foundations they adhere to. Below we give more detail of this analysis.

\subsubsection{Dataset}

We focus on threads with highly conflicting communities of judgment. To do so, we search over all threads in the dataset by \citet{nguyenMappingTopics1002022b} and only keep threads satisfying the following conditions:
\begin{itemize}
	\item Only comments within 18 hours from the time the original post was made are considered.\footnote{This condition is consistent with the way the AITA bot determines the final verdict for each post. We do aim to argue that the comments made after this period are less important.}
	\item Only top-level comments are considered. That is, these comments must reply directly to their original post but not to another comment.
	\item Only comments with a valid judgment (\yta, \nta, \esh, \nah) are kept. We remove posts labeled with \info as they do not present a concrete judgment. Then, \yta and \esh comments are grouped into \YA judgments of negative valence. Similarly, \nta and \nah comments are grouped into \NA judgments of positive valence.
	\item To be considered controversial, each thread must contain at least 50 such judgments. Further, the proportion of \YA or \NA judgments in a thread must not exceed 70\% of all judgments.\footnote{This condition is inspired by the ``AITA Filtered'' subreddit: \url{https://www.reddit.com/r/AITAFiltered}.}
\end{itemize}

Based on these, we obtain a dataset of 2,135 threads with a total of 466,485 judgments. Each thread contains an original post and a median of 110 judgments (max = 3,251).

\subsubsection{Labeling posts and judgments with moral foundations}

We use the \ourmodel models to score every post and judgment on the five moral foundations. Similar to before, we label a post/judgment with a foundation $f$ if it scores in the top 20\% of the dataset. In other words, to contain $f$ a post/judgment must score higher than at least 80\% of all posts or judgments.

%\begin{table}
%	\centering
%	\caption{Number (and percentage) of AITA controversial threads giving statistically significant chi-square results at three levels: $p < 0.05$, $p < 0.01$ and $p < 0.001$.}
%	\begin{tabular}{lccc}
%		\toprule
%		Foundation &   * $p < 0.05$ &   * $p < 0.01$ & * $p < 0.001$ \\
%		\midrule
%		Authority               &  366 (17.1\%) &   197 (9.2\%) &   88 (4.1\%) \\
%		Care                    &  496 (23.2\%) &  315 (14.8\%) &  180 (8.4\%) \\
%		Fairness                &  435 (20.4\%) &  236 (11.1\%) &  126 (5.9\%) \\
%		Loyalty                 &  421 (19.7\%) &  235 (11.0\%) &  129 (6.0\%) \\
%		Sanctity                &  399 (18.7\%) &  232 (10.9\%) &  127 (5.9\%) \\
%		\midrule
%		At least one foundation &  1188 (55.6\%) &  747 (35.0\%) &  412 (19.3\%) \\
%		\bottomrule
%	\end{tabular}
%\end{table}

\subsubsection{Comparing \YA and \NA judgments based on moral foundations}

To estimate the difference in their appeal to moral foundations between \YA and \NA judgments in one thread, we use the odds ratio (OR) defined as
\begin{equation*}
	\text{OR}_f = \frac{N_{f, \texttt{\NA}} \cdot N_{\neg f, \texttt{\NA}}}{N_{\neg f, \texttt{\NA}} \cdot N_{f, \texttt{\YA}}},
\end{equation*}
where
\begin{itemize}
	\item $f$ is one of the five foundations;
	\item $N_{f, \texttt{\NA}}$ is the number of \NA judgments labeled \emph{with} foundation $f$;
	\item $N_{\neg f, \texttt{\YA}}$ is the number of \YA judgments labeled \emph{without} foundation $f$;
	\item $N_{\neg f, \texttt{\NA}}$ is the number of \NA judgments labeled \emph{without} foundation $f$; and
	\item $N_{f, \texttt{\YA}}$ is the number of \YA judgments labeled \emph{with} foundation $f$.
\end{itemize}
A value of OR greater than 1 (resp. less than 1) means that the presence of foundation $f$ in a judgment raises the odds that the judgment is \NA (resp. \YA). 
% In other words, if $\text{OR}_f > 1$, \NA judgments are more associated with foundation $f$ than are \YA judgments. 
For example, if $\text{OR}_{\text{\foundation{loyalty}}} = 2.5$, then the presence of the foundation \foundation{loyalty} raises the odds that a judgment is \NA (i.e., positive toward the author) by 2.5 times. On the other hand, if $\text{OR}_{\text{\foundation{loyalty}}} = 0.2$, then the presence of \foundation{loyalty} raises the odds that a judgment is \YA (i.e., negative toward the author) by $1 / 0.2 = 5$ times. Finally, to establish statistical significance, we can estimate the standard error of the natural logarithm of the OR as
\begin{equation*}
	\text{SE} = \sqrt{ \frac{1}{N_{f, \texttt{\NA}}} + \frac{1}{N_{\neg f, \texttt{\YA}}} + \frac{1}{N_{\neg f, \texttt{\NA}}} + \frac{1}{N_{f, \texttt{\YA}}}}.
\end{equation*}
The log OR can be approximated by the normal distribution $\mathcal{N}(\log{(\text{OR}_f)}, \text{SE}^2)$ and, hence, the 95\% confidence interval for the log OR is $(\log(\text{OR}_f) - 1.96 \text{SE}, \log(\text{OR}_f) + 1.96 \text{SE})$. In presenting the results, we often use the following two conventions:
\begin{itemize}
	\item If $\text{OR}_f > 1$, that is, when $f$ is associated with \NA judgments, we present the OR and its 95\% CI as $\text{OR}_f$ and $(\text{OR}_f - \exp(1.96 \text{SE}), \text{OR}_f + \exp(1.96 \text{SE}))$, respectively.
	\item If $\text{OR}_f < 1$, that is, when $f$ is associated with \YA judgments, we present the OR its 95\% CI as $\frac{1}{\text{OR}_f}$ and $\left( \frac{1}{\text{OR} + \exp(1.96 \text{SE})}, \frac{1}{\text{OR} - \exp(1.96 \text{SE})} \right)$, respectively. In this case, the OR represents the number of times the odds of a \YA judgment is raised in the presence of foundation $f$.
\end{itemize}
If the 95\% CI does not contain 1, then the OR is said to be significant at the 0.05 level.

\section{Moral Foundations and Stance toward Controversial Topics}
\label{appn:twitter_stance}

Stance, in general, can be understood as a person's opinion---in favor, against or neutral---toward a proposition or target topic \citep{mohammadStanceSentimentTweets2017}. In this paper, we analyze the association between people's stances on some controversial topics and moral foundations. We use the dataset by \citet{mohammadSemEval2016TaskDetecting2016}, which contains three components in each of its instances: a topic, a tweet, and the stance of the tweet's author toward the topic. We aim to reproduce some of the results by \citet{rezapourIncorporatingMeasurementMoral2021} using our new method of detecting moral foundations, i.e., \ourmodel. We also present new results in light of our classifiers.

\begin{table}[]
	\centering
	\caption{Example tweets and their authors' stance on some controversial topics.}
	\begin{tabular}{lccp{0.6\linewidth}}
		\toprule
		Topic                              & Stance   & \# Tweets & Example \\ 
		\midrule
		\multirow{2}{*}{Atheism}           & In favor & 124       &   \example{Now that the SCOC has ruled Canadians have freedom from religion, can someone tell Harper to dummy his 'god bless Canada'. \#cdnpoli}      \\
		& Against  & 464       &      \example{dear lord thank u for all of ur blessings forgive my sins lord give me strength and energy for this busy day ahead \#blessed \#hope}   \\
		\midrule
		\multirow{2}{*}{Climate Change}    & In favor & 335       &   \example{We cant deny it, its really happening.}      \\
		& Against  & 26        &     \example{The Climate Change people are disgusting assholes. Money transfer scheme for elite. May you rot. }    \\
		\midrule
		\multirow{2}{*}{Donald Trump}      & In favor & 148       &     \example{Donald Trump isn't afraid to roast everyone.}    \\
		& Against  & 299       &      \example{@ABC Stupid is as stupid does! Showedhis true colors; seems that he ignores that US was invaded, \& plundered,not discovered}   \\
		\midrule
		\multirow{2}{*}{Feminist Movement} & In favor & 268       &     \example{Always a delight to see chest-drumming alpha males hiss and scuttle backwards up the wall when a feminist enters the room. \#manly}    \\
		& Against  & 511       &    \example{If feminists spent 1/2 as much time reading papers as they do tumblr they would be real people, not ignorant sexist bigots.}     \\
		\midrule
		\multirow{2}{*}{Hillary Clinton}   & In favor & 163       &   \example{Hillary is our best choice if we truly want to continue being a progressive nation. \#Ohio}      \\
		& Against  & 565       &    \example{@tedcruz And, \#HandOverTheServer she wiped clean + 30k deleted emails, explains dereliction of duty/lies re \#Benghazi,etc \#tcot}     \\
		\midrule
		\multirow{2}{*}{Abortion}          & In favor & 167       &   \example{@tooprettyclub Are you OK with \#GOP males telling you what you can and can't do with your own body?}      \\
		& Against  & 544       &     \example{Just laid down the law on abortion in my bioethics class. \#Catholic}   \\
		\bottomrule
	\end{tabular}
	\label{table:twitter_stance:examples}
\end{table}

\subsection{Dataset}
\label{appn:twitter_stance:dataset}

This data contains 4,870 instances each with three components: a topic, a tweet, and an annotated stance. Six ``controversial'' topics are considered: \topic{Atheism}, \topic{Climate Change is a Real Concern} (henceforth \topic{Climate Change}), \topic{Donald Trump}, \topic{Feminist Movement}, \topic{Hillary Clinton} and \topic{Legalization of Abortion} (henceforth \topic{Abortion}). Each tweet was annotated with a stance toward the topic it is about: whether the author of the tweet is in favor, against, or neither toward the topic.

For each topic, our goal is to characterize the difference between those in favor of and those against it; therefore, we remove all tweets labeled with ``None'' as its author's stance, resulting in 3,614 tweets left. \Cref{table:twitter_stance:examples} presents some example tweets for each stance in every topic as well as the number of tweets for each stance. In all topics except for \topic{climate change}, there are more tweets against it than there are tweets in favor of it.

Finally, we score every tweet using our five \ourmodel moral foundation classifiers.

\begin{table}[]
	\centering
	\caption{Comparison moral foundation scores (produced by fine-tuned \ourmodel models) between tweets in favor and those against each controversial topic. \textbf{F} \textgreater A indicates that a randomly chosen tweet in favor of a topic scores significantly higher than a randomly chosen tweet against that topic. Similar for F \textless \textbf{A}. The higher-scoring stance is in \textbf{bold}. Statistical significance is established by the two-sided Mann-Whitney U test using the asymptotic method with continuity correction. * $p < 0.05$, ** $p < 0.01$, *** $p < 0.001$. All insignificant results at the 0.05 level are replaced by the ``--'' symbol.}
	\begin{tabular}{l | cc | lllll}
		\toprule
		\multirow{2}{*}{Topic} & \multirow{2}{*}{\# Favor} & \multirow{2}{*}{\# Against} & \multicolumn{5}{c}{Moral foundations}                                                                \\ 
		&                           &                             & Authority           & Care            & Fairness            & Loyalty             & Sanctity         \\
		\midrule
		Atheism           & 124      & 464        & \textbf{F} \textgreater A*   & F \textless \textbf{A}** & \textbf{F} \textgreater A*** & F \textless \textbf{A}***    & F \textless \textbf{A}*** \\
		Climate Change    & 335      & 26         & F \textless \textbf{A}*      & --              & F \textless \textbf{A}*      & --                  & --               \\
		Donald Trump      & 148      & 299        & \textbf{F} \textgreater A*** & F \textless \textbf{A}** & F \textless \textbf{A}**     & \textbf{F} \textgreater A*** & --               \\
		Feminist Movement & 268      & 511        & F \textless \textbf{A}*      & F \textless \textbf{A}*  & --                  & --                  & --               \\
		Hillary Clinton   & 163      & 565        & F \textless \textbf{A}***    & --              & F \textless \textbf{A}***    & --                  & F \textless \textbf{A}*** \\
		Abortion          & 167      & 544        & --                  & F \textless \textbf{A}*  & \textbf{F} \textgreater A*** & --                  & F \textless \textbf{A}*** \\
		\bottomrule
	\end{tabular}
	\label{tab:twitter_stance:mwu}
\end{table}

\subsection{Comparing Conflicting Stances on Each Topic Using Moral Foundation Scores}
\label{appn:twitter_stance:mwu}

Here we examine the difference in foundation scores for tweets in favor of a topic and those against it. Formally, for each topic $t$ and foundation $f$, we compare \ourmodel scores when predicting whether $f$ exists in two groups: tweets in favor of topic $t$ and those against topic $t$. We choose the Mann-Whitney U test to examine statistically significant differences in foundation scores between the groups.

\Cref{tab:twitter_stance:mwu} presents the results. For some topics, there are clear differences between the two conflicting stances. For example, on the topic \topic{atheism}, tweets in favor of it score significantly higher on \foundation{authority} ($p < 0.05$) and \foundation{fairness} ($p < 0.001$). These tweets often criticize the political authority of religion, as in the tweet \example{``Religious leaders are like political leaders - they say what they think people want to hear.  \#freethinker''}; some also focus on discrimination, e.g., \example{``if u discriminate based on ur religion, be ready to be discriminated against for having that religion.''} On the other hand, those against this topic score higher on \foundation{care} ($p < 0.01$), \foundation{loyalty} ($p < 0.001$), and \foundation{sanctity} ($p < 0.001$). They often focus on the virtues of compassion, faithfulness, and holiness such as the tweet \example{``Give us this day our daily bread, and forgive us our sins as we forgive those who sin against us. \#rosary \#God \#teamjesus.''}

For some other topics, we only find one-way difference between the stances. For instance, on the topic \topic{feminist movement}, tweets against it score significantly higher than those in favor of it on two foundations: \foundation{authority} ($p < 0.05$) and \foundation{care} ($p < 0.05$).  Those who voice their disagreement with the movement often mention government authority such as in the tweet \example{``I hate government, but if I were in government i'd want to be a District Attorney or a Judge to hold \#YesAllWomen accountable.''}. However, there is no significant evidence suggesting that tweets in favor of the movement appeal to any foundation more than those against it.

\subsection{Comparison Based on Binary Predictions}
\label{appn:twitter_stance:ORs}

In this subsection, we estimate the effect sizes of the difference between tweets in favor and those against a topic with respect to their adherence to moral foundations. First, we convert all raw scores outputted by \ourmodel into \emph{binary labels} as follows. For each foundation $f$, we give the binary label 1 to the highest-scoring 20\% of the tweets. In other words, to be considered to contain $f$, a tweet must score higher than at least 80\% of all tweets in the dataset. We choose a rather high threshold as we prefer high precision at the cost of a lower recall. Refer to \Cref{appn:roberta_eval:thresholding} for a more detailed discussion on setting binary classification thresholds for \ourmodel; results on the test set suggest that setting the threshold at the 80th percentile is reasonable as it gives both high F-1 and accuracy.

\begin{table}[]
	\centering
	\caption{Results of the chi-square test for the independence of moral foundations and stance (in favor or against) toward a controversial topic. The columns denoted by ``MFD'' give the results presented in \citep[Table 5]{rezapourIncorporatingMeasurementMoral2021}. The columns denoted by ``\ourmodel'' are the results based on the binary labels predicted by our \ourmodel models. * $p < 0.05$, ** $p < 0.01$, *** $p < 0.001$. All insignificant results at the 0.05 level are replaced by the ``--'' symbol.}
	\begin{tabular}{l|cc|cc|cc|cc|cc}
		\toprule
		\multirow{2}{*}{Topic} & \multicolumn{2}{c|}{Authority} & \multicolumn{2}{c|}{Care} & \multicolumn{2}{c|}{Fairness} & \multicolumn{2}{c|}{Loyalty} & \multicolumn{2}{c}{Sanctity} \\
		& MFD         & \ourmodel          & MFD       & \ourmodel       & MFD          & \ourmodel        & MFD        & \ourmodel         & MFD           & \ourmodel       \\ 
		\midrule
		Atheism                & --          & --               & --        & --            & --           & 11.82***       & --         & 47.58***        & --            & 46.22***      \\
		Climate       & --          & --               & --        & --            & --           & 16.55***       & --         & --              & --            & --            \\
		Trump           & --          & 51.77***         & --        & 5.90*         & --           & --             & --         & 34.05***        & --            & --            \\
		Feminist      & --          & --               & --        & --            & --           & --             & --         & --              & --            & --            \\
		Clinton        & --          & --               & --        & --            & --           & 17.55***       & --         & --              & 13.49**       & --            \\
		Abortion               & --          & --               & --        & 12.21***      & 6.84*        & 11.05***       & --         & 5.72*           & --            & 15.13***  \\
		 \bottomrule
	\end{tabular}
	\label{tab:twitter_stance:chisquare}
\end{table}

\subsubsection{Significance of association between stance and moral foundations}

We replicate a prior analysis by \citet{rezapourIncorporatingMeasurementMoral2021} in which the authors used the chi-square test to examine the dependence between two binary variables: whether a foundation is found in a tweet and whether the tweet is in favor or against some topic. Whereas \citet{rezapourIncorporatingMeasurementMoral2021} labeled each tweet with a moral foundation using the MFD,\footnote{The authors used a slightly modified version of the MFD in which each word is annotated with its part of speech \citep{rezapourEnhancingMeasurementSocial2019}.} which has been shown in this paper to produce misleading results, we use the labels produced by threshold \ourmodel prediction scores described above. We use the scipy implementation of this test and invoke Yates' correction for continuity.

The results of this significance test are given in \Cref{tab:twitter_stance:chisquare}. The entries in the ``MFD'' columns are taken from \citep[Table 5]{rezapourIncorporatingMeasurementMoral2021} and those based on our method are in the ``\ourmodel'' columns. We first observe some similar results: for example, \citet{rezapourIncorporatingMeasurementMoral2021} found no significant result for the topic \topic{feminist movement}, which agrees with our analysis. Similarly, the prior finding that there is a significant dependence between the foundation \foundation{fairness} and the stance toward the topic \topic{abortion} agrees with our result, although our result yields a higher level of significance ($p < 0.001$ compared to $p < 0.05$).

However, we find that a lot of significant dependencies discovered by \ourmodel were overlooked by the prior study. For example, in \citep{rezapourIncorporatingMeasurementMoral2021}, the authors found no significant results for the topics \topic{Donald Trump} and \topic{climate change}. Our findings, on the other hand, show that very strong associations in these topics exist. For example, on the topic \topic{climate change}, we find a strong relationship between stance and \foundation{fairness} ($p < 0.001$). On the topic \topic{Donald Trump}, three foundations correlate significantly with stance: \foundation{authority} ($p < 0.001$), \foundation{care} ($p < 0.05$) and \foundation{loyalty} ($p < 0.001$).

\subsubsection{Odds ratios between moral foundations and stance toward a topic} 

Finally, to quantify the association between the presence of a moral foundation in a tweet and the tweet's stance toward a topic, we calculate the odds ratios (ORs) as follows. In each topic, for foundation $f$, the OR between $f$ and stance is
\begin{equation*}
	\text{OR}_f = \frac{N_{f, \texttt{favor}} \cdot N_{\neg f, \texttt{against}}}{N_{\neg f, \texttt{favor}} \cdot N_{f, \texttt{against}}},
\end{equation*}
where
\begin{itemize}
	\item $N_{f, \texttt{favor}}$ is the number of tweets labeled \emph{with} foundation $f$ and are \emph{in favor} of the topic;
	\item $N_{\neg f, \texttt{against}}$ is the number of tweets labeled \emph{without} foundation $f$ and are \emph{against} the topic;
	\item $N_{\neg f, \texttt{favor}}$ is the number of tweets labeled \emph{without} foundation $f$ and are \emph{in favor} of the topic; and
	\item $N_{f, \texttt{against}}$ is the number of tweets labeled \emph{with} foundation $f$ and are \emph{against} the topic.
\end{itemize}
If there is no association between $f$ and stance, the OR equals 1. A value of OR greater than 1  (resp. less than 1) means that the presence of foundation $f$ in a tweet raises the odds that the tweet is in favor of (resp. against) the target topic. The standard error for the natural logarithm of the OR is
\begin{equation*}
	\text{SE} = \sqrt{ \frac{1}{N_{f, \texttt{favor}}} + \frac{1}{N_{\neg f, \texttt{against}}} + \frac{1}{N_{\neg f, \texttt{favor}}} + \frac{1}{N_{f, \texttt{against}}}}.
\end{equation*}
We calculate the 95\% confidence interval (CI) of the OR as $(\text{OR} - \exp(1.96 \text{SE}), \text{OR} + \exp(1.96 \text{SE}))$. If this interval does not contain 1, we consider the strength of the relationship statistically significant. Equivalently, we present the log ORs and their CIs: if an interval contains 0 the OR is significant.

\Cref{fig:twitter_stance:odds_ratios} presents the log ORs with their 95\% CIs for all moral foundations in each topic. Significant values are plotted with colors: blue for positive (i.e., a moral foundation is associated with the ``in favor'' stance) and red for negative (association with the ``against'' stance). We note that the missing value for the foundation \foundation{loyalty} on the topic \topic{climate change} is because no tweet was predicted to contain \foundation{loyalty}.

\newpage
\nociteAP{*}
\bibliographystyleAP{aaai22}
{\bibliographyAP{appendix-bib.bib}}

\end{document}